\documentclass[
    letterpaper,
    twocolumn,
    superscriptaddress,
    floatfix
]{revtex4-2}

\usepackage[utf8]{inputenc}
\usepackage{bm}
\usepackage{multirow,amssymb,amsbsy,amsmath}
\usepackage{graphicx}
\usepackage{float}
\usepackage{verbatim}
\usepackage{color}
\usepackage{pifont}
\usepackage{upgreek}
\usepackage{indentfirst}
\usepackage{soul}
\usepackage{braket}
\usepackage{xcolor}
\usepackage{makecell}
\usepackage{ragged2e}

\usepackage{comment}

\usepackage[
    colorlinks=true,
    linkcolor=blue,
    citecolor=blue,
    urlcolor=blue
]{hyperref}

\newcommand{\Rmnum}[1]{\expandafter\@slowromancap\romannumeral #1@}

\makeatletter

\long\def\@makecaption#1#2{%
    \vskip\abovecaptionskip
    \begingroup
    \small
    \setlength{\parindent}{0pt}%
    \setlength{\parskip}{0pt}%
    \justifying
    \noindent\textbf{#1.}\ #2\par
    \endgroup
    \vskip\belowcaptionskip
}

\makeatother

\makeatletter
\renewcommand{\section}{%
    \@startsection{section}{1}{0mm}
    {-\baselineskip}
    {0.5\baselineskip}
    {\bfseries\raggedright}
}
\makeatother

\begin{document}

\title{Coherent Spin-Photon Interface of Single PL6 Oxygen–Vacancy Centers in 4H-SiC}

\author{Zhen-Xuan He}
\altaffiliation{These authors contributed equally to this work.}
\affiliation{Laboratory of Quantum Information, University of Science and Technology of China, Hefei, Anhui 230026, China}
\affiliation{Anhui Province Key Laboratory of Quantum Network, University of Science and Technology of China, Hefei, Anhui 230026, China}
\affiliation{CAS Center For Excellence in Quantum Information and Quantum Physics, University of Science and Technology of China, Hefei, Anhui 230026, China}
\affiliation{Hefei National Laboratory, University of Science and Technology of China, Hefei, Anhui 230088, China}

\author{Gerg\H{o} Thiering}
\altaffiliation{These authors contributed equally to this work.}
\affiliation{HUN-REN Wigner Research Centre for Physics, P.O.\ Box 49, H-1525 Budapest, Hungary}

\author{Xiangru Han}
\affiliation{HUN-REN Wigner Research Centre for Physics, P.O.\ Box 49, H-1525 Budapest, Hungary}

\author{Rui-Jian Liang}
\affiliation{Laboratory of Quantum Information, University of Science and Technology of China, Hefei, Anhui 230026, China}
\affiliation{Anhui Province Key Laboratory of Quantum Network, University of Science and Technology of China, Hefei, Anhui 230026, China}
\affiliation{CAS Center For Excellence in Quantum Information and Quantum Physics, University of Science and Technology of China, Hefei, Anhui 230026, China}

\author{Ji-Yang Zhou}
\affiliation{Laboratory of Quantum Information, University of Science and Technology of China, Hefei, Anhui 230026, China}
\affiliation{Anhui Province Key Laboratory of Quantum Network, University of Science and Technology of China, Hefei, Anhui 230026, China}
\affiliation{CAS Center For Excellence in Quantum Information and Quantum Physics, University of Science and Technology of China, Hefei, Anhui 230026, China}


\author{Shuo Ren}
\affiliation{Laboratory of Quantum Information, University of Science and Technology of China, Hefei, Anhui 230026, China}
\affiliation{Anhui Province Key Laboratory of Quantum Network, University of Science and Technology of China, Hefei, Anhui 230026, China}
\affiliation{CAS Center For Excellence in Quantum Information and Quantum Physics, University of Science and Technology of China, Hefei, Anhui 230026, China}
\affiliation{Hefei National Laboratory, University of Science and Technology of China, Hefei, Anhui 230088, China}

\author{Wu-Xi Lin}
\affiliation{Laboratory of Quantum Information, University of Science and Technology of China, Hefei, Anhui 230026, China}
\affiliation{Anhui Province Key Laboratory of Quantum Network, University of Science and Technology of China, Hefei, Anhui 230026, China}
\affiliation{CAS Center For Excellence in Quantum Information and Quantum Physics, University of Science and Technology of China, Hefei, Anhui 230026, China}
\affiliation{Hefei National Laboratory, University of Science and Technology of China, Hefei, Anhui 230088, China}
	
\author{Zhi-He Hao}
\affiliation{Laboratory of Quantum Information, University of Science and Technology of China, Hefei, Anhui 230026, China}
\affiliation{Anhui Province Key Laboratory of Quantum Network, University of Science and Technology of China, Hefei, Anhui 230026, China}
\affiliation{CAS Center For Excellence in Quantum Information and Quantum Physics,
University of Science and Technology of China, Hefei, Anhui 230026, China}

\author{Qi-Cheng Hu}
\affiliation{Laboratory of Quantum Information, University of Science and Technology of China, Hefei, Anhui 230026, China}
\affiliation{Anhui Province Key Laboratory of Quantum Network, University of Science and Technology of China, Hefei, Anhui 230026, China}
\affiliation{CAS Center For Excellence in Quantum Information and Quantum Physics, University of Science and Technology of China, Hefei, Anhui 230026, China}
\affiliation{Hefei National Laboratory, University of Science and Technology of China, Hefei, Anhui 230088, China}

\author{Jun-Feng Wang}
\affiliation{College of Physics, Sichuan University, Chengdu, Sichuan 610065, China}

\author{Adam Gali}
\altaffiliation{Email: gali.adam@wigner.hun-ren.hu}
\affiliation{HUN-REN Wigner Research Centre for Physics, P.O.\ Box 49, H-1525 Budapest, Hungary}
\affiliation{Department of Atomic Physics, Institute of Physics, Budapest University of Technology and Economics, 
M\H{u}egyetem rakpart 3., H-1111 Budapest, Hungary}
\affiliation{MTA-WFK Lend\"ulet ``Momentum'' Semiconductor Nanostructures Research Group, P.O.\ Box 49, H-1525 Budapest, Hungary}

\author{Jin-Shi Xu}
\altaffiliation{Email: jsxu@ustc.edu.cn}
\affiliation{Laboratory of Quantum Information, University of Science and Technology of China, Hefei, Anhui 230026, China}
\affiliation{Anhui Province Key Laboratory of Quantum Network, University of Science and Technology of China, Hefei, Anhui 230026, China}
\affiliation{CAS Center For Excellence in Quantum Information and Quantum Physics, University of Science and Technology of China, Hefei, Anhui 230026, China}
\affiliation{Hefei National Laboratory, University of Science and Technology of China, Hefei, Anhui 230088, China}

\author{Chuan-Feng Li}
\altaffiliation{Email: cfli@ustc.edu.cn}
\affiliation{Laboratory of Quantum Information, University of Science and Technology of China, Hefei, Anhui 230026, China}
\affiliation{Anhui Province Key Laboratory of Quantum Network, University of Science and Technology of China, Hefei, Anhui 230026, China}
\affiliation{CAS Center For Excellence in Quantum Information and Quantum Physics, University of Science and Technology of China, Hefei, Anhui 230026, China}
\affiliation{Hefei National Laboratory, University of Science and Technology of China, Hefei, Anhui 230088, China}
 
\author{Guang-Can Guo}
\affiliation{Laboratory of Quantum Information, University of Science and Technology of China, Hefei, Anhui 230026, China}
\affiliation{Anhui Province Key Laboratory of Quantum Network, University of Science and Technology of China, Hefei, Anhui 230026, China}
\affiliation{CAS Center For Excellence in Quantum Information and Quantum Physics, University of Science and Technology of China, Hefei, Anhui 230026, China}
\affiliation{Hefei National Laboratory, University of Science and Technology of China, Hefei, Anhui 230088, China}

\begin{abstract}

The oxygen-vacancy PL6 center in 4H-SiC shows a far higher optically detected magnetic resonance contrast than the divacancy, for reasons that have been unclear. Strain-dependent photoluminescence-excitation spectroscopy of single centers yields its excited-state Hamiltonian and state-resolved intersystem-crossing rates, and first-principles calculations trace the enhanced, more spin-selective intersystem crossing to the oxygen-modified excited state. The centers remain photostable under resonant excitation, enabling 99.7\% initialization, coherent optical and spin control, and 5.7~ms coherence.

\end{abstract}
    
\maketitle
  
\date{\today}

Optically addressable solid-state spins provide a route to quantum interfaces that combine local spin storage and control with optical access. Silicon carbide (SiC) is a promising host for such interfaces because it supports a diverse range of spin-active defects within a mature semiconductor platform. Representative systems include silicon vacancies~\cite{widmann2015coherent,nagy2018quantum,Morioka2022SpinOpticalDA,Liu2024silicon,fuchs2015engineering,simin2017locking}, nitrogen-vacancy centers \cite{Bardeleben2016NVCI,wang2020coherent,mu2020coherent}, transition-metal ions \cite{Diler2019CoherentCA,Wolfowicz2019VanadiumSQ,cilibrizzi2023ultra}, divacancies \cite{koehl2011room,falk2013polytype,christle2015isolated,christle2017isolated,miao2019electrically,Anderson2019ElectricalAO,Bourassa2020EntanglementAC}, and related vacancy complexes \cite{Son2022Modified,Li2020RoomtemperatureCM,yan2020room,He2024robust}. Long spin-coherence times~\cite{christle2015isolated,Bourassa2020EntanglementAC,anderson2022five}, spin-selective optical transitions \cite{christle2017isolated,nagy2019high}, and electrical controllability \cite{miao2019electrically,Anderson2019ElectricalAO,Nishikawa2025coherent,Timo2025Schottky} make these centers attractive building blocks for integrated quantum technologies.

The PL6 center in 4H-SiC is particularly promising because of its high brightness, large optical spin contrast, and robust charge-state stability~\cite{Li2020RoomtemperatureCM,He2024robust}. Recent first-principles calculations and oxygen-isotope-resolved hyperfine spectroscopy have identified PL6 as an axial oxygen–vacancy complex O$_\text{C}$(h)V$_\text{Si}$(h), hereafter denoted hh-OV, consisting of a substitutional oxygen atom on a carbon site adjacent to a silicon vacancy, both at hexagonal (h) lattice sites~\cite{Bai2025,Hu2026,Chen2026Oxygen}. Despite this structural assignment, its excited-state fine structure and spin-dependent optical decay pathways remain unresolved. In particular, the mechanisms underlying its large optical spin contrast and favorable optical cycling have not been quantitatively established. Resolving these questions is essential for connecting the microscopic electronic structure of PL6 to its optical spin preparation, readout, and coherent-interface capabilities.

\begin{figure*}[htbp]
\centering
\includegraphics[scale=1]{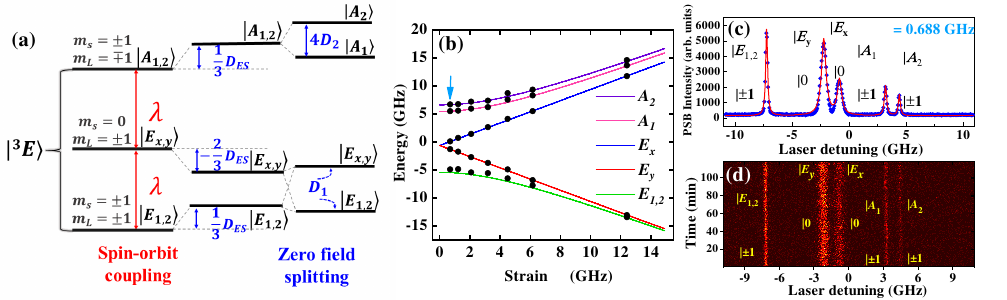}
\caption{\textbf{Excited-state Hamiltonian of the hh-OV PL6 center.}
(a) Energy-level diagram of the \(^{3}\mathrm{E}\) excited state, showing splittings generated by spin-orbit coupling \(\lambda\) and spin-spin interactions \(D_{\mathrm{ES}}\), \(D_1\), and \(D_2\).
(b) Strain-dependent evolution of the excited-state manifold as a function of transverse strain \(\delta_{\perp}\). Solid curves are eigenvalues of the effective Hamiltonian and points are measurements from seven PL6 centers.
(c) Representative PLE spectrum of a low-strain PL6 center with \(\delta_{\perp}=0.688\) GHz.
(d) Time-resolved PLE measurement over 118 min (895 consecutive scans), showing no pronounced long-term spectral jumps or drift.}
\label{fig1}
\end{figure*}

Here we determine the excited-state Hamiltonian and spin-dependent optical dynamics of single PL6 centers. Photoluminescence-excitation (PLE) spectroscopy performed on single PL6 centers spanning a range of transverse strains, together with first-principles calculations, resolves the excited-state fine structure and provides independent spectroscopic support for the oxygen--vacancy assignment. The extracted Hamiltonian identifies a low-strain regime with weak spin mixing, consistent with the measured slow spin-flip dynamics under resonant excitation. State-resolved lifetime measurements and spin-dependent fluorescence dynamics provide a quantitative explanation for the measured pulsed ODMR contrast of -31.9\% under off-resonant excitation. We further demonstrate coherent optical control, resonant spin initialization with 99.69\% fidelity, and polarization-selective transitions. These optical capabilities coexist with coherent ground-state spin control and a spin-coherence time of 5.70 ms under dynamical decoupling. Together, these results connect the excited-state structure of PL6 to its spin-selective optical dynamics and establish a microscopic basis for developing coherent spin–photon interfaces in SiC.



Its six eigenstates are expressed in the symmetry-adapted basis \(A_1,A_2,E_x,E_y,E_1,E_2\)~\cite{Maze2011group,Roger2009time,Doherty2011negatively,Thiering2024Nuclear,Bolalov2009low}, constructed from the orbital-spin product states \(\lvert E_{\pm},m_s\rangle\), as shown in Fig.~\ref{fig1}(a). Here, \(E_{\pm}\) label the orbital-doublet components with effective orbital angular-momentum projection \(m_L=\pm1\), and \(m_s=0,\pm1\) denotes the electron-spin projection along the crystallographic \(c\) axis. We use the effective Hamiltonian:
\begin{equation*}
\hat{H}_{\mathrm{ES}}=D_{\mathrm{ES}}\left(S_z^2-\frac{2}{3}\right)
+\lambda\sigma_z S_z
+\hat{H}_{\mathrm{ss}}(D_1,D_2)
+\hat{H}_{\mathrm{strain}}(\delta_{\perp}).
\end{equation*}
\(S_z\) is the spin-1 operator along the \(c\) axis, and \(\sigma_z\) is the Pauli operator. The parameters \(D_{\mathrm{ES}}\) and \(\lambda\) characterize the axial spin--spin and spin--orbit interactions, respectively; \(\hat{H}_{\mathrm{ss}}(D_1,D_2)\) contains the remaining spin--spin terms, and \(\hat{H}_{\mathrm{strain}}(\delta_\perp)\) describes the transverse strain-induced orbital coupling of strength \(\delta_\perp\). The full matrix representation is provided in the Supplemental Material (SM)~\cite{SupplementalMaterial}.

\begin{table}[!b]
\centering
\caption{Excited-state parameters of the hh-OV PL6 center in 4H-SiC. Values in brackets denote 95\% credible intervals of the experimental posterior. The calculated values include vibronic coupling and the Ham reduction associated with the dynamic Jahn--Teller effect (see SM~\cite{SupplementalMaterial}).}
\label{tab:params}
\begin{tabular}{l c c}
\hline
Parameter & \makecell{PL6 (hh-OV)\\Expt.} & \makecell{PL6 (hh-OV)\\Calc.} \\
\hline
$\lambda$ (GHz)
& \makecell{$5.739$\\$[5.656,5.881]$}
& $3.7$ \\
$D_{\mathrm{ES}}$ (GHz)
& \makecell{$0.932$\\$[0.913,0.957]$}
& $1.22$ \\
$D_1$ (GHz)
& \makecell{$0.026$\\$[0.0004,0.037]$}
& $0.044$ \\
$D_2$ (GHz)
& \makecell{$0.285$\\$[0.272,0.302]$}
& $0.296$ \\
\hline
\end{tabular}
\end{table}

Spin-selective PLE spectroscopy is performed under continuous microwave driving resonant with the ground-state zero-field splitting \(D_{\mathrm{GS}}=1.365\) GHz \cite{koehl2011room,He2024robust}. This allows us to access the predominantly \(m_s=0\) branches \(E_x/E_y\) together with the predominantly \(m_s=\pm1\) branches \(A_{1,2}/E_{1,2}\). Figure~\ref{fig1}(b) summarizes measurements on seven centers spanning \(\delta_{\perp}=0.688-12.416\) GHz. A global Bayesian analysis of the transition energies determines the Hamiltonian parameters and their credible intervals (Table~\ref{tab:params}).

\begin{figure*}[htbp]
\centering
\includegraphics[scale=1]{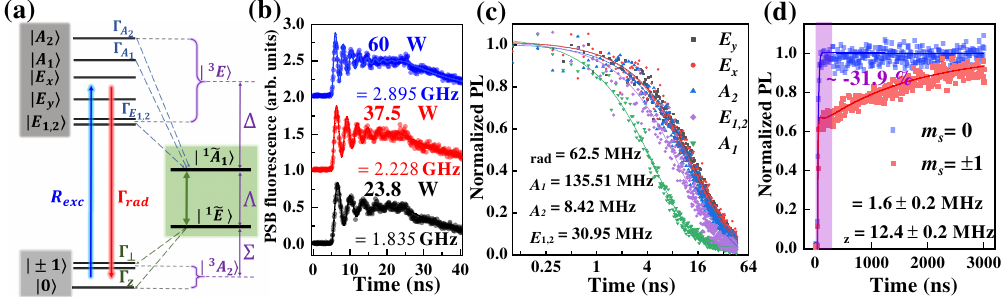}
\caption{\textbf{Spin-selective optical dynamics of hh-OV PL6.}
(a) Spin-dependent optical-cycle model.
\(R_{\mathrm{exc}}\) denotes the optical excitation rate.
Radiative decay from the five excited-state branches occurs at a common rate
\(\Gamma_{\mathrm{rad}}\), while \(\Gamma_i\) denotes the state-dependent
intersystem-crossing (ISC) rate through the upper branch
\(\lvert{}^3E\rangle\rightarrow\lvert{}^1\tilde{A}_1\rangle\).
The lower ISC branch
\(\lvert{}^1\tilde{E}\rangle\rightarrow\lvert{}^3A_2\rangle\)
repopulates \(\lvert \pm1\rangle\) and \(\lvert 0\rangle\)
at rates \(\Gamma_{\perp}\) and \(\Gamma_z\), respectively.
\(\Delta\), \(\Lambda\), and \(\Sigma\) denote the energy separations
\(\lvert{}^3E\rangle\)--\(\lvert{}^1\tilde{A}_1\rangle\),
\(\lvert{}^1\tilde{A}_1\rangle\)--\(\lvert{}^1\tilde{E}\rangle\), and
\(\lvert{}^1\tilde{E}\rangle\)--\(\lvert{}^3A_2\rangle\), respectively;
the tilde denotes vibronic states.
(b) Coherent optical Rabi oscillations on the
$\lvert m_s=0\rangle\leftrightarrow\lvert E_x\rangle$ transition
at three excitation powers.
(c) State-resolved excited-state lifetimes of the five resolved
optical branches, with the extracted radiative decay rate
$\Gamma_{\mathrm{rad}}$ and ISC rates $\Gamma_i$ indicated.
(d) Spin-dependent PL dynamics for initial $m_s=0$ and
$m_s=\pm1$ populations. The shaded region marks the 280-ns detection window,
yielding a pulsed ODMR contrast of $-31.9\%$. Solid curves are master-equation simulations in (b), exponential fits in (c), and rate-equation simulations based on the model in (a) in (d).}
\label{fig2}
\end{figure*}

First-principles calculations for the hh-OV configuration reproduce the measured excited-state fine structure of PL6 (Table~\ref{tab:params}). The transverse spin-spin parameter \(D_2\) is obtained within \(4\%\) of the measured value and \(D_1\) is correctly placed at the few-tens-of-MHz scale, while the axial spin-spin parameter \(D_{\mathrm{ES}}\) is overestimated by about \(30\%\). The calculated spin-orbit parameter, \(\lambda=3.7\) GHz, underestimates the measured \(5.739\) GHz; the deviation originates predominantly from the calculated vibronic reduction factor of the dynamic Jahn-Teller effect, which is the most sensitive quantity in the calculation (see SM \cite{SupplementalMaterial}). These calculations extend the microscopic description of the hh-OV center, established previously by isotope-resolved hyperfine spectroscopy~\cite{Hu2026,Chen2026Oxygen}, to the optically excited manifold.

A representative low-strain PLE spectrum [Fig.~\ref{fig1}(c)] resolves the \(A_1\), \(A_2\), \(E_x\), \(E_y\), and \(E_{1,2}\) branches. At the lowest excitation power investigated (20 nW), the \(A_1/A_2/E_{1,2}\) linewidths narrow to approximately \(90-140\) MHz, whereas \(E_x/E_y\) remain near 300 MHz (see SM \cite{SupplementalMaterial}). The different linewidths are consistent with distinct sensitivities of the excited-state branches to local strain and charge fluctuations. Time-resolved PLE over 118 min shows no pronounced long-term jumps or drift [Fig.~\ref{fig1}(d)], indicating good long-term spectral stability on the timescale of the measurement.

We next connect the excited-state fine structure to the spin-selective optical
cycle [Fig.~\ref{fig2}(a)]. Following resonant excitation, each $^{3}E$ state
either decays radiatively at rate $\Gamma_{\mathrm{rad}}$ or undergoes
state-dependent intersystem crossing (ISC) to the singlet manifold at rate
$\Gamma_i$ ( \(i=A_1,A_2,E_x,E_y,E_{1,2}\) ). Population subsequently returns to $m_s=\pm1$ and $m_s=0$ with
rates $\Gamma_{\perp}$ and $\Gamma_z$, respectively
\cite{Bian2025,Thiering2017}. We first calibrate resonant optical excitation
through coherent Rabi oscillations. As an example of coherent optical control, we drive Rabi oscillations on the
$\lvert m_s=0\rangle\leftrightarrow\lvert E_x\rangle$ transition
[Fig.~\ref{fig2}(b)]. Clear oscillations are observed at resonant optical powers of 23.8, 37.5, and 60~$\mu$W, with the optical Rabi frequency reaching $\Omega=2.895$~GHz at 60~$\mu$W. Using calibrated $\pi$ pulses, we then measure the
lifetimes of all five resolved excited-state branches [Fig.~\ref{fig2}(c)].
Taking the longest-lived $E_y$ branch ($\tau_{E_y}=16.0$ ns) to have
negligible ISC gives $\Gamma_{\mathrm{rad}}=62.5$ MHz \cite{Goldman2015}; subtracting this
radiative contribution yields the state-resolved upper-branch ISC rates.
In particular, $A_1$ exhibits the dominant ISC channel, whereas $A_2$ is
much more weakly coupled to the singlet manifold. 
The stronger ISC of PL6 follows the expected dependence
\(\Gamma_{\mathrm{ISC}}\propto\lambda_\perp^2F(\Delta)\), where
\(\lambda_\perp\) is the transverse spin--orbit coupling and
\(F(\Delta)\) is the vibrational-overlap factor associated with the
triplet--singlet energy separation \(\Delta\)~\cite{Goldman2015,Thiering2018loop}. The relevant \(\lambda_\perp\) is
proportional to the intrinsic spin--orbit parameter rather than to the
spectroscopic \(\lambda\), which is reduced by the vibronic coupling within
the \(^3E\) manifold (see SM \cite{SupplementalMaterial}).

The calculated triplet--singlet energy separation is also smaller
for PL6 (\(202~\mathrm{meV}\)) than for hh-DV PL1
(\(233.62~\mathrm{meV}\)), which may further favor ISC through
the vibrational-overlap contribution \(F(\Delta)\)
(see SM~\cite{SupplementalMaterial}).

The state-dependent ISC rates produce strongly spin-selective optical
dynamics. We initialize the spin in $m_s=0$ or $m_s=\pm1$ and record the
resulting fluorescence under off-resonant excitation [Fig.~\ref{fig2}(d)].
A global fit to the rate-equation model yields
$\Gamma_{\perp}=1.6(2)$ MHz and $\Gamma_z=12.4(2)$ MHz
(see SM \cite{SupplementalMaterial}), revealing preferential relaxation
through the singlet manifold into $m_s=0$. Together with the state-selective
upper-branch ISC, this preferential singlet relaxation establishes the
spin-selective optical cycle and quantitatively accounts for the large
fluorescence contrast of PL6~\cite{Li2020RoomtemperatureCM}. The 280-ns detection window indicated in
Fig.~\ref{fig2}(d) yields an pulsed ODMR contrast of $-31.9\%$, approaching the value of approximately $-43.1\%$ predicted from the measured decay rates (see SM~\cite{SupplementalMaterial}).

It is instructive to compare these figures with those of the
crystallographically analogous divacancy, i.e.\ the hh-DV (PL1).
The measured longitudinal spin-orbit parameter of PL6,
\(\lambda=5.739\) GHz, is substantially larger than the
\(3.538\) GHz of hh-DV PL1~\cite{christle2017isolated}, and
the first-principles calculations give an intrinsic spin-orbit
parameter \(2.6\) times larger for PL6, i.e.\ a factor of
\(6.8\) in \(\lambda_\perp^2\) (see SM \cite{SupplementalMaterial}).
The measured excited-state decay rates of the two centers are
comparable, so the difference in ODMR contrast is not radiative
in origin but follows from this enhancement of the
intersystem-crossing channel. The measured Debye-Waller factor of PL6,\(\approx10\%\), exceeds those of hh-DV PL1 and kk-DV PL2, which are \(3.69\%\) and \(6.11\%\), respectively
\cite{Jin2021}.This comparison shows that PL6
has a larger ZPL fraction than both divacancy configurations.

The two centers also differ in their behavior under resonant
illumination: the zero-phonon-line photon of PL6 lies above
the calculated threshold for reionizing the dark negatively
charged state~\cite{Kobayashi2023}, whereas that of the
divacancy lies below it~\cite{Csore2022}, which accounts for
the stable photoluminescence-excitation signal of
Fig.~\ref{fig1}(d).

\begin{figure}[htbp]
\centering
\includegraphics[scale=1]{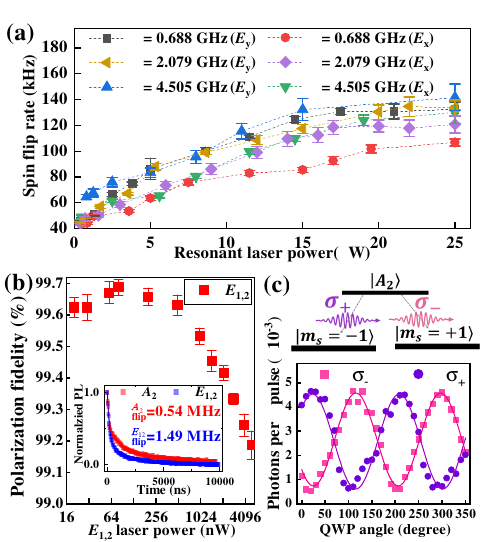}
\caption{\textbf{Spin-selective optical functionality of hh-OV PL6.}
(a) Measured spin-flip rates of the $E_x$ and $E_y$ branches for three
PL6 centers with different transverse strains.
(b) Spin-polarization fidelity under resonant $E_{1,2}$ excitation as a
function of excitation power, reaching $99.69\%$ at the optimum condition.
Inset: fluorescence decay through the $A_2$ and $E_{1,2}$ transitions.
(c) Polarization-selective \(\Lambda\) system formed by the \(m_s=\pm1\) ground states and the common \(A_2\) excited state (schematic), and measured polarization dependence of the corresponding optical transitions, yielding a visibility of approximately \(82\%\). \(\sigma_+\) and \(\sigma_-\) denote left- and right-circular polarizations, respectively, while \(\lvert-1\rangle\) and \(\lvert+1\rangle\) denote the
electron-spin states with \(m_s=-1\) and \(m_s=+1\), respectively.
}
\label{fig3}
\end{figure}

We next examine the functional consequences of the oxygen-modified
excited-state Hamiltonian. Calculations using the experimentally determined
parameters show that the principal \(E_y\) spin-mixing feature shifts from
approximately \(5.83\) GHz in hh-DV PL1 to \(14.3\) GHz in hh-OV PL6
(see SM \cite{SupplementalMaterial}), preserving high spin purity throughout
the experimentally relevant low-strain regime. Consistent with this
prediction, the measured \(E_x/E_y\) spin-flip rates remain low for three PL6
centers spanning \(\delta_\perp=0.688\)--\(4.505\) GHz
[Fig.~\ref{fig3}(a)]. The measured \(E_x\) spin-flip rates remain at or below approximately
\(140~\mathrm{kHz}\) over the investigated strain range, substantially
lower than the \(\sim330~\mathrm{kHz}\) reported for hh-DV
PL1 and kk-DV PL2 \cite{christle2017isolated}. The reduced spin admixture suppresses the intrinsic spin-nonconserving
component of the optical cycle and therefore favors enhanced optical cycling.

At the lowest measured transverse strain
(\(\delta_\perp=0.688~\mathrm{GHz}\)), calculations based on the
experimentally determined Hamiltonians predict approximately
1.8- and 2.9-fold enhancements of the Hamiltonian-limited
\(E_x\) and \(E_y\) optical cycling, respectively, relative
to PL1. Compared with kk-DV PL2, PL6 also exhibits
approximately 1.23- and 1.18-fold higher \(E_x\) and
\(E_y\) cycling photon numbers, respectively
(see SM~\cite{SupplementalMaterial}).

The transition-dependent spin dynamics also enable efficient resonant spin initialization. Under a resonant excitation power of \(30~\mathrm{nW}\), the spin-flip rate is
\(\Gamma_{\mathrm{flip}}^{E_{1,2}}=1.49~\mathrm{MHz}\) under \(E_{1,2}\) excitation,
compared with \(\Gamma_{\mathrm{flip}}^{A_2}=0.54~\mathrm{MHz}\) under \(A_2\) excitation, resulting in preferential optical pumping
into the target spin manifold. Optimizing the \(E_{1,2}\) excitation power yields a
spin-polarization fidelity of \(99.69\pm0.03\%\) [Fig.~\ref{fig3}(b)]. Moreover,
the \(A_2\) state forms a polarization-selective \(\Lambda\) system, in which the
\(m_s=\pm1\) ground states couple to a common excited state through opposite
circular polarizations \(\sigma_{\mp}\). The measured polarization dependence exhibits a
visibility of approximately \(82\%\) [Fig.~\ref{fig3}(c)], providing the
optical selection rules required for a spin--photon interface
\cite{togan2010quantum}.

\begin{figure}[tbp]
\centering
\includegraphics[width=\columnwidth]{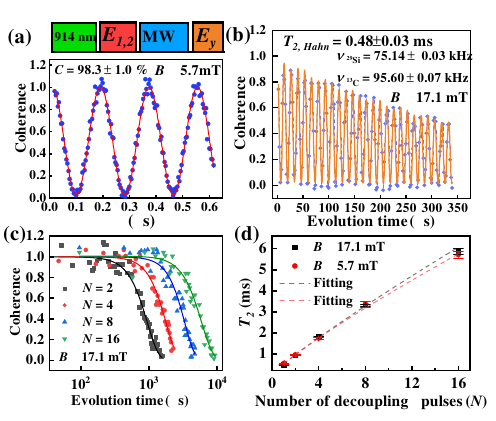}
\caption{\textbf{Coherent spin control and millisecond coherence of hh-OV PL6.}
(a) Single-spin microwave Rabi oscillations of a low-strain PL6 center
(\(\delta_\perp=0.688\) GHz). The upper panel shows the pulse sequence:
914-nm charge-state preparation is followed by resonant \(E_{1,2}\) spin
initialization, microwave (MW) spin control, and resonant \(E_y\) readout. $\tau$ denotes the MW duration time.
The measured Rabi oscillations exhibit \(98.3\pm1.0\%\) contrast.
(b) Hahn-echo decay at \(B\approx17.1\) mT.
(c) XY8 dynamical-decoupling decays for pulse numbers \(N=2,4,8,16\) at
\(B\approx17.1\) mT.
(d) Coherence time \(T_2(N)\) at 5.7 and 17.1 mT.
All measurements are performed at 6.35 K.}
\label{fig4}
\end{figure}

Finally, we verify that the optical properties established above
coexist with coherent ground-state spin control. Resonant optical preparation
and readout of a low-strain center (\(\delta_\perp=0.688\) GHz) yield
microwave Rabi oscillations with \(98.3\pm1.0\%\) contrast at 6.35 K and
an axial magnetic field \(B\) of \(\sim\) 5.7 mT [Fig.~\ref{fig4}(a)]. Together with the
initialization result above, this demonstrates efficient resonant preparation
and measurement of a single PL6 spin.

At \(B\approx17.1\) mT, a Hahn-echo measurement gives \(T_{2,\mathrm{Hahn}}=0.48\pm0.03\) ms [Fig.~\ref{fig4}(b)]. The decay exhibits electron-spin-echo envelope modulation with frequencies consistent with weakly coupled \(^{29}\)Si and \(^{13}\)C nuclear spins \cite{simin2017locking,koehl2011room}. Repeating the Hahn-echo measurement at \(\sim\) 5.7 mT gives a similar coherence time (\(T_{2,\mathrm{Hahn}}=0.54\pm0.02~\mathrm{ms}\); see SM \cite{SupplementalMaterial}), suggesting that the decoherence is not governed solely by a simple field-dependent nuclear-spin-bath mechanism.

Applying XY8 dynamical decoupling with \(N=2,4,8,\) and 16 \(\pi\) pulses extends the coherence time to \(T_2=5.70\pm0.16\) ms for \(N=16\), approximately one order of magnitude beyond the Hahn-echo value [Fig.~\ref{fig4}(c)]. The scaling \(T_2(N)=\alpha N^\beta\) gives \(\beta=0.86\pm0.01\) at \(\sim\) 5.7 mT and \(0.89\pm0.02\) at \(\sim\) 17.1 mT [Fig.~\ref{fig4}(d)]. Longer coherence has been demonstrated in isotopically purified SiC divacancies under more extensive dynamical decoupling \cite{anderson2022five}; here the role of the coherence measurement is to show that the oxygen-containing defect retains millisecond-scale spin coherence while exhibiting the modified optical dynamics identified above.

In conclusion, we have established a microscopic connection
between atomic-scale defect composition, excited-state Hamiltonian, and
spin-selective optical dynamics in the PL6 center of 4H-SiC.
PLE spectroscopy of single PL6 centers with different transverse strains, combined with Bayesian Hamiltonian analysis, determines the previously unresolved \(^{3}E\) fine structure, which is reproduced by first-principles calculations. Comparison with the crystallographically analogous hh-DV PL1 shows that oxygen substitution
increases the intrinsic spin-orbit coupling, enhancing the spin-selective intersystem crossing
and hence the ODMR contrast, while reducing spin mixing in the experimentally relevant low-strain
regime and favoring enhanced intrinsic optical
cycling of the \(E_x/E_y\) branches. State-resolved lifetime and spin-dependent fluorescence
measurements further reveal spin-selective ISC pathways underlying efficient
optical spin polarization and large ODMR contrast. Together, these results
connect microscopic defect configuration and excited-state electronic structure to
quantum-interface functionality, including high-fidelity resonant spin
initialization and readout, a polarization-selective \(\Lambda\) system, and
millisecond spin coherence extending to \(T_2=5.70\pm0.16~\mathrm{ms}\)
under dynamical decoupling. Combined with coherent optical and spin control, these capabilities make PL6 a promising platform for exploring photon-mediated entanglement between solid-state quantum nodes.




This work was supported by the National Natural Science Foundation of China (No. W2411001 and No.\ 92365205), the Quantum Science and Technology-National Science and Technology Major Project (No.\ 2021ZD0301400 and No.\ 2021ZD0301200), and USTC Major Frontier Research Program (No.\ LS2030000002). A.G.\ gratefully acknowledges the support of the Quantum Information National Laboratory of Hungary, funded by the National Research, Development, and Innovation Office of Hungary (NKFIH) under Grant No.\ 2022-2.1.1-NL-2022-00004 and funding from the European Commission for the QuSPARC (Grant No.\ 101186889) and SPINUS (Grant No.\ 101135699) projects. G.T.\ was supported by the J\'anos Bolyai Research Scholarship of the Hungarian Academy of Sciences and by NKFIH under Grant No.\ STARTING 150113. This work was partially performed at the University of Science and Technology of China Center for Micro and Nanoscale Research and Fabrication.  


  


\end{document}


\title{Supplementary Materials: Coherent Spin-Photon Interface of Single PL6 Oxygen–Vacancy Centers in 4H-SiC}

\author{Zhen-Xuan He}
\altaffiliation{These authors contributed equally to this work.}
\affiliation{Laboratory of Quantum Information, University of Science and Technology of China, Hefei, Anhui 230026, China}
\affiliation{Anhui Province Key Laboratory of Quantum Network, University of Science and Technology of China, Hefei, Anhui 230026, China}
\affiliation{CAS Center For Excellence in Quantum Information and Quantum Physics, University of Science and Technology of China, Hefei, Anhui 230026, China}
\affiliation{Hefei National Laboratory, University of Science and Technology of China, Hefei, Anhui 230088, China}

\author{Gerg\H{o} Thiering}
\altaffiliation{These authors contributed equally to this work.}
\affiliation{HUN-REN Wigner Research Centre for Physics, P.O.\ Box 49, H-1525 Budapest, Hungary}

\author{Xiangru Han}
\affiliation{HUN-REN Wigner Research Centre for Physics, P.O.\ Box 49, H-1525 Budapest, Hungary}

\author{Rui-Jian Liang}
\affiliation{Laboratory of Quantum Information, University of Science and Technology of China, Hefei, Anhui 230026, China}
\affiliation{Anhui Province Key Laboratory of Quantum Network, University of Science and Technology of China, Hefei, Anhui 230026, China}
\affiliation{CAS Center For Excellence in Quantum Information and Quantum Physics, University of Science and Technology of China, Hefei, Anhui 230026, China}

\author{Ji-Yang Zhou}
\affiliation{Laboratory of Quantum Information, University of Science and Technology of China, Hefei, Anhui 230026, China}
\affiliation{Anhui Province Key Laboratory of Quantum Network, University of Science and Technology of China, Hefei, Anhui 230026, China}
\affiliation{CAS Center For Excellence in Quantum Information and Quantum Physics, University of Science and Technology of China, Hefei, Anhui 230026, China}


\author{Shuo Ren}
\affiliation{Laboratory of Quantum Information, University of Science and Technology of China, Hefei, Anhui 230026, China}
\affiliation{Anhui Province Key Laboratory of Quantum Network, University of Science and Technology of China, Hefei, Anhui 230026, China}
\affiliation{CAS Center For Excellence in Quantum Information and Quantum Physics, University of Science and Technology of China, Hefei, Anhui 230026, China}
\affiliation{Hefei National Laboratory, University of Science and Technology of China, Hefei, Anhui 230088, China}

\author{Wu-Xi Lin}
\affiliation{Laboratory of Quantum Information, University of Science and Technology of China, Hefei, Anhui 230026, China}
\affiliation{Anhui Province Key Laboratory of Quantum Network, University of Science and Technology of China, Hefei, Anhui 230026, China}
\affiliation{CAS Center For Excellence in Quantum Information and Quantum Physics, University of Science and Technology of China, Hefei, Anhui 230026, China}
\affiliation{Hefei National Laboratory, University of Science and Technology of China, Hefei, Anhui 230088, China}
	
\author{Zhi-He Hao}
\affiliation{Laboratory of Quantum Information, University of Science and Technology of China, Hefei, Anhui 230026, China}
\affiliation{Anhui Province Key Laboratory of Quantum Network, University of Science and Technology of China, Hefei, Anhui 230026, China}
\affiliation{CAS Center For Excellence in Quantum Information and Quantum Physics,
University of Science and Technology of China, Hefei, Anhui 230026, China}

\author{Qi-Cheng Hu}
\affiliation{Laboratory of Quantum Information, University of Science and Technology of China, Hefei, Anhui 230026, China}
\affiliation{Anhui Province Key Laboratory of Quantum Network, University of Science and Technology of China, Hefei, Anhui 230026, China}
\affiliation{CAS Center For Excellence in Quantum Information and Quantum Physics, University of Science and Technology of China, Hefei, Anhui 230026, China}
\affiliation{Hefei National Laboratory, University of Science and Technology of China, Hefei, Anhui 230088, China}

\author{Jun-Feng Wang}
\affiliation{College of Physics, Sichuan University, Chengdu, Sichuan 610065, China}

\author{Adam Gali}
\altaffiliation{Email: gali.adam@wigner.hun-ren.hu}
\affiliation{HUN-REN Wigner Research Centre for Physics, P.O.\ Box 49, H-1525 Budapest, Hungary}
\affiliation{Department of Atomic Physics, Institute of Physics, Budapest University of Technology and Economics, M\H{u}egyetem rakpart 3., H-1111 Budapest, Hungary}
\affiliation{MTA-WFK Lend\"ulet ``Momentum'' Semiconductor Nanostructures Research Group, P.O.\ Box 49, H-1525 Budapest, Hungary}

\author{Jin-Shi Xu}
\altaffiliation{Email: jsxu@ustc.edu.cn}
\affiliation{Laboratory of Quantum Information, University of Science and Technology of China, Hefei, Anhui 230026, China}
\affiliation{Anhui Province Key Laboratory of Quantum Network, University of Science and Technology of China, Hefei, Anhui 230026, China}
\affiliation{CAS Center For Excellence in Quantum Information and Quantum Physics, University of Science and Technology of China, Hefei, Anhui 230026, China}
\affiliation{Hefei National Laboratory, University of Science and Technology of China, Hefei, Anhui 230088, China}

\author{Chuan-Feng Li}
\altaffiliation{Email: cfli@ustc.edu.cn}
\affiliation{Laboratory of Quantum Information, University of Science and Technology of China, Hefei, Anhui 230026, China}
\affiliation{Anhui Province Key Laboratory of Quantum Network, University of Science and Technology of China, Hefei, Anhui 230026, China}
\affiliation{CAS Center For Excellence in Quantum Information and Quantum Physics, University of Science and Technology of China, Hefei, Anhui 230026, China}
\affiliation{Hefei National Laboratory, University of Science and Technology of China, Hefei, Anhui 230088, China}
 
\author{Guang-Can Guo}
\affiliation{Laboratory of Quantum Information, University of Science and Technology of China, Hefei, Anhui 230026, China}
\affiliation{Anhui Province Key Laboratory of Quantum Network, University of Science and Technology of China, Hefei, Anhui 230026, China}
\affiliation{CAS Center For Excellence in Quantum Information and Quantum Physics, University of Science and Technology of China, Hefei, Anhui 230026, China}
\affiliation{Hefei National Laboratory, University of Science and Technology of China, Hefei, Anhui 230088, China}

\maketitle
\date{\today}
\clearpage
\tableofcontents

\newpage
\section*{Supplementary Note 1. Group-theory analysis of the excited state of
hh-OV PL6}

\subsection*{Symmetry-adapted basis and excited-state Hamiltonian}

The hh-OV PL6 center in 4H-SiC has \(C_{3v}\) point-group symmetry.
Its optically excited manifold is an orbitally degenerate spin triplet
(\(S=1\)) of \(^{3}E\) symmetry. Following the conventions of
Ref.~\cite{Thiering2024Nuclear}, we define the six symmetry-adapted
basis states as 

\[
\begin{aligned}
|A_{1}\rangle &= \frac{1}{\sqrt{2}}
\left(|E_{-}^{1}\rangle-|E_{+}^{-1}\rangle\right), \\
|A_{2}\rangle &= \frac{1}{\sqrt{2}}
\left(|E_{-}^{1}\rangle+|E_{+}^{-1}\rangle\right), \\
|E_{y}\rangle &= \frac{i}{\sqrt{2}}
\left(|E_{+}^{0}\rangle-|E_{-}^{0}\rangle\right), \\
|E_{x}\rangle &= -\frac{1}{\sqrt{2}}
\left(|E_{+}^{0}\rangle+|E_{-}^{0}\rangle\right), \\
|E_{1}\rangle &= \frac{1}{\sqrt{2}}
\left(|E_{-}^{-1}\rangle-|E_{+}^{1}\rangle\right), \\
|E_{2}\rangle &= -\frac{i}{\sqrt{2}}
\left(|E_{-}^{-1}\rangle+|E_{+}^{1}\rangle\right).
\end{aligned}
\]

Here, \(E_{\pm}\) denote the two orbital components of the \(E\)
representation, and the superscripts indicate the spin projections
\(m_s=1,0,-1\) along the crystallographic \(c\) axis.
The effective excited-state Hamiltonian is given by

\[
\hat H_{\mathrm{ES}}
=
D_{\mathrm{ES}}\left(\hat S_z^2-\frac{2}{3}\right)
+\lambda\,\hat{\sigma}_z\hat S_z
+\hat H_{\mathrm{ss}}(D_1,D_2)
+\hat H_{\mathrm{strain}}(\delta_\perp).
\]

The operator \(\hat S_z\) is the spin-1 operator along the \(c\) axis,
and \(\hat{\sigma}_z\) is the Pauli operator acting in the
orbital subspace spanned by \(E_{\pm}\).
The parameters \(D_{\mathrm{ES}}\) and \(\lambda\) characterize
the axial spin--spin and longitudinal spin--orbit interactions,
respectively. The axial spin--spin term contributes
\(D_{\mathrm{ES}}/3\) to the \(m_s=\pm1\) components and
\(-2D_{\mathrm{ES}}/3\) to the \(m_s=0\) components.
The orbitally dependent spin--spin interactions are parameterized
by \(D_1\) and \(D_2\) and take the form

\[
\begin{aligned}
\hat H_{\mathrm{ss}}(D_1,D_2) ={}&
-D_1\left[
\left(\hat S_-\hat S_z+\hat S_z\hat S_-\right)\hat{\sigma}_-
+
\left(\hat S_+\hat S_z+\hat S_z\hat S_+\right)\hat{\sigma}_+
\right] \\
&+D_2\left[
\hat S_+^2\hat{\sigma}_-
+\hat S_-^2\hat{\sigma}_+
\right].
\end{aligned}
\]

The operators \(\hat S_{\pm}=\hat S_x\pm i\hat S_y\) are the
spin-1 raising and lowering operators, while
\(\hat{\sigma}_{\pm}=(\hat{\sigma}_x\pm i\hat{\sigma}_y)/2\)
are the orbital raising and lowering operators in the
\(\{E_+,E_-\}\) subspace, with
\(\hat{\sigma}_+=|E_+\rangle\langle E_-|\) and
\(\hat{\sigma}_-=|E_-\rangle\langle E_+|\).
The \(D_2\) term produces the \(A_1\)--\(A_2\) splitting,
whereas the \(D_1\) term couples the
\(\{E_x,E_y\}\) basis states with \(m_s=0\) to the
\(\{E_1,E_2\}\) basis states with \(m_s=\pm1\).
Finally, \(\hat H_{\mathrm{strain}}(\delta_\perp)\) describes
the transverse strain-induced orbital coupling.
We choose a transverse coordinate axis aligned with the effective
transverse-strain direction, allowing this coupling to be
parameterized by a single strength \(\delta_\perp\).

In the ordered basis
\((A_1,A_2,E_x,E_y,E_1,E_2)\), the Hamiltonian used throughout this work
takes the form

\[
\hat H_{\mathrm{ES}}=
\begin{pmatrix}
\frac{1}{3}D_{\mathrm{ES}}+\lambda-2D_{2}
&0&0&0&\delta_{\perp}&0 \\

0&
\frac{1}{3}D_{\mathrm{ES}}+\lambda+2D_{2}
&0&0&0&\delta_{\perp} \\

0&0&
-\frac{2}{3}D_{\mathrm{ES}}+\delta_{\perp}
&0&0&\sqrt{2}D_{1} \\

0&0&0&
-\frac{2}{3}D_{\mathrm{ES}}-\delta_{\perp}
&-\sqrt{2}iD_{1}&0 \\

\delta_{\perp}&0&0&
\sqrt{2}iD_{1}&
\frac{1}{3}D_{\mathrm{ES}}-\lambda&0 \\

0&\delta_{\perp}&\sqrt{2}D_{1}&0&0&
\frac{1}{3}D_{\mathrm{ES}}-\lambda
\end{pmatrix}.
\]

This matrix makes the relevant coupling structure explicit. The
\(A_1\) and \(A_2\) states are split by \(4D_2\) and couple to
\(E_1\) and \(E_2\), respectively, through transverse strain.
The \(E_x\) and \(E_y\) states are shifted oppositely by
\(\delta_\perp\) and couple to \(E_2\) and \(E_1\), respectively,
through \(D_1\). At zero transverse strain, \(E_1\) and \(E_2\)
remain degenerate, while the spin-orbit interaction \(\lambda\)
sets their energetic separation from the \(A_{1,2}\) manifold.
The Energy-level diagram of the \(^{3}\mathrm{E}\) excited state is shown in Fig. 1(a) of the main text.

\subsection*{Connection to the strain-dependent spectroscopy}

We diagonalize \(\hat H_{\mathrm{ES}}\) numerically to obtain the
strain-dependent excited-state eigenenergies used to analyze the photoluminescence-excitation (PLE)
measurements. The calculated eigenenergies reproduce the evolution of
the five resolved optical branches measured for seven individual PL6
centers over \(\delta_\perp=0.688\)--\(12.416\) GHz
[Fig.~1(b) of the main text and Supplementary Fig.~S\ref{fig:strain_spectra}].
The parameters \(D_{\mathrm{ES}}\), \(\lambda\), \(D_1\), and \(D_2\)
are determined by the Bayesian Markov chain Monte Carlo (MCMC) analysis described in
Supplementary Note~4.

\subsection*{Relation between different parameter conventions}

The spin-spin parameters used in the present work are related to the
\(\Delta_1\) and \(\Delta_2\) convention introduced for the nitrogen vacancy (NV) centers in
diamond by Maze \textit{et al.}~\cite{Maze2011group}.
Comparison of the corresponding matrix elements gives

\[
\Delta_2=\sqrt{2}D_1,\qquad
\Delta_1=2D_2.
\]

In particular, \(\sqrt{2}D_1\) is the matrix element coupling
\(E_x\) with \(E_2\) (and \(E_y\) with \(E_1\)), whereas \(4D_2\)
is the total \(A_1\)--\(A_2\) splitting. In the convention by Maze \textit{et al.},
the corresponding quantities are \(\Delta_2\) and \(2\Delta_1\),
respectively. A numerical comparison in both conventions is provided in
Supplementary Note~4.

\clearpage
\section*{Supplementary Note 2: Experimental methods and details}

\subsection*{Sample fabrication}

A commercial N-type 4H-SiC wafer with a \(10\ \mu\mathrm{m}\) epitaxial
layer and a nitrogen doping concentration of
\(5\times10^{14}\ \mathrm{cm}^{-3}\) was used.
 Single PL6 color centers were created by focused helium-ion-beam implantation using a helium ion microscope (Zeiss Orion NanoFab) operated at 30 keV. The microscope provides an imaging resolution of approximately 0.5 nm. A dose of 300 ions per implantation site was applied. After implantation, the sample
underwent a two-step annealing procedure: first at \(500^{\circ}\mathrm{C}\)
for \(2\ \mathrm{h}\), followed by a second annealing at
\(900^{\circ}\mathrm{C}\) for \(1\ \mathrm{h}\). In both steps, the
temperature was ramped at \(5^{\circ}\mathrm{C}\,\mathrm{min}^{-1}\), and
the sample was allowed to cool naturally to room temperature after each
annealing stage.

\subsection*{Low-temperature setup}

Optical measurements were performed in a closed-cycle cryostat
(Montana Cryostation S200) at a base temperature of \(6.35\ \mathrm{K}\).
A low-temperature piezo-driven nanopositioner (Attocube ANC350) was used to
position individual color centers within the diffraction-limited focus of
the confocal microscope.

Excitation and photoluminescence collection were performed using an Olympus
objective with a numerical aperture of \(0.85\).

Photoluminescence was collected via a single-mode optical fiber (1060-XP) and
detected by a superconducting nanowire single-photon detector (PHOTEC).

Unless otherwise specified, resonant measurements were performed by
collecting phonon-sideband (PSB) emission while rejecting the resonant
excitation light.

\subsection*{Spin control}

Microwave signals for spin manipulation were generated by a synthesized signal
generator (Mini-Circuits SSG-6000 RC). The output was split by a
\(90^{\circ}\) phase shifter (Mini-Circuits ZX10Q-2-19-S+) and routed through
two independently controlled microwave switches
(Mini-Circuits ZASWA-2-50DR+). The two channels were recombined using a power
splitter (Mini-Circuits ZAPD30-S+), amplified by a microwave amplifier
(Mini-Circuits ZHL-25W-272+), and delivered to a
\(50\ \mu\mathrm{m}\)-wide copper wire positioned approximately
\(80\ \mu\mathrm{m}\) above the sample surface.

An off-resonant \(914\ \mathrm{nm}\) continuous-wave laser
(Changchun New Tech.) was employed for charge-state reinitialization.
Spin initialization via the \(E_{1,2}\) transition and readout via the
\(E_y\) transition were performed using two tunable diode lasers
(Toptica DL Pro and CTL). To suppress laser-induced spin decoherence during
free evolution, all resonant laser beams were gated by fiber-coupled
acousto-optical modulators (CETC SGTF200) providing an extinction ratio exceeding
\(60\ \mathrm{dB}\). The timing of all microwave pulses, optical gates, and
detection windows was synchronized by a digital pulse generator
(SpinCore PBESRPRO500).

\subsection*{PLE spectroscopy with continuous microwave driving and linewidth analysis}

\begin{figure}[htbp]
    \centering
    \includegraphics[scale=1]{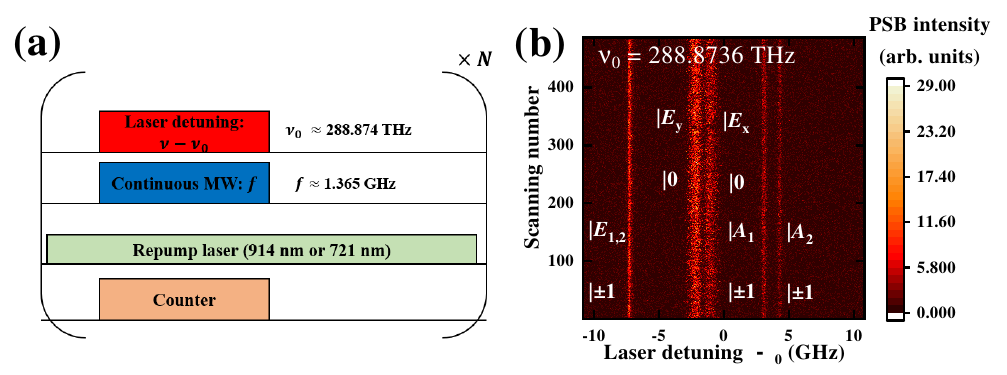}
    \caption{
    \textbf{PLE spectroscopy with continuous microwave driving.}
    (a) Pulse sequence for measuring PLE spectra under continuous microwave
    driving.
    (b) Time-resolved PLE spectra recorded over 486 scanning cycles.
    A \(721\ \mathrm{nm}\) repump laser (\(200\ \mathrm{nW}\)) is applied
    to maintain charge-state stability. \(\nu_0=288.8736 ~\mathrm{THz}\) represents the zero-detuning laser frequency.
    }
    \label{fig:ple_microwave}
\end{figure}

Supplementary Fig.~S\ref{fig:ple_microwave}(a) depicts the pulse sequence used to acquire
photoluminescence-excitation (PLE) spectra under continuous microwave (MW)
driving. The absolute laser wavelength was monitored with a high-precision
wavemeter (HighFinesse WS7-30, absolute accuracy \(\pm3\ \mathrm{MHz}\)).
To suppress spurious optical pumping and spectral diffusion during the
measurements, the resonant laser was gated by a fiber-coupled acousto-optical
modulator (CETC SGTF-200) with an extinction ratio exceeding \(60\ \mathrm{dB}\).
A continuous-wave repump laser (\(914\ \mathrm{nm}\) or
\(721\ \mathrm{nm}\), \(200\ \mathrm{nW}\)) was applied to mitigate
charge-state ionization.

The resonant-laser frequency, referenced to a nominal value
\(\nu_0\simeq288.874\ \mathrm{THz}\), was swept by applying a triangular
voltage waveform from an arbitrary function generator
(Tektronix AFG31252) to the piezo-tuning element of the laser. At each
frequency step, the resonant laser and microwave field
(\(f_{\mathrm{MW}}=1.365\ \mathrm{GHz}\), power \(-25\ \mathrm{dBm}\))
were simultaneously enabled for \(8\ \mathrm{ms}\), during which phonon sideband (PSB)
fluorescence was recorded. The two repump wavelengths yielded similar
spectral stability and suppression of ionization.

The time-resolved PLE spectra in Supplementary Fig.~S\ref{fig:ple_microwave}(b) were acquired using
the \(721\ \mathrm{nm}\) repump laser on the same low-strain PL6 center
(\(\delta_\perp=0.688\ \mathrm{GHz}\)) characterized in
Figs.~1(c) and 1(d) of the main text. The measurement shown in main-text
Fig.~1(d) instead used a \(914\ \mathrm{nm}\) repump laser.

\begin{figure}[htbp]
    \centering
    \includegraphics[scale=1]{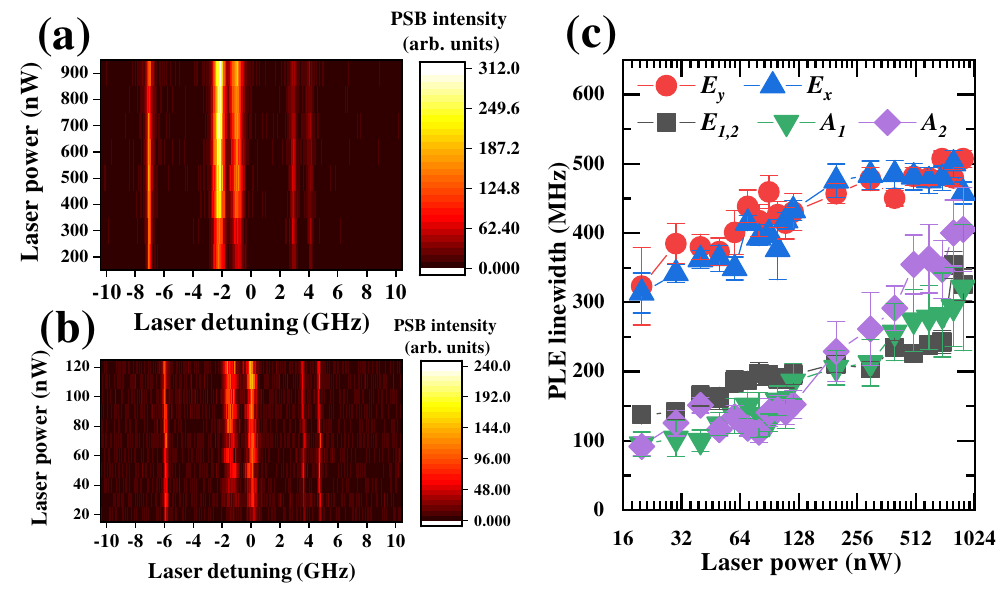}
    \caption{
    \textbf{Excitation-power-dependent PLE linewidths.}
    (a) PLE spectra at excitation powers of \(200\)--\(900\ \mathrm{nW}\)
    and (b) \(20\)--\(120\ \mathrm{nW}\) under continuous microwave driving.
    (c) Extracted linewidths of the five optical transitions as a function of
    resonant excitation power. Error bars represent fitting uncertainties from
    multi-peak Lorentzian fits.
    }
    \label{fig:ple_microwave2}
\end{figure}

To investigate power broadening and the low-power optical linewidths, we
measure power-resolved PLE spectra on the same low-strain PL6 center over
resonant excitation powers of \(20\)--\(900\ \mathrm{nW}\)
[Supplementary Figs.~S\ref{fig:ple_microwave2}(a) and S\ref{fig:ple_microwave2}(b)]. The linewidths of the \(A_1\),
\(A_2\), \(E_x\), \(E_y\), and \(E_{1,2}\) transitions are extracted from
multi-peak Lorentzian fits.

Supplementary Fig.~S\ref{fig:ple_microwave2}(c) summarizes the extracted linewidths as a function
of excitation power. All five transitions broaden with increasing power,
with the increase becoming weaker as the excitation approaches the saturation
regime near \(\sim450\ \mathrm{nW}\). The \(E_x/E_y\) transitions remain
systematically broader than the \(A_1\), \(A_2\), and \(E_{1,2}\)
transitions over the measured power range. At \(20\ \mathrm{nW}\), the
\(E_x/E_y\) linewidths are approximately \(300\ \mathrm{MHz}\), whereas
the narrowest \(A_1/A_2/E_{1,2}\) linewidths are approximately
\(90\)--\(140\ \mathrm{MHz}\). The persistence of this difference at the
lowest measured powers indicates that it cannot be attributed solely to
power broadening.

The systematically broader \(E_x/E_y\) transitions are closely related to
their larger first-order susceptibility to transverse strain. As shown by the
strain-dependent excited-state energies in Fig.~1(b) of the main text and
Supplementary Fig.~S\ref{fig:strain_spectra}, the \(E_x\) and \(E_y\) branches exhibit an
approximately linear dependence on \(\delta_\perp\) in the low-strain regime.
Consequently, fluctuations of the local strain, or other perturbations that
couple to the orbital degree of freedom in a strain-like manner, are efficiently
converted into fluctuations of the \(E_x/E_y\) optical transition frequencies,
producing additional spectral broadening. In contrast, the \(A_1\), \(A_2\),
and \(E_{1,2}\) branches exhibit substantially weaker strain dependence in the
same regime, resembling a ``clock-like'' response with reduced first-order
strain susceptibility. Their reduced strain susceptibility provides a natural
explanation for the narrower linewidths observed at low excitation power.
The calculated strain dependence underlying this interpretation is discussed
further in Supplementary Note~3.

Even the narrowest measured linewidths of approximately
\(90\)--\(140\ \mathrm{MHz}\) remain broader than the transform-limited
linewidth
\(\Gamma_{\mathrm{rad}}/(2\pi)\simeq10\ \mathrm{MHz}\)
corresponding to the measured \(E_y\) lifetime of \(16\ \mathrm{ns}\)
[main-text Fig.~2(c)]. We attribute this residual broadening primarily to
spectral diffusion caused by fluctuations of the local charge environment.
For a residual nitrogen concentration of approximately
\(5\times10^{14}\ \mathrm{cm}^{-3}\), comparable optical broadening has
been reported for divacancy centers in 4H-SiC
\cite{Bourassa2020EntanglementAC}. Implantation-induced defects may provide
an additional source of local charge fluctuations.

Higher-purity material, electrical stabilization of the local charge
environment, and further optimization of implantation and annealing may
therefore provide routes toward narrower optical transitions
\cite{miao2019electrically,Anderson2019ElectricalAO}. Such improvements
would be beneficial for photon indistinguishability and photon-mediated
entanglement protocols based on PL6.

\subsection*{Coherent resonant optical control}

Coherent optical Rabi oscillations shown in main-text Fig.~2(b) are measured
on the
\(\lvert m_s=0\rangle\leftrightarrow\lvert E_x\rangle\)
transition. Resonant optical pulses are generated using an amplitude
electro-optic modulator (Exail NIR-MX-LN-10) driven by an arbitrary function
generator (Tektronix AFG31252). The modulator provides an extinction ratio
of approximately \(30\ \mathrm{dB}\) and a rise time of \(35\ \mathrm{ps}\).

The pulse sequence is illustrated in Supplementary Fig.~S\ref{fig:OpticalPulseSequence}(a).
The spin is first initialized into \(\lvert m_s=0\rangle\), after which a
resonant optical pulse of \(\sim20\ \mathrm{ns}\) duration drives the
\(\lvert m_s=0\rangle\leftrightarrow\lvert E_x\rangle\)
transition. Measurements are performed at excitation powers of
\(23.8\), \(37.5\), and \(60.0\ \mu\mathrm{W}\), corresponding to the
three data sets in main-text Fig.~2(b). The resulting PSB fluorescence is
recorded with a Swabian Instruments Time Tagger 20 with a timing resolution
of \(100\ \mathrm{ps}\). The time-tagger is triggered by the arbitrary
function generator at the rising edge of each optical pulse, and the overall
sequence is synchronized by the SpinCore digital pulse generator.

The observed oscillations demonstrate coherent resonant driving of the
optical transition. Their excitation-power dependence is reproduced by
master-equation simulations that include coherent optical driving and
excited-state decay, as described in Supplementary Note 6. The calibrated Rabi frequencies are subsequently used to generate the short resonant optical \(\pi\) pulses required for the
state-resolved lifetime measurements described below.

\begin{figure}[htbp]
\centering
\includegraphics[scale=0.5]{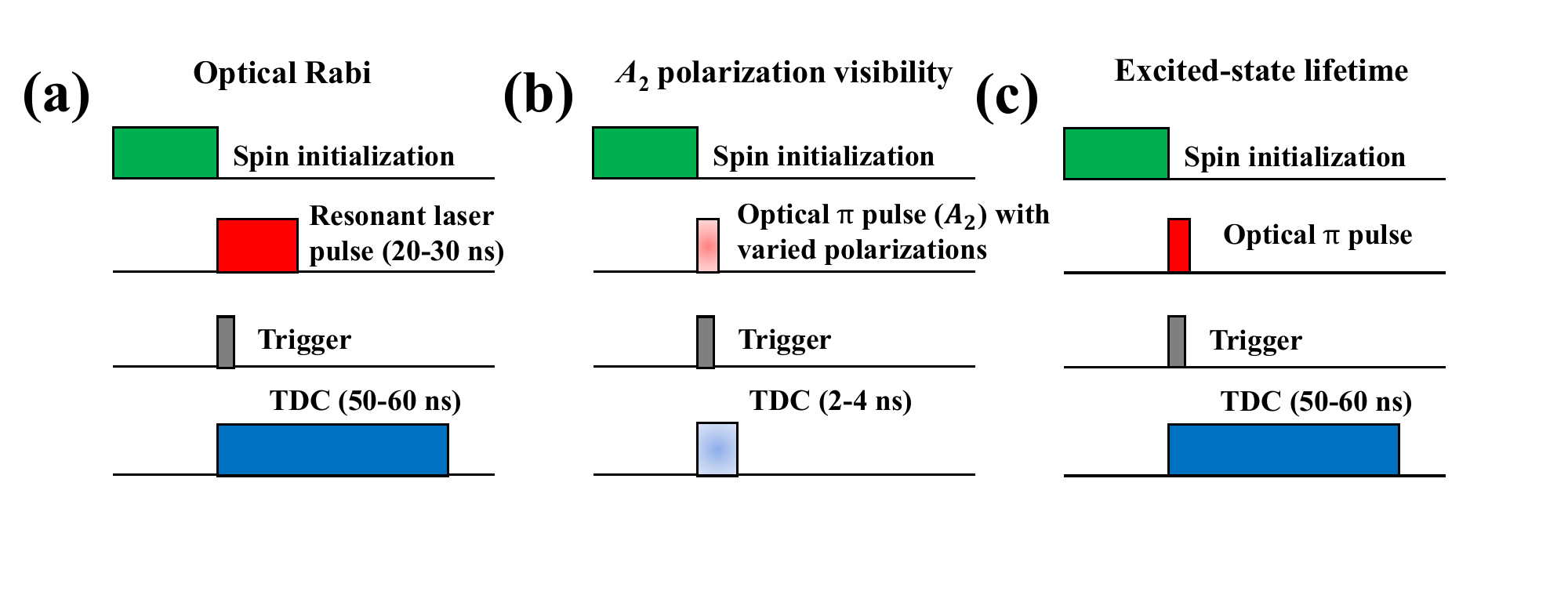}
\caption{
\textbf{Experimental pulse sequences for resonant optical measurements.}
(a) Coherent optical Rabi oscillations: following spin initialization, a
variable-duration resonant pulse drives the
\(\lvert m_s=0\rangle\leftrightarrow\lvert E_x\rangle\) transition,
while PSB fluorescence is recorded.
(b) \(A_2\) polarization measurement: after preparation of
\(\lvert m_s=+1\rangle\) or \(\lvert m_s=-1\rangle\) , a resonant optical \(\pi\) pulse with variable
polarization is applied and the PSB fluorescence is recorded.
(c) Excited-state lifetime measurement: a short resonant \(\pi\) pulse
excites the selected branch, followed by time-resolved detection of the
PSB fluorescence decay. The optical-pulse waveform and time-tagger trigger
are generated by the AFG, and all sequences are synchronized by the
SpinCore pulse generator.
}
\label{fig:OpticalPulseSequence}
\end{figure}

\subsection*{Polarization dependence of the \(\lvert A_2\rangle\) transition}

We characterize the polarization selection rules of the
\(\lvert A_2\rangle\) transition following the approach of
Togan \textit{et al.}~\cite{togan2010quantum}. The \(A_2\) excited state
provides a polarization-selective \(\Lambda\) system in which the
\(\lvert m_s=+1\rangle\) and \(\lvert m_s=-1\rangle\) ground states couple to the common \(A_2\)
state through opposite circular polarizations. Ideally, spontaneous
emission from this excited state therefore produces the spin--photon
correlation

\[
{
|\Psi\rangle
=
\frac{1}{\sqrt{2}}
\left(
|\sigma_-\rangle|m_s=+1\rangle
+
|\sigma_+\rangle|m_s=-1\rangle
\right),
}
\]

where the relative phase \(\phi\) depends on the optical convention and
evolution.  \(\sigma_+\) and \(\sigma_-\) denote left- and right-circular polarizations, respectively.

For the polarization measurement, an axial magnetic field of
\(5.7\ \mathrm{mT}\) lifts the \(\lvert m_s=\pm1\rangle\) degeneracy, allowing the
electron spin to be prepared in either \(\lvert m_s=+1\rangle\) or \(\lvert m_s=-1\rangle\).
A linear polarizer and quarter-wave plate are used to vary the polarization
of a resonant \(A_2\) pulse. The PSB fluorescence during the optical pulse is
recorded as a function of quarter-wave-plate angle and accumulated over
approximately \(10^9\) repetitions. The resulting polarization dependence
[main-text Fig.~3(c)] exhibits a visibility of approximately \(82\%\),
demonstrating strong polarization selectivity of the two optical legs.
The corresponding sequence is shown in Supplementary Fig.~S\ref{fig:OpticalPulseSequence}(b).

\subsection*{State-resolved excited-state lifetime measurements}

We measure the lifetimes of the five resolved excited-state branches
(\(i=\)\(E_x\), \(E_y\), \(E_{1,2}\), \(A_1\), and \(A_2\)) using resonant
pulsed excitation. Short optical \(\pi\) pulses with a full width at
half maximum of approximately \(2\ \mathrm{ns}\) excite the selected
transition. Unlike the \(A_2\) polarization measurement, the lifetime
measurements are performed at zero magnetic field, where the
\(\lvert m_s=\pm1\rangle\) ground states are degenerate. Following excitation, the PSB
fluorescence decay is recorded over a \(50\ \mathrm{ns}\) acquisition
window with \(100\ \mathrm{ps}\) timing resolution
[Supplementary Fig.~S\ref{fig:OpticalPulseSequence}(c)]. The decay of each branch is well described by a
single exponential, from which the lifetime \(\tau_i\) is extracted.

The total decay rate of branch \(i\) is

\[
{
\frac{1}{\tau_i}
=
\Gamma_{\mathrm{rad}}
+
\Gamma_{\mathrm{ISC},i},
}
\]

where \(\Gamma_{\mathrm{rad}}\) is the radiative decay rate and
\(\Gamma_{\mathrm{ISC},i}\) is the upper-branch intersystem-crossing
rate. Taking the longest-lived \(E_y\) branch
(\(\tau_{E_y}=16.0\ \mathrm{ns}\)) to have negligible ISC gives
\(\Gamma_{\mathrm{rad}}=62.5\ \mathrm{MHz}\) \cite{Goldman2015}. The ISC rates of the
remaining branches are then obtained from

\[
{
\Gamma_{\mathrm{ISC},i}
=
\frac{1}{\tau_i}-\Gamma_{\mathrm{rad}}.
}
\]

The extracted state-resolved lifetimes are shown in main-text Fig.~2(c).
Among the resolved branches, \(A_1\) exhibits the strongest ISC, whereas
\(A_2\) is substantially more weakly coupled to the singlet manifold.

\subsection*{Measurement of resonant spin-flip dynamics}

To measure the resonant spin-flip dynamics, the ground-state spin is first
initialized into the spin manifold addressed by the selected optical
transition. The resonant laser is then switched on using a fiber-coupled
acousto-optical modulator (CETC SGTF200) with an extinction ratio exceeding \(60\ \mathrm{dB}\), and the
time-dependent PSB fluorescence is recorded. A trigger generated by the
arbitrary function generator starts the time-tagger at the rising edge of
the resonant pulse.

The acquisition window is adjusted according to the characteristic
spin-pumping time. For the slowly decaying \(E_x\) and \(E_y\)
transitions, acquisition windows extend to \(200\ \mu\mathrm{s}\) or
longer, whereas shorter windows are sufficient for the \(A_1\), \(A_2\),
and \(E_{1,2}\) transitions. The extraction of intrinsic spin-flip rates
from these fluorescence transients, including the correction for finite
optical excitation rate, is described in Supplementary Note~8.

\subsection*{Optical saturation measurements}

\begin{figure}[htbp]
\centering
\includegraphics[scale=0.6]{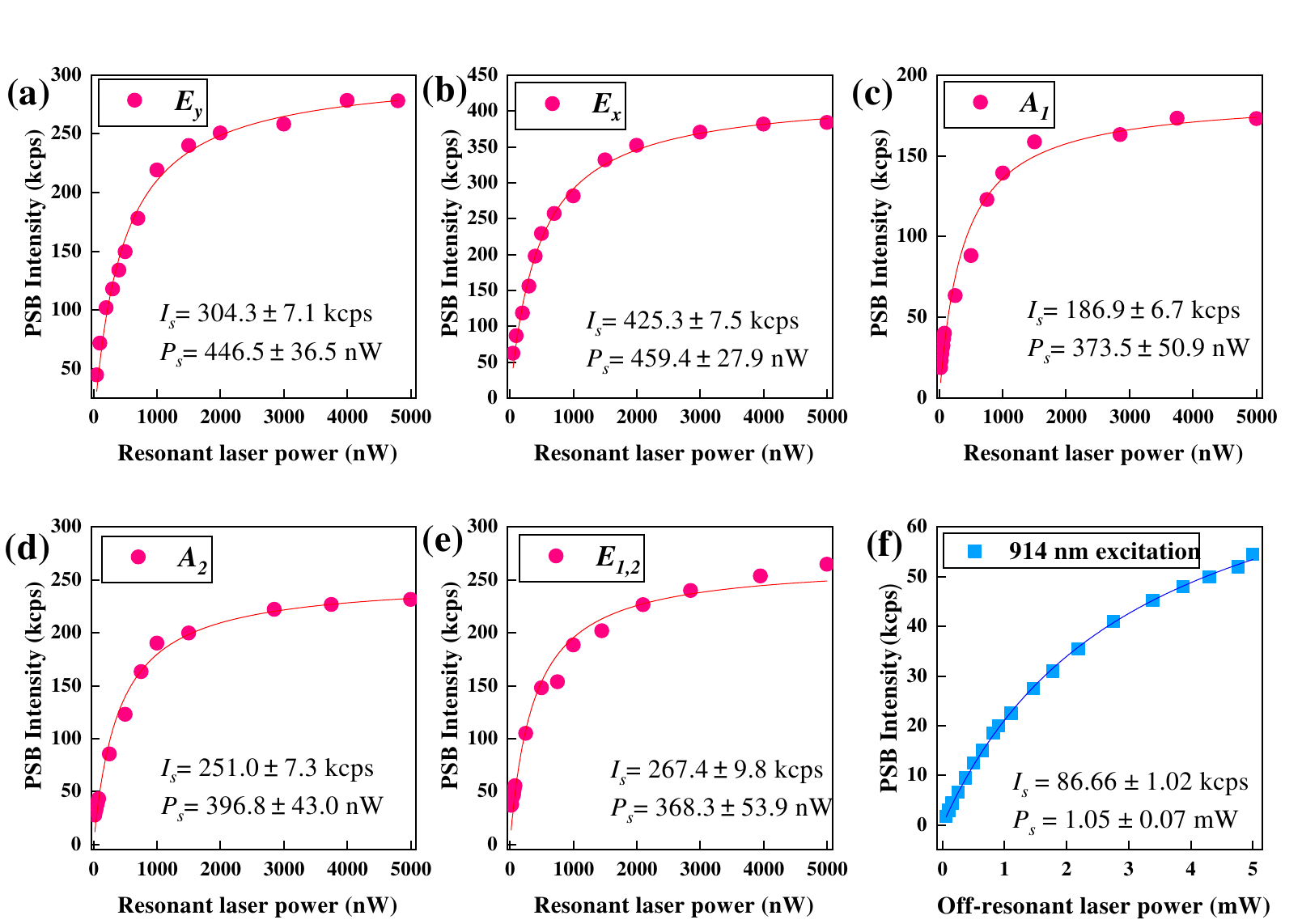}
\caption{
\textbf{Power-dependent PSB fluorescence under resonant and off-resonant excitation.}
(a)--(e) PSB count rate as a function of resonant excitation power for the
\(E_y\), \(E_x\), \(A_1\), \(A_2\), and \(E_{1,2}\) transitions,
respectively. (f) PSB count rate under \(914\ \mathrm{nm}\) off-resonant
excitation. Solid curves are fits to
\(I(P)=I_SP/(P+P_S)\). The resonant saturation parameters are summarized
in Supplementary Table~1; the off-resonant fit gives
\(P_S=1.05\pm0.07\ \mathrm{mW}\) and
\(I_S=86.66\pm1.02\ \mathrm{kcps}\).
}
\label{fig:SA}
\end{figure}

The resonant saturation curves are obtained from the power-dependent
spin-flip measurements. For each optical transition, the PSB count rate is
extracted from the accumulated fluorescence and fitted to

\[
{
I(P)=I_S\frac{P}{P+P_S},
}
\]

where \(P_S\) and \(I_S\) denote the saturation power and saturation count
rate, respectively. Supplementary Figs.~S\ref{fig:SA}(a)--S\ref{fig:SA}(e) show the resonant
saturation curves for the five resolved optical transitions, and the
extracted parameters are summarized in Supplementary Table~\ref{tab:SA}.

\begin{table}[htbp]
\centering
\caption{
Saturation parameters of the five resolved PL6 optical transitions.
All data are fitted with \(I(P)=I_SP/(P+P_S)\).
}
\label{tab:SA}
\begin{tabular}{l c c}
\hline
State & Saturation power (nW) & Saturation count rate (kcps) \\
\hline
\(E_y\)     & \(446.5 \pm 36.5\) & \(304.3 \pm 7.1\) \\
\(E_x\)     & \(459.4 \pm 27.9\) & \(425.3 \pm 7.5\) \\
\(A_1\)     & \(373.5 \pm 50.9\) & \(186.9 \pm 6.7\) \\
\(A_2\)     & \(396.8 \pm 43.0\) & \(251.0 \pm 7.3\) \\
\(E_{1,2}\) & \(368.3 \pm 53.9\) & \(267.4 \pm 9.8\) \\
\hline
\end{tabular}
\end{table}

The saturation powers of the \(E_x\) and \(E_y\) transitions agree with
our previously reported values obtained from power-resolved PLE measurements
\cite{He2024robust}. The \(A_1\) transition exhibits the lowest saturation
count rate among the five branches, consistent with its large nonradiative
ISC rate determined from the lifetime measurements [main-text Fig.~2(c)].

We additionally characterize the PSB fluorescence under off-resonant
\(914\ \mathrm{nm}\) excitation [Supplementary Fig.~S\ref{fig:SA}(f)]. Fitting the
same saturation model gives
\(P_S=1.05\pm0.07\ \mathrm{mW}\) and
\(I_S=86.66\pm1.02\ \mathrm{kcps}\). The substantially larger saturation
power compared with resonant excitation reflects the less efficient
off-resonant excitation pathway.

\subsection*{Debye--Waller factor of hh-OV PL6}

To quantify the fraction of optical emission into the zero-phonon line
(ZPL), we measure the Debye--Waller (DW) factor of individual hh-OV PL6
centers under off-resonant excitation at \(6.35\ \mathrm{K}\). We define

\[
{
\mathrm{DW}
=
\frac{I_{\mathrm{ZPL}}}
     {I_{\mathrm{ZPL}}+I_{\mathrm{PSB}}},
     }
\]

where \(I_{\mathrm{ZPL}}\) and \(I_{\mathrm{PSB}}\) are the spectrally
integrated detected intensities associated with the ZPL and phonon
sideband, respectively.

The spectra are acquired under \(914\ \mathrm{nm}\) continuous-wave
excitation using a home-built scanning single-photon spectrometer consisting
of a blazed grating (LBTEK, BG25-600-1000), a piezo-driven rotation stage
(Newport, CONEX-AGP), and a superconducting nanowire single-photon detector.
To obtain the integrated ZPL intensity more accurately, the ZPL is fitted
with a Gaussian profile, as shown in the insets of Supplementary
Fig.~S\ref{fig:DWfactor}, and \(I_{\mathrm{ZPL}}\) is determined from the integrated area
of the fitted Gaussian peak. The total emission intensity,
\(I_{\mathrm{ZPL}}+I_{\mathrm{PSB}}\), is obtained by numerical integration
of the full measured spectrum.

For the PL6 center labeled hh-OV1, with a transverse strain parameter
\(\delta_\perp=4.505~\mathrm{GHz}\)
[Supplementary Fig.~S\ref{fig:DWfactor}(a)], this analysis yields
\(\mathrm{DW}=10.4\%\).
For a second PL6 center, labeled hh-OV2, with
\(\delta_\perp=0.688~\mathrm{GHz}\)
[Supplementary Fig.~S\ref{fig:DWfactor}(b)], we obtain
\(\mathrm{DW}=10.0\%\). The similar values suggest no appreciable
variation of the measured DW factor over the strain range examined here. However, the spectral range accessible with our spectrometer grating limits the measurements to wavelengths below 1100 nm. Consequently, phonon-sideband emission beyond 1100 nm is excluded from the integrated total emission intensity. The extracted DW factors should therefore be regarded as upper bounds on their absolute values.

The measured DW factor of hh-OV PL6 is considerably larger than that of the
structurally analogous hh divacancy, for which a value of \(3.69\%\) has been
reported~\cite{Jin2021} and multimode dynamic Jahn--Teller theory gives
\(7.04\%\)~\cite{Zalandauskas2025}. This is consistent with the weaker
electron--phonon coupling obtained from first principles for the neutral
\(\mathrm{O_CV_{Si}}\) complex, a total Huang--Rhys factor of \(S=2.01\) with a
mass-weighted displacement of
\(\Delta Q=0.66~\mathrm{\AA}\,\mathrm{amu}^{1/2}\)~\cite{Kobayashi2023},
compared with \(S=3.21\) and
\(\Delta Q=0.823~\mathrm{\AA}\,\mathrm{amu}^{1/2}\) for the hh
divacancy~\cite{Zalandauskas2025}.

\begin{figure}[htbp]
    \centering
    \includegraphics[scale=1]{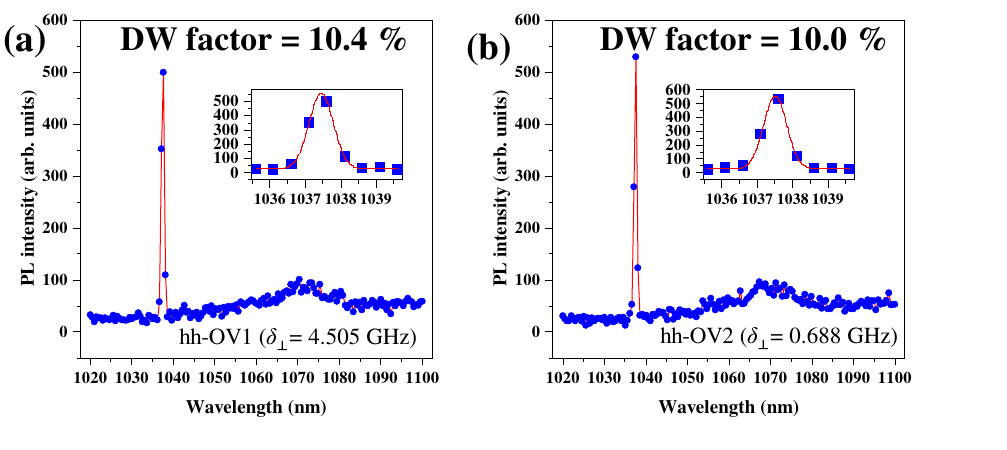}
    \caption{
    \textbf{Debye--Waller factor of hh-OV PL6.}
    (a) Emission spectrum of hh-OV1
    (\(\delta_\perp=4.505\ \mathrm{GHz}\)) under \(914\ \mathrm{nm}\)
    excitation at \(6.35\ \mathrm{K}\), showing the ZPL and PSB.
    The inset shows an enlarged view of the ZPL and its Gaussian fit.
    The integrated ZPL intensity is obtained from the fitted Gaussian
    peak, yielding a DW factor of \(10.4\%\).
    (b) Corresponding spectrum and ZPL fit of hh-OV2
    (\(\delta_\perp=0.688\ \mathrm{GHz}\)), yielding
    \(\mathrm{DW}=10.0\%\).
    }
    \label{fig:DWfactor}
\end{figure}


\clearpage
\section*{Supplementary Note 3: Strain-dependent PLE spectra of hh-OV PL6}

This section presents the complete set of photoluminescence-excitation (PLE)
spectra used to determine the strain-dependent excited-state structure of
single hh-OV PL6 centers. The data points shown in Fig.~1(b) of the main text
are extracted from multi-peak Lorentzian fits to the spectra shown in
Supplementary Figs.~S\ref{fig:strain_spectra}(a)-(f), together with the low-strain spectrum at
\(\delta_{\perp}=0.688\ \mathrm{GHz}\) shown in Fig.~1(c) of the main text.
The corresponding time-resolved PLE measurement of this low-strain center is
shown in Fig.~1(d).

The transverse-strain parameter \(\delta_{\perp}\) is determined directly from
the measured \(E_x\) and \(E_y\) transition frequencies according to
\[
\delta_{\perp}
=
\frac{1}{2}
\left|
\nu_{E_x}-\nu_{E_y}
\right|,
\]
following the definition of the transverse-strain term in the excited-state
Hamiltonian given in Supplementary Note~1. \(\nu_{E_x}\) and \(\nu_{E_y}\) denote the frequencies of the optical transitions connecting the ground state \(|m_s=0\rangle\) to the excited states \(E_x\) and \(E_y\), respectively. All spectra were acquired under
continuous-wave microwave driving resonant with the ground-state spin
transition
(\(f_{\mathrm{MW}}=1.365\ \mathrm{GHz}\),
\(P_{\mathrm{MW}}=-25\ \mathrm{dBm}\)), together with a
\(914\ \mathrm{nm}\) repump laser (\(200\ \mathrm{nW}\)) for charge-state
stabilization.

\begin{figure}[htbp]
    \centering
    \includegraphics[scale=1]{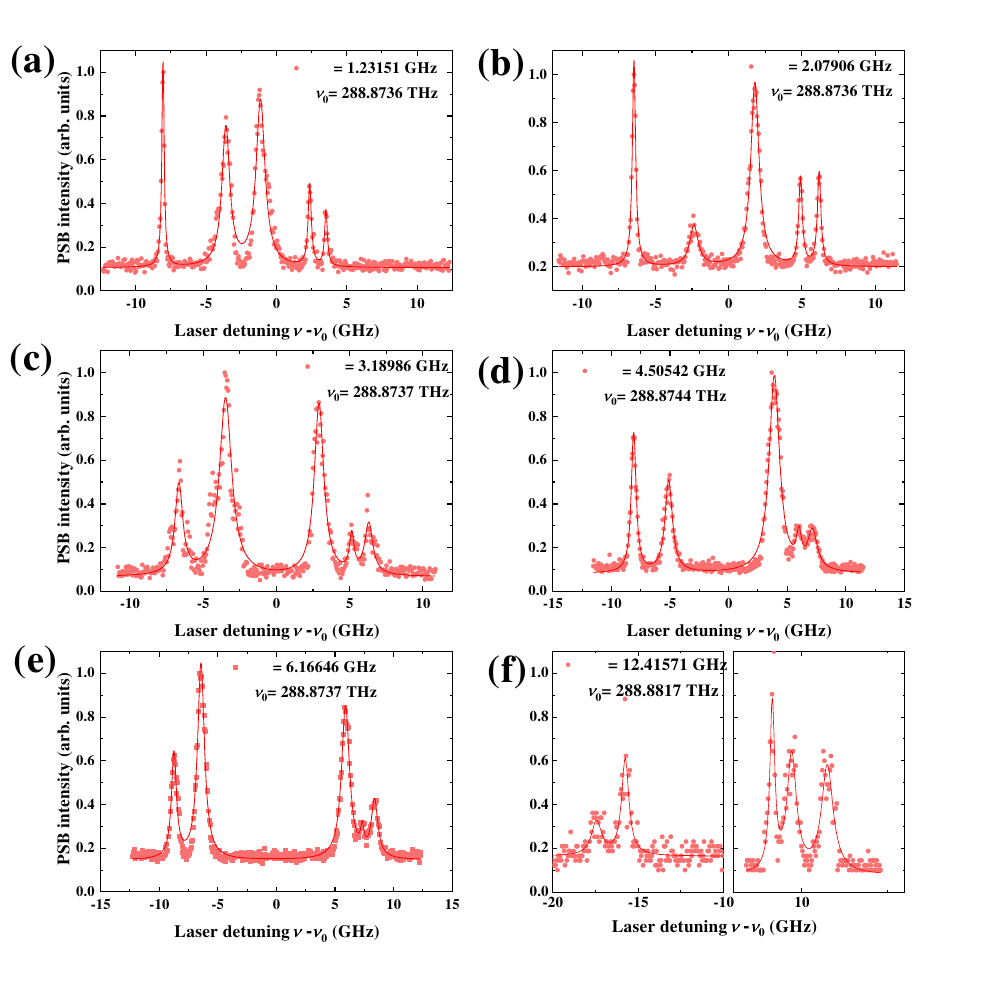}
    \caption{
    \textbf{Strain-dependent PLE spectra of single hh-OV PL6 centers.}
    PLE spectra of individual centers with transverse-strain parameters
    \(\delta_{\perp}\) of
    (a) \(1.23151\ \mathrm{GHz}\),
    (b) \(2.07906\ \mathrm{GHz}\),
    (c) \(3.18986\ \mathrm{GHz}\),
    (d) \(4.50542\ \mathrm{GHz}\),
    (e) \(6.11646\ \mathrm{GHz}\), and
    (f) \(12.41571\ \mathrm{GHz}\).
    Solid curves are multi-peak Lorentzian fits used to extract the five
    resolved excited-state transition frequencies.
    }
    \label{fig:strain_spectra}
\end{figure}

The evolution of the optical spectrum with increasing
\(\delta_{\perp}\) is quantitatively reproduced by the excited-state
Hamiltonian introduced in Supplementary Note~1. Transverse strain shifts and
mixes the symmetry-adapted excited states, producing the systematic evolution
of the five observed optical branches shown in Fig.~1(b) of the main text.

An important consequence of this level structure is the strongly
state-dependent strain susceptibility. In the low-strain regime, the
predominantly \(E_x\)- and \(E_y\)-like branches exhibit approximately linear
energy shifts with transverse strain and therefore possess appreciable
first-order susceptibilities
\(\partial E_{E_x,E_y}/\partial\delta_\perp\).
In contrast, the \(A_1\), \(A_2\), and \(E_{1,2}\) branches vary much more
weakly with \(\delta_\perp\) in this regime, exhibiting a comparatively
``clock-like'' response with reduced first-order strain susceptibility.
This difference provides a direct connection between the strain-dependent
level structure and the optical-linewidth measurements in Supplementary
Fig.~S\ref{fig:ple_microwave2}: local strain fluctuations, or perturbations coupling to the orbital
degree of freedom in a strain-like manner, are expected to produce larger
frequency fluctuations for the strain-sensitive \(E_x/E_y\) transitions than
for the more weakly strain-sensitive \(A_1\), \(A_2\), and \(E_{1,2}\)
transitions.

For the lower-strain centers
(\(\delta_{\perp}=0.688\), \(1.23151\), \(2.07906\), and
\(3.18986\ \mathrm{GHz}\)), \(\delta_{\perp}\) remains smaller than the
measured longitudinal spin-orbit interaction
\(\lambda=5.739\ \mathrm{GHz}\). The excited-state structure therefore
retains its characteristic spin-orbit-dominated level ordering, while the
different slopes of the five branches already reveal their strongly
state-dependent strain susceptibilities
[main-text Figs.~1(b) and 1(c) and Supplementary
Figs.~S\ref{fig:strain_spectra}(a), S\ref{fig:strain_spectra}(b), and S\ref{fig:strain_spectra}(c)].

As \(\delta_{\perp}\) becomes comparable to or larger than \(\lambda\)
(\(\delta_{\perp}=4.50542\), \(6.11646\), and
\(12.41571\ \mathrm{GHz}\)), transverse strain increasingly modifies both
the excited-state eigenenergies and their spin-orbital compositions
[Supplementary Figs.~S\ref{fig:strain_spectra}(d), S\ref{fig:strain_spectra}(e), and S\ref{fig:strain_spectra}(f)]. Nevertheless, the five
branches remain spectrally distinguishable over the measured range.
In particular, with increasing strain the predominantly \(E_x\)-like branch
approaches the \(A_1\)-like branch, increasing the possibility of
off-resonant excitation of transitions associated with the
\(m_s=\pm1\) manifold when addressing the nominal
\(m_s=0\rightarrow E_x\) transition.

Despite the substantial variation in transverse strain, the fitted optical frequency varies only weakly among the measured centers and
lies near \(\nu_0=288.874\ \mathrm{THz}\). This relatively small center-to-center
variation of \(\nu_0\) coexists with the pronounced strain dependence of the
fine-structure splittings, demonstrating that transverse strain primarily
redistributes the excited-state fine structure rather than producing a
comparable shift of the overall optical transition frequency.

A quantitative prediction of the optical linewidth as a function of strain
would additionally require the magnitude and spectral characteristics of the
local strain fluctuations, which are not independently determined in the
present experiment. The Hamiltonian analysis nevertheless determines the
state-dependent strain susceptibility and therefore explains why the
\(E_x/E_y\) transitions are more sensitive to strain-like fluctuations than
the comparatively clock-like \(A_1\), \(A_2\), and \(E_{1,2}\) branches.
The measured spectra and calculated eigenenergies validate the
excited-state Hamiltonian and establish the microscopic basis for the
linewidth differences discussed in Supplementary Note~2.

\clearpage

\section*{Supplementary Note 4: Bayesian MCMC analysis of the strain-dependent excited-state fine structure}

The fine-structure parameters
\((D_{\mathrm{ES}},D_1,D_2,\lambda)\)
of the PL6 excited-state Hamiltonian are determined from the
strain-dependent photoluminescence-excitation (PLE) spectra using a Bayesian
Markov chain Monte Carlo (MCMC) analysis.

The Bayesian treatment provides posterior probability distributions and
credible intervals for the Hamiltonian parameters.

To reduce the influence of occasional poorly described spectral points, we
use a two-component Gaussian likelihood with a broad component for potential
outliers.

\subsection*{Experimental transition energies and uncertainties}

The experimental transition energies are extracted from the PLE spectra of
seven individual PL6 centers spanning transverse-strain parameters
\(\delta_{\perp}=0.688\)--\(12.416\ \mathrm{GHz}\)
[main-text Fig.~1(b) and Supplementary Fig.~S\ref{fig:strain_spectra}]. Each optical peak is obtained
from a multi-peak Lorentzian fit, with typical fitting uncertainties of
\(8\)--\(20\ \mathrm{MHz}\).

The spectra are measured under continuous microwave driving resonant with the
ground-state zero-field splitting,
\(D_{\mathrm{GS}}=1.365\ \mathrm{GHz}\). The \(E_x\) and \(E_y\)
transitions originate predominantly from the \(m_s=0\) ground-state
sublevel, whereas the \(A_1\), \(A_2\), and \(E_{1,2}\) transitions
originate predominantly from the \(m_s=\pm1\) manifold. Because the
\(m_s=\pm1\) ground states lie \(D_{\mathrm{GS}}\) above the \(m_s=0\)
ground state, the measured optical frequencies of the
\(A_1\), \(A_2\), and \(E_{1,2}\) branches are shifted by
\(-D_{\mathrm{GS}}\) relative to excited-state energies referenced to
\(m_s=0\). We therefore add \(D_{\mathrm{GS}}\) to these measured optical
frequencies before comparison with the excited-state Hamiltonian.

\subsection*{Bayesian likelihood and sample-dependent energy offsets}

For center \(i\) and optical branch \(j\), we denote the corrected
experimental transition energy and its fitting uncertainty by
\(E_{\mathrm{exp}}^{(i,j)}\) and
\(\sigma_{\mathrm{exp}}^{(i,j)}\), respectively. For a given set of
Hamiltonian parameters
\[
\boldsymbol{\theta}
=
(D_{\mathrm{ES}},D_1,D_2,\lambda),
\]

the theoretical transition energies
\(E_{\mathrm{th}}^{(i,j)}(\boldsymbol{\theta})\)
are obtained by numerical diagonalization of the excited-state Hamiltonian
in Supplementary Note~1 using the experimentally determined transverse-strain
parameter \(\delta_{\perp,i}\) of the corresponding center. The values of
\(\delta_{\perp,i}\) are treated as fixed inputs in the MCMC analysis.

A sample-dependent offset \(\xi_i\) is included to account for the overall
optical-frequency shift of each individual center. Because such an offset
acts equally on all five transitions of a given center, it does not alter the
fine-structure splittings that determine
\(D_{\mathrm{ES}},D_1,D_2\), and \(\lambda\).

The likelihood for each measured transition is taken as a two-component Gaussian mixture:
\[
\begin{aligned}
p\!\left(
E_{\mathrm{exp}}^{(i,j)}
\mid\boldsymbol{\theta},\xi_i,\eta
\right)
={}&
(1-\eta)\,
\mathcal{N}\!\left[
E_{\mathrm{exp}}^{(i,j)}-\xi_i;
E_{\mathrm{th}}^{(i,j)}(\boldsymbol{\theta}),
\sigma_{\mathrm{good}}^2
\right]
\\
&+
\eta\,
\mathcal{N}\!\left[
E_{\mathrm{exp}}^{(i,j)}-\xi_i;
E_{\mathrm{th}}^{(i,j)}(\boldsymbol{\theta}),
\sigma_{\mathrm{bad}}^2
\right].
\end{aligned}
\]

Here, \(\mathcal{N}(x;\mu,\sigma^2)\) denotes a Gaussian probability
density with mean \(\mu\) and variance \(\sigma^2\), and \(\eta\) is the
probability that a given data point belongs to the broader error component.
We use
\(\sigma_{\mathrm{good}}=50\ \mathrm{MHz}\) and
\(\sigma_{\mathrm{bad}}=500\ \mathrm{MHz}\).
The narrow component accounts for typical residual model--experiment
deviations, whereas the broad component prevents a small number of anomalous
points from dominating the posterior. The Lorentzian fitting uncertainties
\(\sigma_{\mathrm{exp}}^{(i,j)}\) are used in the subsequent normalized
residual and goodness-of-fit analyses but are not included in the
Gaussian-mixture likelihood used for MCMC sampling.

The priors used in the analysis are
\[
\begin{aligned}
\lambda
&\sim \mathcal{N}(5.8,\ 0.5^2)\ \mathrm{GHz}, &
D_{\mathrm{ES}}
&\sim \mathcal{N}(0.94,\ 0.05^2)\ \mathrm{GHz}, \\
D_1
&\sim \mathcal{N}(0,\ 0.02^2)\ \mathrm{GHz}, &
D_2
&\sim \mathcal{N}(0.28,\ 0.05^2)\ \mathrm{GHz}, \\
\xi_i
&\sim \mathcal{N}(-0.6,\ 0.3^2)\ \mathrm{GHz}, &
\eta
&\sim \mathrm{Beta}(2,38).
\end{aligned}
\]
The prior on \(\eta\) has a mean of \(0.05\), corresponding to a weak
prior expectation that only a small fraction of the measured transitions may
require the broader error component.

\subsection*{MCMC sampling and convergence}

The posterior distribution is sampled using a random-walk Metropolis
algorithm. Two independent chains are used, each with \(1000\) burn-in
steps followed by \(2000\) retained samples. During the first half of the
burn-in period, the Gaussian proposal widths are adjusted every \(100\)
steps: they are reduced by \(10\%\) when the recent acceptance probability
falls below \(20\%\) and increased by \(10\%\) when it exceeds \(50\%\).
The proposal widths are subsequently held fixed, and the retained samples
from the two chains are combined for posterior analysis.

The experimental data set contains five resolved transition energies from
each of seven PL6 centers, giving a total of 35 measured energies. For each
MCMC step, the excited-state Hamiltonian is constructed and diagonalized for
each center, the calculated eigenenergies are associated with the five
experimentally observed branches, and the joint likelihood is evaluated over
all measured transition energies.

The two chains are initialized independently by applying small random
perturbations to the starting parameter values. The posterior estimates
reported below are obtained from the combined samples retained after the
burn-in period.

\subsection*{Posterior estimates of the excited-state parameters}

The Bayesian analysis gives the following posterior means and
\(95\%\) credible intervals, with the interval bounds corresponding to the
\(2.5\%\) and \(97.5\%\) posterior quantiles:

\[
\boxed{
{
\begin{aligned}
\lambda
&=5.739\ \mathrm{GHz}
\quad [5.656,\ 5.881]\ \mathrm{GHz},\\
D_{\mathrm{ES}}
&=0.932\ \mathrm{GHz}
\quad [0.913,\ 0.957]\ \mathrm{GHz},\\
D_1
&=0.026\ \mathrm{GHz}
\quad [0.0004,\ 0.037]\ \mathrm{GHz},\\
D_2
&=0.285\ \mathrm{GHz}
\quad [0.272,\ 0.302]\ \mathrm{GHz}.
\end{aligned}
}
}
\]

The longitudinal spin--orbit parameter \(\lambda\) is particularly well
resolved by the strain-dependent spectra and is substantially larger than the
corresponding value reported for the crystallographically analogous
hh-DV PL1. The parameter \(D_2\), which determines the
\(A_1\)--\(A_2\) splitting, is also tightly constrained.
By contrast, \(D_1\) is much smaller and has a strongly asymmetric posterior
extending close to zero, reflecting the relatively weak spectroscopic
sensitivity to the associated coupling between the
\(\{E_x,E_y\}\) and \(\{E_1,E_2\}\) subspaces.

\subsection*{Model validation}

The posterior-mean Hamiltonian reproduces the measured strain dependence of
the five excited-state branches over the full
\(\delta_\perp=0.688\)--\(12.416\ \mathrm{GHz}\) range
[main-text Fig.~1(b)].

The residuals are evaluated by comparing the posterior-mean theoretical
energies with the offset-corrected experimental energies. The Lorentzian
peak-fitting uncertainties are used to calculate the normalized residuals,
the weighted coefficient of determination, and the corresponding
goodness-of-fit quantities.

\subsection*{Comparison with PL1, PL2, and the NV center}

To facilitate a direct comparison with related spin defects, we express the
excited-state Hamiltonian parameters of PL6, together with those of the
hh-DV PL1, the kk-DV PL2, and the NV center in diamond,
using the convention introduced for the NV center by Maze
\textit{et al.}~\cite{Maze2011group}. The parameters used throughout this work
are related to the convention by

\[
\Delta_2=\sqrt{2}D_1,\qquad
\Delta_1=2D_2.
\]

As summarized in Supplementary Table~\ref{tab:params_maze_notation}, the most
pronounced difference between PL6 and the crystallographically analogous
hh-DV PL1 is the longitudinal spin-orbit interaction:
\(\lambda=5.739~\mathrm{GHz}\) for PL6, compared with
\(3.538~\mathrm{GHz}\) for PL1. The PL6 value is instead comparable in
magnitude to those reported for the kk-DV PL2
(\(6.090~\mathrm{GHz}\)) and the NV center
(\(\sim5.3~\mathrm{GHz}\)). Because PL6 and PL1 occupy analogous
hh configurations but differ in their local defect chemistry, this enhanced
spin-orbit interaction provides
a clear spectroscopic indication of a substantial modification
of the excited-state electronic structure associated with oxygen substitution.

In contrast, the spin--spin parameters listed in Supplementary
Table~\ref{tab:params_maze_notation} show much smaller differences between
PL6 and either PL1 or PL2. The excited-state zero-field splitting of PL6,
\(D_{\mathrm{ES}}=0.932~\mathrm{GHz}\), is moderately larger than the values
reported for PL1 (\(0.855~\mathrm{GHz}\)) and PL2
(\(0.852~\mathrm{GHz}\)), while remaining substantially smaller than the
NV-center value of approximately \(1.42~\mathrm{GHz}\).
Similarly, \(\Delta_1=0.570~\mathrm{GHz}\) for PL6 is close to the
reported values for PL1 and PL2.

\begin{table*}[htbp]
\centering
\caption{Excited-state Hamiltonian parameters of PL6 (hh-OV), PL1 (hh-DV),
and PL2 (kk-DV) in 4H-SiC and the NV center in diamond, expressed in the
Maze convention~\cite{Maze2011group}. All Hamiltonian parameters are in GHz.
The PL6 values are converted from the experimentally determined parameters
reported in the main text using
\(\Delta_2=\sqrt{2}D_1\), \(\Delta_1=2D_2\).
PL1 and PL2 values are from Christle \textit{et al.}~\cite{christle2017isolated}.
The NV values are from Batalov \textit{et al.}~\cite{Bolalov2009low}.}
\label{tab:params_maze_notation}
{
\begin{tabular}{lcccc}
\hline
Parameter
& PL6 (hh-OV)
& PL1 (hh-DV)
& PL2 (kk-DV)
& NV \\
\hline
\(\lambda\)
& \(5.739\,[5.656,5.881]\)
& \(3.538\pm0.052\)
& \(6.090\pm0.052\)
& \(\sim5.3\) \\
\(D_{\mathrm{ES}}\)
& \(0.932\,[0.913,0.957]\)
& \(0.855\pm0.017\)
& \(0.852\pm0.012\)
& \(\sim1.42\) \\
\(\Delta_1\)
& \(0.570\,[0.544,0.604]\)
& \(0.577\pm0.019\)
& \(0.584\pm0.012\)
& \(\sim1.55\) \\
\(\Delta_2\)
& \(0.0368\,[0.0006,0.0523]\)
& \(0.031_{-0.031}^{+0.050}\)
& \(0.044_{-0.044}^{+0.046}\)
& \(\sim0.2\) \\
ZPL (nm)
& \(\sim1038\)
& \(\sim1132\)
& \(\sim1131\)
& \(\sim637\) \\
\hline
\end{tabular}
}
\end{table*}

The coupling parameter \(\Delta_2\) is also small for all three SiC defects
(Supplementary Table~\ref{tab:params_maze_notation}:
\(0.0368~\mathrm{GHz}\) for PL6, approximately
\(0.031~\mathrm{GHz}\) for PL1, and approximately
\(0.044~\mathrm{GHz}\) for PL2, compared with
\(\sim0.2~\mathrm{GHz}\) for the NV center).
The small \(\Delta_2\) reduces the direct coupling between the predominantly
\(\{E_x,E_y\}\) states and the predominantly \(\{E_1,E_2\}\) states.
The resulting spin admixture, however, is also governed by
the energy separations between the coupled states, which depend on
\(\lambda\), \(D_{\mathrm{ES}}\), \(\Delta_1\), and transverse strain.
Together, these parameters determine the spin purity of the low-strain
\(E_x/E_y\) branches discussed in Supplementary Note~9.

Overall, the comparison in Supplementary Table~\ref{tab:params_maze_notation}
separates two aspects of the PL6 excited-state structure. Its spin--spin
interaction parameters remain broadly similar to those of the SiC
divacancies, whereas its longitudinal spin-orbit interaction is strongly
enhanced relative to the structurally analogous PL1 center. This distinction
is consistent with a substantial modification of the local electronic
structure produced by oxygen substitution while retaining the overall
\(C_{3v}\) triplet fine-structure framework.

The silicon-vacancy centers V1 and V2 in 4H-SiC also possess
\(C_{3v}\) symmetry, but their relevant electronic states are spin quartets
(\(S=3/2\)) rather than the orbitally degenerate \(S=1\) triplet considered
here. Their effective Hamiltonians therefore have a different structure and
cannot be compared directly through the
\(\lambda\), \(\Delta_1\), and \(\Delta_2\) parameters used above
\cite{Nagy2019high,Liu2024silicon,Morioka2020spin}.

\clearpage

\section*{Supplementary Note 5: First-principles calculation of the excited-state parameters of the hh-OV PL6}

\subsection*{Computational methodology}

First-principles calculations were performed within density functional theory (DFT) using the Vienna \textit{Ab initio} Simulation Package (VASP)~\cite{kresse1996efficient} with the projector augmented wave (PAW) method~\cite{blochl1994projector}. A \(6\times6\times2\) supercell of 4H-SiC containing 576 atoms was used with \(\Gamma\)-point sampling of the Brillouin zone and a plane-wave cutoff of 420~eV. The convergence criteria for the total energy and the atomic forces were \(10^{-4}\)~eV and \(0.01\)~eV/\AA, respectively.

The dynamic Jahn--Teller (DJT) effect in the excited \(^3E\) state was treated by constrained-occupation DFT (CDFT)~\cite{Gali2009} with the Heyd--Scuseria--Ernzerhof (HSE06) functional~\cite{heyd2003hybrid}. The symmetry-broken \(C_{1h}\) minimum was obtained by structural relaxation with integer occupation of the two orbitals derived from the \(e\) doublet, and half-half occupations were used in the fixed-geometry calculations along the Jahn--Teller distortion coordinate to construct the harmonic reference potential, following Ref.~\cite{mohseni2025pressure}. The resulting potential-energy surface was analyzed with the Jahn--Teller dynamics code of Ref.~\cite{Toth2026}, yielding a Ham reduction factor \(p=0.077\), which quantifies the partial quenching of the orbital operators by vibronic coupling in the \(^3E\) manifold~\cite{Thiering2018}. The Perdew--Burke--Ernzerhof (PBE) functional~\cite{perdew1996generalized} was used for the zero-field-splitting tensor \(\bm{D}\), the hyperfine tensor \(\bm{A}\), and the nuclear quadrupole tensor \(\bm{P}\), following Ref.~\cite{Thiering2024Nuclear}. The excited-state parameters \(D_1\), \(D_2\), \(A_1\), \(A_2\), \(P_1\), and \(P_2\) were scaled by the vibronic reduction factor \(q=(1+p)/2=0.54\). The calculated magnetic parameters are collected in Supplementary Table~\ref{tab:dft_magnetic}.

\begin{table}[htbp]
\centering
\caption{
Calculated magnetic parameters of the \(^3E\) excited state of the hh-OV PL6 center. \(D_1\), \(D_2\), \(A_1\), \(A_2\), \(P_1\), and \(P_2\) include the HSE06-derived vibronic reduction factor \(q=0.54\).
}
\label{tab:dft_magnetic}
\begin{tabular}{cccccccccc}
\hline
\(D_{\mathrm{ES}}\) & \(D_1\) & \(D_2\) & \(A_{\parallel}\) & \(A_{\perp}\) & \(A_1\) & \(A_2\) & \(P\) & \(P_1\) & \(P_2\) \\
(GHz) & (GHz) & (GHz) & (MHz) & (MHz) & (kHz) & (kHz) & (MHz) & (kHz) & (kHz) \\
\hline
1.216 & 0.044 & 0.296 & 7.376 & 4.232 & 38.005 & 107.523 & 6.398 & 3.445 & 17.842 \\
\hline
\end{tabular}
\end{table}

\subsection*{Spin--orbit coupling and finite-size scaling}

The intrinsic axial spin--orbit coupling (SOC) parameter \(\lambda_{z0}\) was obtained within the noncollinear formalism with the spin quantization axis along the defect symmetry axis (\(z\parallel c\)). Half-half occupation was applied to the \(e\) doublet within the CDFT procedure and \(\lambda_{z0}\) was extracted from the SOC-induced splitting of the two degenerate levels~\cite{Thiering2018,Thiering2018loop}. As shown in Ref.~\cite{Bian2025}, the standard 576-atom supercell is not necessarily sufficient to converge \(\lambda_{z0}\) for defects in 4H-SiC, so an explicit finite-size analysis was carried out with supercells containing up to 2048 atoms. These SOC calculations used the PBE functional with total-energy and force convergence criteria of \(10^{-6}\)~eV and \(10^{-3}\)~eV/\AA.

Because of the anisotropic hexagonal lattice, the convergence was examined separately along the \(c\) axis and within the basal plane. For an \(m\times m\times n\) supercell, \(m\) and \(n\) denote the expansion factors along the two basal-plane lattice vectors and the \(c\) axis, and the corresponding supercell dimensions are denoted by \(L_{ab}\) and \(L_{c}\). The axial dependence at fixed \(m\) was fitted to
\begin{equation}
\lambda_{z}^{(c)}=\lambda_{z0}^{(c)}+A_c\exp\!\left(-\frac{L_{c}}{\eta_c}\right),
\label{eq:soc-c-scaling}
\end{equation}

where \(A_c\) and \(\eta_c\) are fitting parameters. As shown in Supplementary Fig.~S\ref{fig:soc-scaling}(a), increasing \(n\) from 2 to 3 changes \(\lambda_{z0}^{(c)}\) by less than 0.5~GHz for all basal-plane sizes considered, whereas a pronounced dependence on \(m\) remains. We therefore fixed \(n=2\) for the basal-plane analysis and fitted
\begin{equation}
\lambda_{z}(ab)=\lambda_{z0}+A_{ab}\exp\!\left(-\frac{L_{ab}}{\eta_{ab}}\right).
\label{eq:soc-ab-scaling}
\end{equation}

for \(m=7\)--\(11\) [Supplementary Fig.~S\ref{fig:soc-scaling}(b)], obtaining \(\lambda_{z0}=48.12\pm0.60\)~GHz. The extrapolation beyond the largest computed supercell amounts to only \(3.4\%\), and including the two smallest basal-plane cells in the fit returns \(\lambda_{z0}=52.8\pm3.2\)~GHz, consistent within the fit uncertainty; we therefore quote \(\lambda_{z0}=48\pm3\)~GHz. The complete set of PBE calculations used in the two convergence analyses is summarized in Supplementary Table~\ref{tab:soc_scaling}. Applying the HSE06-derived Ham reduction factor gives the spectroscopic axial SOC parameter \(\lambda=p\lambda_{z0}=3.71\)~GHz, to be compared with the measured value of \(5.739\)~GHz reported in the main text. The residual underestimation originates predominantly from the calculated Ham reduction factor, which is the most sensitive quantity in this chain.

\begin{table}[htbp]
\centering
\caption{
PBE-calculated intrinsic axial SOC parameter \(\lambda_{z0}\) of the hh-OV PL6 center for the supercells used in the finite-size analysis.
}
\label{tab:soc_scaling}
\begin{tabular}{cccc}
\hline
Supercell & \(L_{ab}\) (\AA) & \(L_{c}\) (\AA) & \(\lambda_{z0}\) (GHz) \\
\hline
\(5\times5\times1\)   & 15.44 & 10.11 & 27.57 \\
\(5\times5\times2\)   & 15.44 & 20.22 & 21.52 \\
\(5\times5\times3\)   & 15.44 & 30.33 & 21.04 \\
\(5\times5\times4\)   & 15.44 & 40.44 & 20.92 \\
\(6\times6\times1\)   & 18.53 & 10.11 & 30.59 \\
\(6\times6\times2\)   & 18.53 & 20.22 & 27.44 \\
\(6\times6\times3\)   & 18.53 & 30.33 & 27.08 \\
\(6\times6\times4\)   & 18.53 & 40.44 & 26.84 \\
\(7\times7\times1\)   & 21.62 & 10.11 & 36.51 \\
\(7\times7\times2\)   & 21.62 & 20.22 & 35.18 \\
\(7\times7\times3\)   & 21.62 & 30.33 & 35.06 \\
\(7\times7\times4\)   & 21.62 & 40.44 & 34.94 \\
\(8\times8\times1\)   & 24.71 & 10.11 & 41.35 \\
\(8\times8\times2\)   & 24.71 & 20.22 & 40.62 \\
\(8\times8\times3\)   & 24.71 & 30.33 & 40.50 \\
\(8\times8\times4\)   & 24.71 & 40.44 & 40.50 \\
\(9\times9\times2\)   & 27.80 & 20.22 & 43.89 \\
\(10\times10\times2\) & 30.88 & 20.22 & 46.06 \\
\(11\times11\times2\) & 33.97 & 20.22 & 46.55 \\
\hline
\end{tabular}
\end{table}

\begin{figure}[htbp]
\centering
\includegraphics[width=\columnwidth,height=0.58\textheight,keepaspectratio]{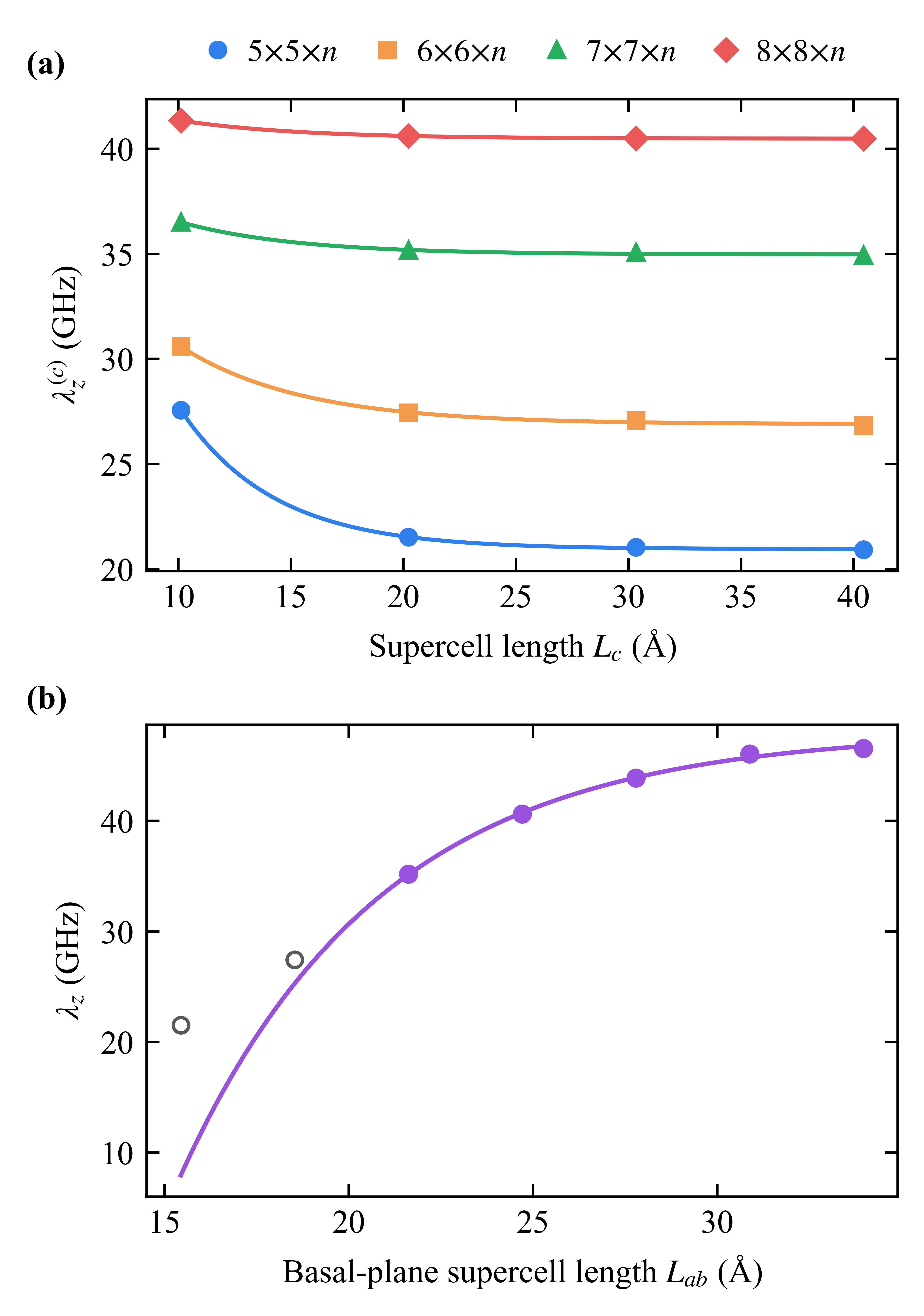}
\caption{
\textbf{Finite-size convergence of the intrinsic axial spin--orbit parameter of hh-OV PL6.}
Exponential convergence of \(\lambda_{z0}\) with supercell size within the \(\Gamma\)-point approximation for the excited state of the OV defect (hh configuration) with the PBE functional.
(a) Dependence on the supercell length along the \(c\) direction for fixed \(5\times5\), \(6\times6\), \(7\times7\), and \(8\times8\) basal-plane dimensions. The solid curves are independent fits to Eq.~(\ref{eq:soc-c-scaling}).
(b) Dependence on the basal-plane dimension for a fixed supercell length along the \(c\) direction of \(L_{c}=20.22\)~\AA, corresponding to \(n=2\). The solid curve is a fit of the five filled purple circles to Eq.~(\ref{eq:soc-ab-scaling}). The two gray open circles, which show the largest deviations from the overall convergence trend, were excluded from the fit; including them returns \(\lambda_{z0}=52.8\pm3.2\)~GHz, consistent with the quoted value within the fit uncertainty.
}
\label{fig:soc-scaling}
\end{figure}

\subsection*{Radiative rate}

The optical transition between the triplet excited \(^3E\) state and the \(^3A_2\) ground state is a spin-conserving electric-dipole transition. The radiative rate is calculated as
\begin{equation}
k_{\mathrm{rad}}=\frac{n E_{\mathrm{ZPL}}^{3}|\mu|^{2}}{3\pi\epsilon_{0}\hbar^{4}c^{3}},
\label{eq:radiative-rate}
\end{equation}

where \(\hbar\) is the reduced Planck constant, \(\epsilon_0\) is the vacuum permittivity, \(c\) is the speed of light in vacuum, and \(n=2.647\) is the refractive index of 4H-SiC~\cite{Bian2025}. \(E_{\mathrm{ZPL}}\) is the zero-phonon-line energy, obtained by CDFT as the total-energy difference between the Jahn--Teller-distorted \(C_{1h}\) minimum of the excited \(^3E\) state and the \(^3A_2\) ground state, and the transition dipole moment (TDM) \(\mu\) was evaluated from the pseudo wave functions of the relevant \(a\) and \(e\) Kohn--Sham defect levels at the distorted excited-state geometry. For the hh-OV center this yields \(k_{\mathrm{rad}}=61.64\)~MHz, in close agreement with the \(62.5\)~MHz deduced from the measured \(E_y\) excited-state lifetime [Fig.~2(c) of the main text]. This agreement is obtained without any adjustable parameter.

\subsection*{Comparison of the calculated properties of hh-OV PL6 and hh-DV PL1}

\begin{table}[htbp]
\centering
\caption{
Calculated properties of the hh-DV (PL1) and hh-OV (PL6) centers obtained within the same first-principles framework. \(\lambda_{z0}\) is the intrinsic axial spin--orbit parameter, \(E_{\mathrm{JT}}\) the Jahn--Teller distortion energy, and \(\Delta\) the energy separation between the \(^3E\) state at the high-symmetry \(C_{3v}\) geometry and the \(^1A_1\) singlet. TDM denotes the optical transition dipole moment. The values in parentheses are the measured zero-phonon-line energies of the respective centers. The PL1 values are taken from Ref.~\cite{Bian2025}.
}
\label{tab:PL1vsPL6_theory}
\begin{tabular}{lcc}
\hline
Parameter & PL1 (hh-DV) & PL6 (hh-OV) \\
\hline
\(\lambda_{z0}\) (GHz)         & 18.50  & 48.12 \\
\(E_{\mathrm{JT}}\) (meV)      & 73.62  & 91.08 \\
\(\Delta\) (meV)               & 233.62 & 202 \\
TDM (D)                        & 7.36   & 8.62 \\
\(E_{\mathrm{ZPL}}\) (eV)      & 1.15 (1.095) & 1.24 (1.195) \\
\hline
\end{tabular}
\end{table}

Supplementary Table~\ref{tab:PL1vsPL6_theory} collects the calculated quantities that govern the radiative and nonradiative decay channels of the two centers. Both sets were obtained within the same framework, so the comparison is internally consistent.

The upper-branch intersystem crossing is governed by the \emph{intrinsic}, unquenched spin--orbit parameter. The vibronic reduction that converts \(\lambda_{z0}\) into the spectroscopic \(\lambda_z=p\lambda_{z0}\) applies to the orbital operators within the \(^3E\) manifold, whereas the non-adiabatic triplet--singlet matrix element is proportional to the intrinsic parameter, \(\lambda_{\perp}\propto\lambda_{z0}\)~\cite{Bian2025,Thiering2017}, with the electron--phonon physics carried by the vibrational overlap function \(F(\Delta)\) instead. The calculated \(\lambda_{z0}\) of PL6 is \(2.6\) times larger than that of PL1~\cite{Bian2025}, corresponding to a factor of \(6.8\) in \(\lambda_{\perp}^{2}\), and the smaller triplet--singlet separation \(\Delta\) of PL6 acts in the same direction. We do not attempt to compute an absolute ISC rate here, because it depends exponentially on a triplet--singlet energy gap that cannot presently be determined to the required accuracy, and on a vibrational overlap function that is correspondingly sensitive to it.

\subsection*{Charge-state stability under resonant illumination}

The photostability of the two centers under resonant illumination is set by the position of the \((0/-)\) charge-transition level relative to the conduction-band minimum (CBM), which defines the threshold for reionizing the negatively charged, optically dark state. For the hh divacancy this level lies \(1.245\)~eV below the CBM, consistent with the measured cryogenic threshold of \(1.310\)~eV~\cite{Csore2022}, so that its \(1.095\)~eV zero-phonon-line photon cannot reionize \(\mathrm{V_CV_{Si}^-}\) and resonant driving progressively transfers the center into its dark charge state. For the hh oxygen--vacancy complex, in contrast, the corresponding level is calculated at \(2.15\)~eV above the valence-band maximum within a HSE06 gap of \(3.17\)~eV~\cite{Kobayashi2023}, i.e.\ approximately \(1.0\)~eV below the CBM and below the \(1.195\)~eV zero-phonon-line photon energy of PL6. Resonant excitation of PL6 therefore repumps its own negative charge state, which is consistent with the stable photoluminescence-excitation signal observed over 118~min in Fig.~1(d) of the main text.

\clearpage
\section*{Supplementary Note 6: Theoretical model and experimental results for optical Rabi oscillations}

\subsection*{Theoretical model for optical Rabi oscillations}

The optical coherence dynamics are modeled using the Lindblad master-equation formalism. The time evolution of the density matrix \(\rho\) is given by
\begin{equation}
    \frac{d\rho}{dt}
    =
    -i[\hat{H},\rho]
    +
    \sum_k \mathcal{D}[\hat{C}_k](\rho),
\end{equation}
where
\begin{equation}
    \mathcal{D}[\hat{C}](\rho)
    =
    \hat{C}\rho\hat{C}^{\dagger}
    -
    \frac{1}{2}
    \left(
    \hat{C}^{\dagger}\hat{C}\rho
    +
    \rho\hat{C}^{\dagger}\hat{C}
    \right)
\end{equation}
is the Lindblad dissipator.

For a resonantly driven optical transition between an addressed ground state
\(\lvert g\rangle\) and an excited state \(\lvert e\rangle\), the Hamiltonian under the rotating-wave approximation is
\begin{equation}
    \hat{H}_{\mathrm{pulse}}
    =
    \begin{pmatrix}
        0 & \Omega/2 & 0\\
        \Omega/2 & 0 & 0\\
        0 & 0 & 0
    \end{pmatrix},
\end{equation}
in the basis
\(\{\lvert g\rangle,\lvert e\rangle,\lvert s\rangle\}\),
where \(\Omega\) is the optical Rabi frequency and
\(\lvert s\rangle\) denotes an undriven ground state when required.
After the resonant optical pulse, the Hamiltonian is set to
\(\hat{H}_{\mathrm{off}}=0\).

Population decay from the excited state back to the optically addressed
ground state is described by
\begin{equation}
    \hat{C}_1
    =
    \frac{\lvert g\rangle\langle e\rvert}{\sqrt{T_1}},
\end{equation}
and optical dephasing is described by
\begin{equation}
    \hat{C}_2
    =
    \frac{
    \lvert e\rangle\langle e\rvert
    -
    \lvert g\rangle\langle g\rvert
    }{\sqrt{T_2^*}},
\end{equation}
where \(T_1\) and \(T_2^*\) are the population-decay and dephasing
parameters, respectively, used in the numerical fits.

For the \(E_x\) and \(E_y\) transitions, spin-changing dynamics occur
on a timescale much longer than the nanosecond optical Rabi dynamics
and are therefore neglected, reducing the model to an effective
two-level system.

For the \(A_2\) transition, the polarization-selective
\(\Lambda\)-type level structure requires the opposite
\(\lvert m_s=\pm1\rangle\) ground-state branch to be retained explicitly.
For clarity, we denote the relaxation parameter associated with
population return to the optically addressed ground-state branch as
\(T_g^{\mathrm{fit}}\), such that

\begin{equation}
    \hat{C}_1
    =
    \frac{\lvert g\rangle\langle A_2\rvert}
    {\sqrt{T_g^{\mathrm{fit}}}}.
\end{equation}

Population transfer into the undriven opposite-spin ground-state
branch is represented by an additional collapse operator

\begin{equation}
    \hat{C}_3
    =
    \frac{\lvert s\rangle\langle A_2\rvert}
    {\sqrt{T_s^{\mathrm{fit}}}}.
\end{equation}

Here \(T_g^{\mathrm{fit}}\) and \(T_s^{\mathrm{fit}}\) are effective
relaxation parameters introduced to reproduce the damping and population
redistribution of the driven \(A_2\) Rabi dynamics. Because both
collapse operators act on the same excited-state population in this
reduced three-level description, the fitted parameters are not
interpreted individually as microscopic excited-state lifetimes or
branching lifetimes. In particular, they are not used to determine the
independently measured total \(A_2\) excited-state lifetime.

The master equation is solved numerically using MATLAB's
\texttt{ode45} solver with adaptive step-size control. The density matrix is vectorized for numerical integration, with relative and absolute tolerances of \(10^{-6}\) and \(10^{-9}\), respectively. The fitting parameters are optimized using MATLAB's \texttt{fmincon} function with the interior-point algorithm by minimizing the squared deviation between the measured signal and the simulated excited-state population \(\rho_{ee}(t)\). All simulations are initialized in the optically addressed ground state,
\(\rho_0=\lvert g\rangle\langle g\rvert\).

\subsection*{Spin-dependent Rabi oscillations:
\(\lvert m_s=\pm1\rangle\leftrightarrow\lvert A_2\rangle\)}

\begin{figure}[htbp]
    \centering
    \includegraphics[scale=0.8]{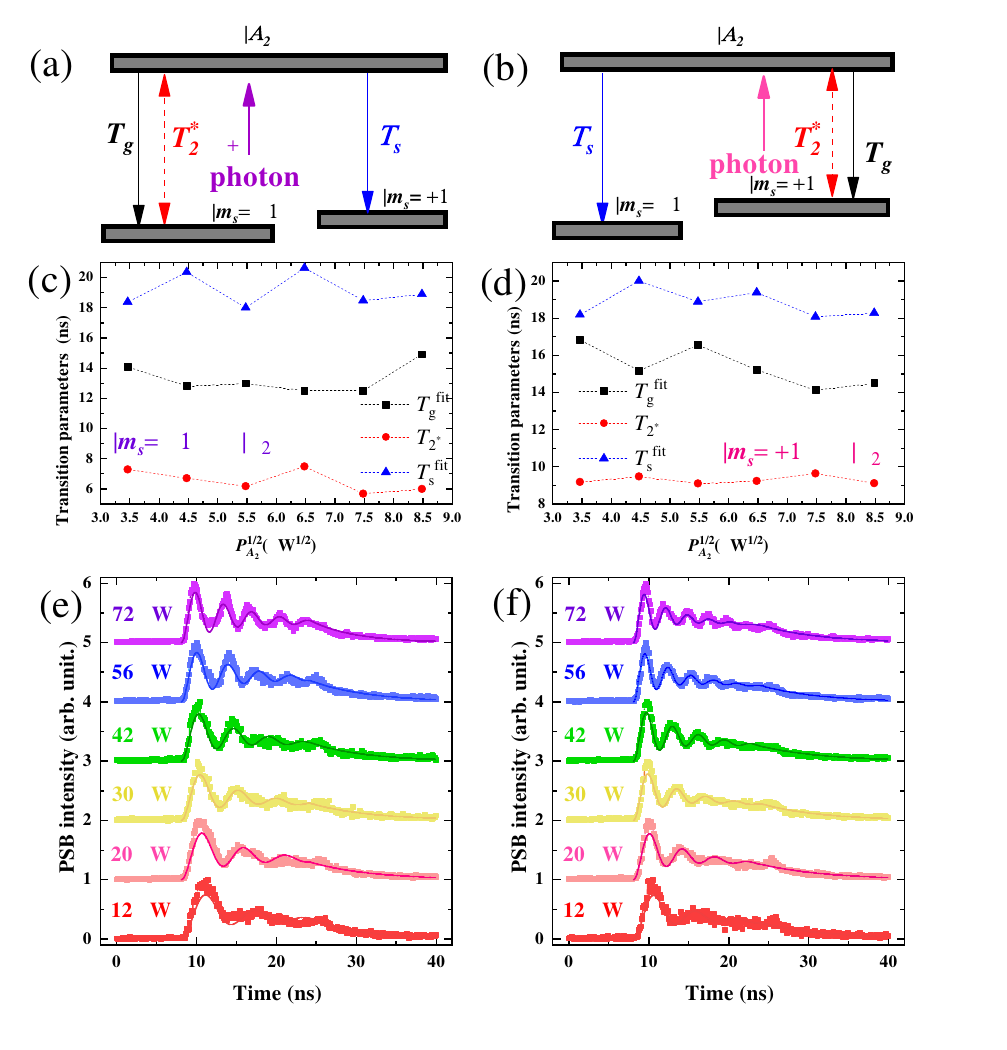}
    \caption{
    \textbf{Spin-selective optical Rabi oscillations between
    \(\lvert m_s=\pm1\rangle\) and \(\lvert A_2\rangle\).}
    (a) Effective three-level model for initialization in
    \(\lvert m_s=-1\rangle\). A \(\sigma_{+}\)-polarized laser
    (purple arrow) resonantly drives
    \(\lvert m_s=-1\rangle\leftrightarrow\lvert A_2\rangle\).
    Population relaxation into the addressed and opposite-spin
    ground-state branches is represented by the effective parameters
    \(T_g^{\mathrm{fit}}\) and \(T_s^{\mathrm{fit}}\), respectively.
    (b) Corresponding three-level model for initialization in
    \(\lvert m_s=+1\rangle\), where a
    \(\sigma_{-}\)-polarized laser (pink arrow) drives
    \(\lvert m_s=+1\rangle\leftrightarrow\lvert A_2\rangle\),
    with the roles of the two ground-state spin sublevels interchanged.
    (c,d) Extracted effective parameters
    \(T_g^{\mathrm{fit}}\), \(T_2^*\), and
    \(T_s^{\mathrm{fit}}\) from the Rabi fits for initialization in
    \(\lvert m_s=-1\rangle\) (c) and
    \(\lvert m_s=+1\rangle\) (d).
    }
    (e,f) Power-dependent optical Rabi oscillations for the two
    spin-selective \(\lvert m_s=-1\rangle\leftrightarrow\lvert A_2\rangle\) (e) and \(\lvert m_s=+1\rangle\leftrightarrow\lvert A_2\rangle\) (f) transitions. Dots represent experimental data
    acquired at resonant laser powers of 12, 20, 30, 42, 56, and
    \(72~\mu\mathrm{W}\). Solid curves are numerical simulations based
    on the effective three-level master-equation model.
    \label{fig:A2Rabi}
\end{figure}

The \(\lvert A_2\rangle\) excited state is optically connected to both
\(\lvert m_s=\pm1\rangle\) ground-state spin sublevels through polarization-selective transitions. We investigate coherent optical driving for the two complementary spin-initialization conditions illustrated in Figs.~S\ref{fig:A2Rabi}(a) and S\ref{fig:A2Rabi}(b).

When the ground-state spin is initialized in
\(\lvert m_s=-1\rangle\), a left-circularly polarized (\(\sigma_{+}\)) laser resonantly drives
\(\lvert m_s=-1\rangle\leftrightarrow\lvert A_2\rangle\).
For initialization in \(\lvert m_s=+1\rangle\), right-circularly polarized (\(\sigma_{-}\)) laser drives
\(\lvert m_s=+1\rangle\leftrightarrow\lvert A_2\rangle\).

For each initialization condition, the initially prepared spin state is
identified as the optically addressed ground state \(\lvert g\rangle\),
while the opposite-spin state is identified as the undriven state
\(\lvert s\rangle\). Thus, for initialization in \(\lvert m_s=-1\rangle\),
\[
\lvert g\rangle=\lvert-1\rangle,\qquad
\lvert s\rangle=\lvert+1\rangle,
\]
whereas for initialization in \(\lvert m_s=+1\rangle\),
\[
\lvert g\rangle=\lvert+1\rangle,\qquad
\lvert s\rangle=\lvert-1\rangle.
\]

Figures~S\ref{fig:A2Rabi}(e) and S\ref{fig:A2Rabi}(f) show the power-dependent optical Rabi oscillations for the two spin-selective \(A_2\) transitions. The experimental results are in good agreement with the numerical simulations over the measured range of laser powers.

For both spin-selective \(A_2\) transitions, the extracted effective
relaxation parameters remain on the \(10\)--\(20~\mathrm{ns}\) scale
over the measured excitation-power range. For excitation from
\(\lvert m_s=-1\rangle\), \(T_g^{\mathrm{fit}}\) ranges from approximately
\(12.5\) to \(14.9~\mathrm{ns}\), while
\(T_s^{\mathrm{fit}}\) ranges from approximately
\(18.0\) to \(20.6~\mathrm{ns}\). For excitation from
\(\lvert m_s=+1\rangle\), \(T_g^{\mathrm{fit}}\) ranges from approximately
\(14.1\) to \(16.8~\mathrm{ns}\), whereas
\(T_s^{\mathrm{fit}}\) lies between approximately
\(18.1\) and \(20.0~\mathrm{ns}\), as shown in Figs.~S\ref{fig:A2Rabi}(c) and \ref{fig:A2Rabi}(d).
These quantities characterize the effective damping and population
redistribution required to reproduce the Rabi oscillations within the
reduced three-level model and are not assigned microscopic branching
lifetimes.

The parameter \(T_s^{\mathrm{fit}}\) introduced here should not be
confused with the spin-flip rate
\(\Gamma_{\mathrm{flip}}^{A_2}\) extracted from the
longer-timescale resonant optical-pumping measurements discussed in
Supplementary Note~8. The latter is obtained from a separate
rate-equation analysis that explicitly accounts for the  intersystem-crossing--singlet
pathway and therefore describes a different dynamical quantity.

\subsection*{Polarization-dependent Rabi oscillations:
\(\lvert m_s=0\rangle\leftrightarrow\lvert E_x\rangle\) and
\(\lvert m_s=0\rangle\leftrightarrow\lvert E_y\rangle\)}

We investigate optical Rabi oscillations between the
\(\lvert m_s=0\rangle\) ground state and the
\(\lvert E_x\rangle\) and \(\lvert E_y\rangle\) excited states.
The ground-state spin is initialized in \(\lvert m_s=0\rangle\).
A linear polarized (X) resonant laser drives
\(\lvert m_s=0\rangle\leftrightarrow\lvert E_x\rangle\),
whereas a orthogonal linear polarized (Y) resonant laser drives
\(\lvert m_s=0\rangle\leftrightarrow\lvert E_y\rangle\).

\begin{figure}[htbp]
    \centering
    \includegraphics[scale=0.8]{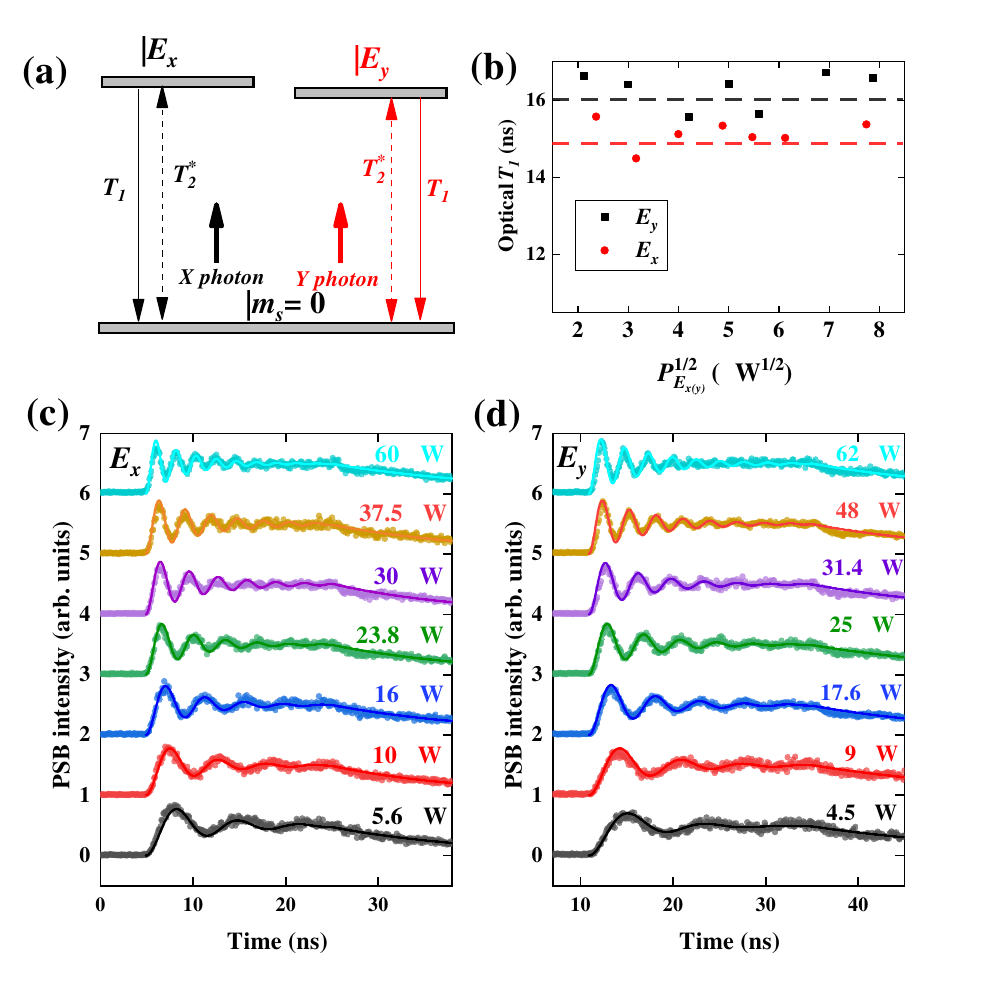}
    \caption{
    \textbf{Optical Rabi oscillations between
    \(\lvert m_s=0\rangle\) and \(\lvert E_{x(y)}\rangle\).}
    (a) Effective two-level model for excitation of
    \(\lvert E_x\rangle\) (black) and
    \(\lvert E_y\rangle\) (red). The spin is initialized in
    \(\lvert m_s=0\rangle\), and linearly polarized resonant light drives
    the selected optical transition.
    Spin-changing dynamics are neglected in these fits because their
    characteristic timescale is much longer than that of the optical Rabi
    oscillations.
    (b) Extracted \(T_1\) as a function of the square root of resonant
    laser power. The horizontal dashed lines indicate the independently
    measured lifetimes in Fig.~2(c) of the main text, approximately
    \(16.0~\mathrm{ns}\) for \(E_y\) and
    \(14.9~\mathrm{ns}\) for \(E_x\).
    (c,d) Power-dependent Rabi oscillations for the \(E_x\) (c) and
    \(E_y\) (d) transitions. Dots represent experimental data measured at
    laser powers of 5.6, 10, 16, 23.8, 30, 37.5, and
    \(60~\mu\mathrm{W}\) for \(E_x\), and 4.5, 9, 17.6, 25, 31.4,
    48, and \(62~\mu\mathrm{W}\) for \(E_y\).
    Solid curves are numerical simulations based on the effective
    two-level master-equation model.}
    \label{fig:ExEyRabi}
\end{figure}

The measured spin-changing dynamics of the \(E_x/E_y\) states occur on a microsecond timescale, much slower than the nanosecond Rabi dynamics considered here. The corresponding spin-flip channel is therefore set to zero in the Rabi fits, and the \(E_x/E_y\) transitions are modeled as effective two-level systems characterized by optical Rabi frequency \(\Omega\), effective optical lifetime \(T_1\), and optical dephasing time \(T_2^*\) , as shown in Fig. ~S\ref{fig:ExEyRabi}(a).

The power-dependent Rabi oscillations are shown in Figs.~S\ref{fig:ExEyRabi}(c) and S\ref{fig:ExEyRabi}(d).
The experimental results are in good agreement with the numerical simulations over the measured range of laser powers. The clear oscillations and their systematic power dependence demonstrate coherent resonant optical driving of both orbital branches.

The fitted \(T_1\) values are shown as a function of the square root of the pumping laser power,
\(P_{E_{x(y)}}^{1/2}\) in Fig.~S\ref{fig:ExEyRabi}(b). The extracted values are approximately
\(16.0~\mathrm{ns}\) for \(E_y\) and
\(14.9~\mathrm{ns}\) for \(E_x\), consistent with the independent lifetime measurements in Fig.~2(c) of the main text. A representative
\(E_x\) optical Rabi measurement is shown in Fig.~2(b) of the main text.

\subsection*{Laser-power dependence of the optical Rabi frequency}

For an electric-dipole transition, the optical Rabi frequency is
\begin{equation}
    \Omega
    =
    \frac{\mathbf{d}\cdot\mathbf{E}}{\hbar},
\end{equation}
where \(\mathbf{d}\) is the transition dipole moment and
\(\mathbf{E}\) is the electric-field amplitude of the resonant laser \cite{Robledo2010Control}.
Since the optical intensity satisfies
\(I\propto|\mathbf{E}|^2\) and is proportional to the incident laser power for a fixed optical mode, the Rabi frequency is expected to scale as
\begin{equation}
    \Omega\propto\sqrt{P}.
\end{equation}

\begin{figure}[htbp]
    \centering
    \includegraphics[scale=0.7]{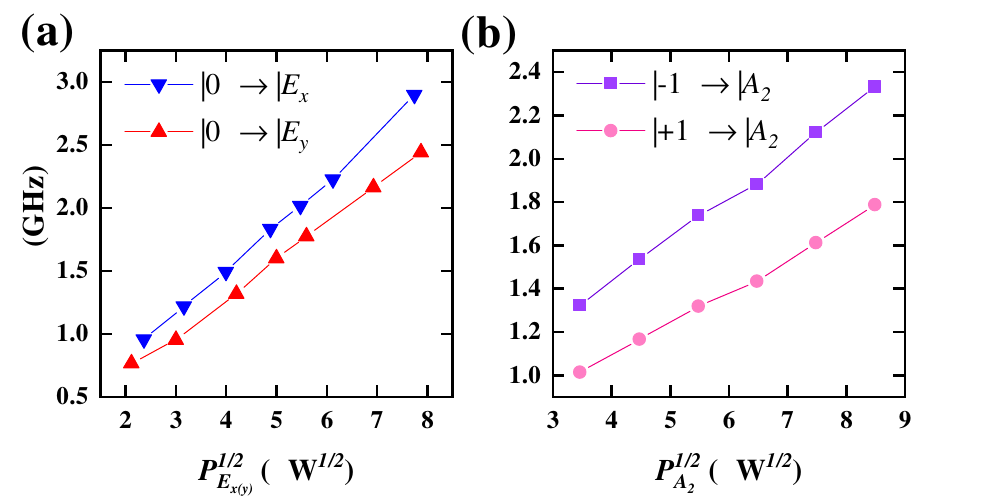}
    \caption{
    \textbf{Laser-power dependence of the optical Rabi frequency.}
    (a) Rabi frequency as a function of \(\sqrt{P}\) for the
    \(\lvert m_s=0\rangle\leftrightarrow\lvert E_x\rangle\) and
    \(\lvert m_s=0\rangle\leftrightarrow\lvert E_y\rangle\)
    transitions, extracted from the fits in Figs.~S\ref{fig:ExEyRabi}(c) and S\ref{fig:ExEyRabi}(d).
    (b) Corresponding results for the
    \(\lvert m_s=\pm1\rangle\leftrightarrow\lvert A_2\rangle\)
    transitions, extracted from Figs.~S\ref{fig:A2Rabi}(e) and S\ref{fig:A2Rabi}(f).
    The linear scaling with \(\sqrt{P}\) is consistent with coherent
    electric-dipole driving. 
    }
    \label{fig:RabiPower}
\end{figure}

Figure~S\ref{fig:RabiPower}(a) shows the fitted Rabi frequency as a function of
\(\sqrt{P}\) for the \(E_x\) and \(E_y\) transitions. Both transitions
exhibit a clear linear dependence, consistent with the expected
\(\sqrt{P}\) scaling for coherent electric-dipole excitation.

Figure~S\ref{fig:RabiPower}(b) presents the corresponding results for the spin-selective
\(\lvert m_s=\pm1\rangle\leftrightarrow\lvert A_2\rangle\)
transitions. A linear dependence on \(\sqrt{P}\) is likewise observed for
both spin configurations.

The linear power dependence observed for the \(E_x\), \(E_y\), and
\(A_2\) transitions provides an independent consistency check of coherent
resonant optical driving.

\clearpage

\section*{Supplementary Note 7: Rate-equation modeling and ODMR contrast discussion}

To determine the singlet relaxation rates \(\Gamma_{\perp}\) and
\(\Gamma_z\) shown in Fig. 2(a) of the main text, we develop a five-level rate-equation model describing
the population dynamics under off-resonant 914-nm pulsed excitation.
In contrast to resonant excitation, for which the individual
excited-state branches are spectrally resolved [Fig.~2(a) of the main
text], off-resonant excitation populates multiple excited states.
We therefore group the excited states associated predominantly with
the \(m_s=0\) and \(m_s=\pm1\) manifolds into two effective excited
levels, denoted \(E_0\) and \(E_{\pm}\), respectively. This reduced model
captures the spin-dependent population dynamics relevant to the
off-resonant excitation experiment.

\subsection*{Five-level rate-equation model}

The five effective levels are the ground-state manifolds \(m_s=0\)
and \(m_s=\pm1\), with populations \(n_0\) and \(n_{\pm}\);
the corresponding effective excited states, with populations
\(n_{E0}\) and \(n_{E\pm}\); and the singlet manifold, with
population \(n_S\). The effective \(m_s=0\) excited manifold
comprises \(E_x\) and \(E_y\), whereas the effective
\(m_s=\pm1\) manifold comprises \(A_1\), \(A_2\), and
\(E_{1,2}\).

Under 914-nm pulsed excitation at a power of \(50~\mu\mathrm{W}\),
the system is pumped from the ground-state manifolds to the
corresponding effective excited-state manifolds at a rate
\(R_{\mathrm{exc}}\), determined from the measured 914-nm saturation
curve in Fig.~S\ref{fig:SA}(f). The excited states decay radiatively to the
ground-state triplet with rate \(\Gamma_{\mathrm{rad}}\), or undergo
intersystem crossing (ISC) to the singlet manifold with effective rates
\(\Gamma_{\mathrm{ISC},0}\) and
\(\Gamma_{\mathrm{ISC},\pm1}\). Population in the singlet manifold
subsequently returns to \(\lvert m_s=0\rangle\) and \(\lvert m_s=\pm1\rangle\) with rates
\(\Gamma_z\) and \(\Gamma_{\perp}\), respectively. This lower branch of the
optical spin-polarization loop is governed by the combined spin--orbit and
vibronic couplings of the singlet and triplet
manifolds~\cite{Thiering2018loop,Bian2025}.

The corresponding rate equations are
\begin{equation}
\begin{aligned}
\dot n_0
&=
-R_{\mathrm{exc}}n_0
+\Gamma_{\mathrm{rad}}n_{E0}
+\Gamma_z n_S,
\\
\dot n_{\pm}
&=
-R_{\mathrm{exc}}n_{\pm}
+\Gamma_{\mathrm{rad}}n_{E\pm}
+\Gamma_{\perp}n_S,
\\
\dot n_{E0}
&=
R_{\mathrm{exc}}n_0
-
\left(
\Gamma_{\mathrm{rad}}
+\Gamma_{\mathrm{ISC},0}
\right)n_{E0},
\\
\dot n_{E\pm}
&=
R_{\mathrm{exc}}n_{\pm}
-
\left(
\Gamma_{\mathrm{rad}}
+\Gamma_{\mathrm{ISC},\pm1}
\right)n_{E\pm},
\\
\dot n_S
&=
\Gamma_{\mathrm{ISC},0}n_{E0}
+
\Gamma_{\mathrm{ISC},\pm1}n_{E\pm}
-
\left(
\Gamma_z+\Gamma_{\perp}
\right)n_S.
\end{aligned}
\end{equation}

Here,
\(\Gamma_{\mathrm{ISC},0}\) and
\(\Gamma_{\mathrm{ISC},\pm1}\) are the effective upper-branch ISC
rates for the \(m_s=0\) and \(m_s=\pm1\) excitation manifolds,
respectively. For the \(m_s=0\) manifold, the relevant excited states
are \(E_x\) and \(E_y\), giving
\begin{equation}
\Gamma_{\mathrm{ISC},0}
=
\frac{1}{2}
\left(
\Gamma_{\mathrm{ISC}}^{E_x}
+
\Gamma_{\mathrm{ISC}}^{E_y}
\right).
\end{equation}

For the \(m_s=\pm1\) manifold, the relevant excited states are
\(A_1\), \(A_2\), and the two degenerate \(E_{1,2}\) states, giving
\begin{equation}
\Gamma_{\mathrm{ISC},\pm1}
=
\frac{1}{4}
\left(
\Gamma_{\mathrm{ISC}}^{A_1}
+
\Gamma_{\mathrm{ISC}}^{A_2}
+
2\Gamma_{\mathrm{ISC}}^{E_{1,2}}
\right).
\end{equation}

These effective rates provide a reduced description of the unresolved
excited-state manifold under off-resonant excitation. The
state-resolved ISC rates entering these averages are independently
determined from the excited-state lifetime measurements.

\subsection*{Extraction of the lower-singlet relaxation rates}

We globally fit the fluorescence transients measured for initialization
in \(\lvert m_s=0\rangle\) and \(\lvert m_s=\pm1\rangle\), with \(\Gamma_z\) and
\(\Gamma_{\perp}\) as free parameters. The relative excitation weights
of the different excited-state branches are determined from their
independently measured saturation curves and incorporated into the
simulation. The fitting is performed using MATLAB's
\texttt{lsqnonlin} routine with experimental noise weighting.

The best-fit results yield
\begin{equation}
\Gamma_z=12.4(2)~\mathrm{MHz},
\qquad
\Gamma_{\perp}=1.6(2)~\mathrm{MHz},
\end{equation}

where the uncertainties represent \(1\sigma\) standard errors. The
model reproduces the fluorescence rise during the excitation pulse,
the steady-state response, and the decay following pulse termination.

The resulting ratio
\begin{equation}
\frac{\Gamma_z}{\Gamma_{\perp}}
\simeq 7.75
\end{equation}

shows that relaxation through the singlet manifold preferentially
repopulates the \(\lvert m_s=0\rangle\) ground state. This spin selectivity provides
the nonradiative pathway required for efficient optical spin
polarization and contributes to the large spin-dependent fluorescence
contrast of PL6.

\subsection*{Comparison of spin--optical dynamics and properties among PL6, PL1, and PL2}

To compare the spin-dependent nonradiative dynamics of PL6 with those
of the structurally analogous PL1 (hh-DV) center, we summarize the
measured radiative and ISC-related parameters of PL6 in Supplementary
Table~\ref{tab:ISC_compare}; the state-resolved upper-branch ISC rates are
obtained from the excited-state lifetime measurements in Fig.~2(c) of
the main text.

Measured properties of PL6, PL1 (hh-DV), and PL2 (kk-DV) are
compared in Supplementary Table~\ref{tab:expt_compare}, providing
a broader benchmark of their optical and spin properties.

The corresponding
state-resolved ISC rates and lower-singlet relaxation rates of PL1 have not
been determined experimentally; the calculated values of Ref.~\cite{Bian2025}
are quoted below where a comparison is required, and the calculated
quantities that govern the two decay channels are collected in Supplementary
Note 5.

The upper-branch ISC is substantially more spin selective in PL6.
In particular, the effective difference between the two spin manifolds is
\begin{equation}
\Gamma_{\mathrm{ISC},\pm1}
-
\Gamma_{\mathrm{ISC},0}
=
49.15~\mathrm{MHz}
\end{equation}

for PL6. The enhancement is particularly pronounced for the \(A_1\)
state, whose measured ISC rate of \(135.51~\mathrm{MHz}\) exceeds that of every
other branch by more than a factor of four. This large difference between the
\(m_s=0\) and \(m_s=\pm1\) nonradiative decay channels provides the microscopic
basis for the strong spin-dependent fluorescence of PL6. The corresponding
state-resolved rates of PL1 have not been measured, so the comparison between
the two centers is made below at the level of the measured ODMR contrasts and of
the calculated spin--orbit parameters, rather than between a measured and a
calculated rate.

\begin{table}[htbp]
\centering
\setlength{\tabcolsep}{4pt}
\small
\caption{
Measured radiative and ISC-related parameters of PL6 (hh-OV). All rates are in
MHz and are obtained experimentally in this work. The corresponding calculated
quantities for PL1 (hh-DV) are given in Supplementary Note 5 and are quoted in
the text where needed.
}
\begin{tabular}{l|c}
\hline
Parameter & PL6 (hh-OV) Expt. \\
\hline
\(\Gamma_{\mathrm{rad}}\) & \(62.5\) \\
\(\Gamma_{\mathrm{ISC}}^{E_x}\) & \(4.61\) \\
\(\Gamma_{\mathrm{ISC}}^{E_y}\) & \(0\) \\
\(\Gamma_{\mathrm{ISC}}^{A_1}\) & \(135.51\) \\
\(\Gamma_{\mathrm{ISC}}^{A_2}\) & \(8.42\) \\
\(\Gamma_{\mathrm{ISC}}^{E_{1,2}}\) & \(30.95\) \\
\hline
\(\Gamma_{\mathrm{ISC},0}\) & \(2.305\) \\
\(\Gamma_{\mathrm{ISC},\pm1}\) & \(51.46\) \\
\hline
\(K_0=\Gamma_{\mathrm{rad}}+\Gamma_{\mathrm{ISC},0}\) & \(64.805\) \\
\(K_{\pm1}=\Gamma_{\mathrm{rad}}+\Gamma_{\mathrm{ISC},\pm1}\) & \(113.96\) \\
\hline
\(\Gamma_z\) & \(12.4\) \\
\(\Gamma_{\perp}\) & \(1.6\) \\
\(\Gamma_z/\Gamma_{\perp}\) & \(7.75\) \\
\hline
Upper bound of pulsed ODMR contrast
\(\frac{K_0-K_{\pm1}}{K_{\pm1}}\) & \(-43.13\%\) \\
\hline
\end{tabular}
\label{tab:ISC_compare}
\end{table}

\begin{table}[htbp]
\centering
\setlength{\tabcolsep}{3pt}
\small
\caption{
Measured properties of PL6 (hh-OV), PL1 (hh-DV), and PL2 (kk-DV)
in 4H-SiC. PL6 values are obtained in this work, whereas PL1 and PL2
values are taken from the literature.
\(^{a}\) Deduced from the reported lifetime-limited linewidth of
\(\approx11\)~MHz.
\(^{b}\) Value tabulated in Ref.~\cite{Zalandauskas2025}, which gives
\(7.04\%\) from multimode dynamic Jahn--Teller theory for comparison.
\(^{c}\) Single hh divacancy; \(9-15\%\) across the four configurations.
\(^{d}\) \(E_y\) transition at \(\delta_\perp=14\)~GHz.
\(^{e}\) Measured at \(\delta_\perp\le12.4\)~GHz for PL6 and at
\(\delta_\perp=14\)~GHz for PL1; spin mixing is strain dependent, so the two
values are not measured under equivalent conditions.\quad
\(^{f}\) The PL2 excited-state lifetime is taken from the
time-resolved measurements of Ref.~\cite{Falk2014}.
\(^{g}\) The PL2 Debye--Waller factor is taken from the experimental
value for the kk divacancy reported in Ref.~\cite{Jin2021}.
\(^{h}\) The positive off-resonant spin-dependent fluorescence
contrast of PL2 reflects the opposite contrast polarity of the kk
divacancy.
\(^{i}\) The resonant spin-readout contrast of PL2 is taken from
Ref.~\cite{Bourassa2020EntanglementAC}.
\(^{j}\) \(E_x\) and \(E_y\) transitions at \(\delta_\perp=6.8\)~GHz.
}
\begin{tabular}{l|c|c|c}
\hline
Quantity & PL6 (hh-OV) & PL1 (hh-DV) &
PL2 (kk-DV) \\
\hline
Zero-phonon-line energy (eV)
& \(1.195\)
& \(1.095\)~\cite{Csore2022}
& \(1.096\)~\cite{christle2017isolated} \\

\(m_s=0\) excited-state lifetime (ns)
& \(16.0\)
& \(\approx14.5\)$^{a}$~\cite{christle2017isolated}
& \(15\pm3\)$^{f}$~\cite{Falk2014} \\

Debye--Waller factor
& \(\approx10\%\)
& \(3.69\%\)$^{b}$~\cite{Jin2021}
& \(6.11\%\)$^{g}$~\cite{Jin2021} \\

Pulsed ODMR contrast, off-resonant
& \(-31.9\%\)
& \(\approx-9\%\)$^{c}$~\cite{christle2017isolated}
& \(\approx+15\%\)$^{h}$~\cite{Bourassa2020EntanglementAC} \\

Resonant readout contrast
& \(98.3\%\)
& \(94\%\)$^{d}$~\cite{christle2017isolated}
& \(>99\%\)$^{i}$~\cite{Bourassa2020EntanglementAC} \\

Saturated spin-flip rate (kHz)
& \(\approx140\)
& \(\approx330\)$^{e}$~\cite{christle2017isolated}
& \(\approx300\)$^{j}$~\cite{christle2017isolated} \\

Reionization threshold (eV)
& ---
& \(1.310\)~\cite{Csore2022}
& \(1.310\)~\cite{Csore2022} \\
\hline
\end{tabular}
\label{tab:expt_compare}
\end{table}

The positive off-resonant spin-dependent fluorescence contrast
reported for PL2 (\(\approx+15\%\)), in contrast to the negative
contrasts observed for PL6 and PL1, further highlights the distinct
optical response of the kk configuration. Together with its reported
resonant spin-readout contrast exceeding \(99\%\), this provides an
additional benchmark for evaluating the optical spin properties
of PL6. These contrast values are used here as comparative
experimental benchmarks and are not taken to imply identical
underlying ISC pathways.

\subsection*{Spin--orbit coupling and the enhanced ISC of PL6}

Within Fermi's golden-rule description, the upper-branch ISC rate can be
written schematically as
\begin{equation}
\Gamma_{\mathrm{ISC}}
\propto
\lambda_{\perp}^{2}F(\Delta),
\end{equation}

where \(\lambda_{\perp}\) is the transverse spin--orbit coupling between the
triplet and singlet manifolds and \(F(\Delta)\) is the vibrational overlap
function associated with the triplet--singlet energy separation \(\Delta\)
\cite{Goldman2015b,Thiering2018loop,Bian2025}. The original formulation of this
expression did not include the dynamic Jahn--Teller effect~\cite{Goldman2015b};
the treatment used here follows the vibronic theory of
Refs.~\cite{Thiering2018loop,Bian2025}, in which the Jahn--Teller dynamics of
the \(^3E\) manifold is taken into account explicitly.

It is important to distinguish the spin--orbit parameter that enters this
expression from the one determined spectroscopically. The measured
\(\lambda=5.739[5.656,5.881]~\mathrm{GHz}\) of PL6, like the
\(\lambda=3.538[3.486,3.590]~\mathrm{GHz}\) of PL1~\cite{christle2017isolated},
is the vibronically reduced axial parameter \(\lambda_z=p\lambda_{z0}\), where
\(p\) is the Ham reduction factor of the dynamic Jahn--Teller \(^3E\) manifold.
The transverse coupling that mediates the triplet--singlet transition is
instead governed by the intrinsic parameter, to which it is proportional,
\(\lambda_{\perp}\propto\lambda_{z0}\), the electron--phonon physics being
carried by \(F(\Delta)\) rather than by an additional reduction of the matrix
element. The proportionality constant of \(1.2\) was established
experimentally for the NV center in diamond and adopted for the axial
divacancy in 4H-SiC in Ref.~\cite{Bian2025}; only the ratio between the two
centers enters the present argument, so that constant cancels.
First-principles calculations within a single framework give
\(\lambda_{z0}=48.12~\mathrm{GHz}\) for PL6 (Supplementary Note 5) and
\(18.50~\mathrm{GHz}\) for PL1~\cite{Bian2025}, i.e.\ a factor of \(2.6\),
corresponding to a factor of \(6.8\) in \(\lambda_{\perp}^{2}\).

This provides a quantitative account of the enhanced ISC of PL6 without
requiring an absolute rate calculation, and without comparing a measured rate
with a calculated one. Within the same framework the smaller triplet--singlet
separation of PL6 (\(\Delta=202\) against \(233.62~\mathrm{meV}\),
Supplementary Note 5) acts through \(F(\Delta)\) in the same direction, so the
predicted enhancement of the upper-branch ISC exceeds the factor of \(6.8\)
supplied by \(\lambda_{\perp}^{2}\) alone. On the experimental side, the
measured excited-state decay rates of the two centers are
comparable---\(62.5~\mathrm{MHz}\) for PL6 and \(\approx69~\mathrm{MHz}\) for
PL1, the latter from its reported lifetime-limited linewidth of
\(\approx11~\mathrm{MHz}\)~\cite{christle2017isolated}---so the difference in
contrast is not radiative in origin.

Since the ODMR contrast is controlled by the branching ratio
\(\Gamma_{\mathrm{ISC}}/\Gamma_{\mathrm{rad}}\), and the radiative rates of the
two centers are comparable, the branching into the singlet manifold is
enhanced for PL6 by essentially the full \(\lambda_{\perp}^{2}\) factor. The
larger intrinsic spin--orbit coupling of the oxygen--vacancy center, and not
the modest change in the spectroscopic \(\lambda\), is therefore the
microscopic origin of its larger ODMR contrast.

\subsection*{Upper bound of the pulsed ODMR contrast}

The strong spin dependence of the upper ISC branch sets an upper bound
on the pulsed ODMR contrast. We define the effective total
excited-state decay rates of the two spin manifolds as
\begin{equation}
K_0
=
\Gamma_{\mathrm{rad}}
+
\Gamma_{\mathrm{ISC},0},
\qquad
K_{\pm1}
=
\Gamma_{\mathrm{rad}}
+
\Gamma_{\mathrm{ISC},\pm1}.
\end{equation}

Because
\(\Gamma_{\mathrm{ISC},\pm1}\gg\Gamma_{\mathrm{ISC},0}\),
the \(m_s=\pm1\) excited-state manifold is depleted much more rapidly
through the nonradiative ISC channel than the \(m_s=0\) manifold.
When only the spin selectivity of the upper ISC branch is considered
and the effects of the lower singlet branch are neglected, the
corresponding upper bound of the pulsed ODMR contrast is
\begin{equation}
C_{\mathrm{upper}}
=
\frac{K_0-K_{\pm1}}{K_{\pm1}}.
\end{equation}

Using the values in Supplementary Table~\ref{tab:ISC_compare} gives
\(C_{\mathrm{upper}}=-43.13\%\) for PL6. This bound is set entirely by the
measured decay rates of PL6 and requires no input from any other center.

\subsection*{From the upper bound to the experimental contrast: role of the lower singlet branch}

The experimentally observed pulsed ODMR contrast of PL6,
\(-31.9\%\), is lower than the upper-branch limit because the complete
optical cycle also includes relaxation from the singlet manifold back
to the ground-state triplet. The lower-singlet relaxation is strongly
spin selective~\cite{Thiering2018loop}, with
\(\Gamma_z=12.4(2)~\mathrm{MHz}\) and
\(\Gamma_{\perp}=1.6(2)~\mathrm{MHz}\), corresponding to
\(\Gamma_z/\Gamma_{\perp}\simeq7.75\).
The dominant \(\Gamma_z\) pathway preferentially repopulates the
\(\lvert m_s=0\rangle\) ground state and therefore enables efficient optical spin
polarization. However, the finite \(\Gamma_{\perp}\) pathway provides
a competing return channel to the \(\lvert m_s=\pm1\rangle\) manifold, preventing
perfectly selective repopulation of \(\lvert m_s=0\rangle\). Consequently, the
lower singlet branch reduces the experimentally accessible ODMR
contrast from the upper-branch limit.

The corresponding rates of PL1 have not
been determined experimentally. The comparison between the two centers can
nevertheless be made on measured quantities alone: the off-resonant pulsed ODMR
contrast of PL6, \(-31.9\%\), is about three times the \(\approx9\%\) reported
for a single hh divacancy~\cite{christle2017isolated}, while their measured
excited-state decay rates are comparable. The difference is therefore not
radiative in origin but originates in the nonradiative channel, consistent with
the larger intrinsic spin--orbit coupling of PL6 obtained from first principles
(Supplementary Note 5). Thus, PL6 combines a strong spin dependence of
the upper ISC branch with selective relaxation through the lower
singlet branch, both of which favor efficient optical spin
polarization and high ODMR contrast.

In summary, the large ODMR contrast of PL6 arises from the combined
spin selectivity of the upper and lower nonradiative pathways. The
substantially enhanced difference between
\(\Gamma_{\mathrm{ISC},\pm1}\) and
\(\Gamma_{\mathrm{ISC},0}\) establishes an idealized upper contrast
bound of approximately \(-43\%\). The experimentally observed contrast
of \(-31.9\%\) is lower than this upper bound because the complete
optical cycle includes the finite branching selectivity of the lower
singlet relaxation, while experimental imperfections can further
reduce the measured contrast. Nevertheless, the measured
\(\Gamma_z/\Gamma_{\perp}=7.75\), substantially larger than the
theoretical PL1 value of \(3.30\), indicates highly selective
singlet-mediated repopulation of the \(\lvert m_s=0\rangle\) ground state. The larger
measured spin--orbit interaction of PL6 is consistent with enhanced
triplet--singlet coupling and provides a microscopic connection between
its distinct oxygen-containing local electronic structure and the
observed spin-selective optical dynamics.

\clearpage
\section*{Supplementary Note 8. Spin-flip dynamics and extraction of spin-flip rates}

\subsection*{Spin-flip dynamics under resonant excitation}

\begin{figure}[htbp]
    \centering
    \includegraphics[scale=0.5]{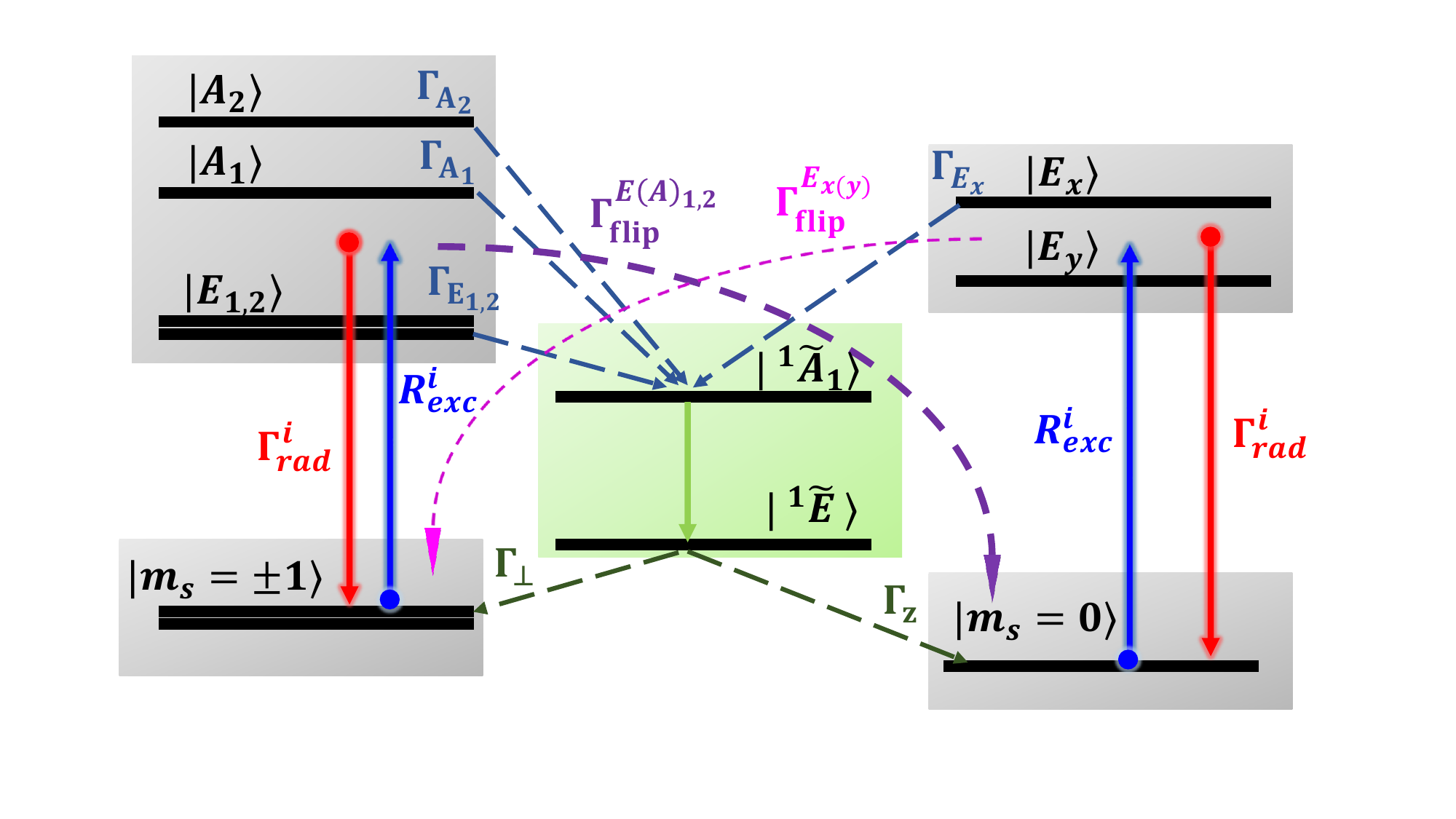}
    \caption{
    \textbf{Spin-flip dynamics and ISC pathways under resonant excitation.}
    The ground-state spin sublevels
    \(\lvert m_s=0\rangle\) and \(\lvert m_s=\pm1\rangle\) are
    resonantly excited with rates \(R_{\mathrm{exc}}^i\) obtained from
    the corresponding saturation curves. Radiative decay with rate
    \(\Gamma_{\mathrm{rad}}\) returns population to the initially
    addressed ground-state spin manifold, while upper-branch ISC occurs
    with state-dependent rates \(\Gamma_i\). The lower singlet state
    repopulates \(\lvert m_s=\pm1\rangle\) and
    \(\lvert m_s=0\rangle\) with rates \(\Gamma_{\perp}\) and
    \(\Gamma_z\), respectively. Direct spin-changing transitions from
    the excited states to the opposite-spin ground-state manifold are
    denoted by \(\Gamma_{\mathrm{flip}}^i\).
    }
    \label{fig:spinflip_scheme}
\end{figure}

As illustrated in Fig.~S\ref{fig:spinflip_scheme}, the spin-flip dynamics under resonant
excitation are governed by the interplay between resonant optical cycling,
spin-conserving radiative decay, intersystem crossing (ISC), and
spin-changing optical transitions. The ground-state spin sublevels
\(\lvert m_s=0\rangle\) and \(\lvert m_s=\pm1\rangle\) are selectively
addressed through the corresponding excited-state branches with excitation
rates \(R_{\mathrm{exc}}^i\). For each branch \(i\),
\(R_{\mathrm{exc}}^i\) is determined from the independently measured
fluorescence saturation curve according to
\(R_{\mathrm{exc}}^i=\Gamma_{\mathrm{rad}}P/P_{\mathrm{sat}}^i\).

Following resonant excitation, radiative decay returns population
predominantly to the initially addressed ground-state spin manifold with
rate \(\Gamma_{\mathrm{rad}}\), while upper-branch ISC transfers population
to the singlet manifold with a state-dependent rate \(\Gamma_i\). The
lower singlet state subsequently repopulates
\(\lvert m_s=0\rangle\) and \(\lvert m_s=\pm1\rangle\) with rates
\(\Gamma_z\) and \(\Gamma_{\perp}\), respectively. In addition, direct
spin-changing optical transitions transfer population from the excited
state to the opposite-spin ground-state manifold with rate
\(\Gamma_{\mathrm{flip}}^i\).

We denote the populations of the
\(\lvert m_s=\pm1\rangle\) and \(\lvert m_s=0\rangle\)
ground-state manifolds by \(N_{\pm1}\) and \(N_0\), respectively, the
population of an addressed excited-state branch \(i\) by \(N_i\), and
the effective singlet population by \(N_S\). The independently measured
lower-branch ISC rates are
\(\Gamma_z=12.4(2)~\mathrm{MHz}\) and
\(\Gamma_{\perp}=1.6(2)~\mathrm{MHz}\).

For resonant excitation from the
\(\lvert m_s=\pm1\rangle\) manifold through
\(k\in\{A_1,A_2,E_{1,2}\}\), the population dynamics are described by
\begin{equation}
\label{eq:rate_pm1}
\begin{aligned}
\dot{N}_{\pm1}
&=
-R_{\mathrm{exc}}^k N_{\pm1}
+\Gamma_{\mathrm{rad}}N_k
+\Gamma_{\perp}N_S,
\\
\dot{N}_0
&=
\Gamma_zN_S
+\Gamma_{\mathrm{flip},k}N_k,
\\
\dot{N}_k
&=
R_{\mathrm{exc}}^kN_{\pm1}
-
\left(
\Gamma_{\mathrm{rad}}
+\Gamma_k
+\Gamma_{\mathrm{flip},k}
\right)N_k,
\\
\dot{N}_S
&=
\Gamma_kN_k
-
\left(
\Gamma_z+\Gamma_{\perp}
\right)N_S.
\end{aligned}
\end{equation}

Similarly, for resonant excitation from the
\(\lvert m_s=0\rangle\) manifold through
\(l\in\{E_x,E_y\}\),

\begin{equation}
\label{eq:rate_0}
\begin{aligned}
\dot{N}_0
&=
-R_{\mathrm{exc}}^lN_0
+\Gamma_{\mathrm{rad}}N_l
+\Gamma_zN_S,
\\
\dot{N}_{\pm1}
&=
\Gamma_{\perp}N_S
+\Gamma_{\mathrm{flip},l}N_l,
\\
\dot{N}_l
&=
R_{\mathrm{exc}}^lN_0
-
\left(
\Gamma_{\mathrm{rad}}
+\Gamma_l
+\Gamma_{\mathrm{flip},l}
\right)N_l,
\\
\dot{N}_S
&=
\Gamma_lN_l
-
\left(
\Gamma_z+\Gamma_{\perp}
\right)N_S.
\end{aligned}
\end{equation}

These equations specify the microscopic population-transfer pathways
represented in Fig.~S\ref{fig:spinflip_scheme}. Experimentally, the
spin-flip dynamics are inferred from time-resolved fluorescence under
continuous resonant excitation. Because the relative contributions of
radiative cycling, ISC, and spin-changing population transfer differ
among the excited-state branches, the experimentally accessible
signatures of \(\Gamma_{\mathrm{flip}}^i\) are analyzed separately below.

\subsection*{Extraction of the \(E_x\) and \(E_y\) spin-flip rates in Fig.~3(a)}

For the \(E_x\) and \(E_y\) transitions, the upper-branch ISC contribution
is negligible or weak compared with radiative decay:
\(\Gamma_{E_y}\simeq0\) and
\(\Gamma_{E_x}\simeq4.6~\mathrm{MHz}\ll
\Gamma_{\mathrm{rad}}=62.5~\mathrm{MHz}\).
The singlet population therefore remains small, and the long-timescale
dynamics can be approximated by a resonantly driven ground--excited
two-level system with a weak spin-changing loss channel.

Because the optical excitation and radiative decay are much faster than
the measured spin-flip dynamics, the excited-state population follows the
ground-state population adiabatically. For
\(l\in\{E_x,E_y\}\), setting \(\dot{N}_l\simeq0\) gives
\begin{equation}
N_l
=
\frac{R_{\mathrm{exc}}^l}
{\Gamma_{\mathrm{rad}}
+R_{\mathrm{exc}}^l
+\Gamma_{\mathrm{flip},l}}
N_0.
\end{equation}

Since
\(\Gamma_{\mathrm{flip},l}\ll\Gamma_{\mathrm{rad}}\),

\begin{equation}
\label{eq:adiabatic_ExEy}
N_l
\simeq
\frac{R_{\mathrm{exc}}^l}
{\Gamma_{\mathrm{rad}}+R_{\mathrm{exc}}^l}
N_0.
\end{equation}

The loss of population from the resonantly addressed spin manifold due
to the spin-changing channel is therefore
\begin{equation}
\dot{N}_0
\simeq
-
\Gamma_{\mathrm{flip},l}N_l
=
-
\frac{R_{\mathrm{exc}}^l}
{\Gamma_{\mathrm{rad}}+R_{\mathrm{exc}}^l}
\Gamma_{\mathrm{flip},l}N_0.
\end{equation}

Accordingly, the fluorescence envelope follows
\begin{equation}
I(t)
=
I_0
\exp
\left(
-\Gamma_{\mathrm{eff}}^l t
\right),
\end{equation}

where

\begin{equation}
\label{eq:Gammaeff}
\Gamma_{\mathrm{eff}}^l
=
\frac{R_{\mathrm{exc}}^l}
{\Gamma_{\mathrm{rad}}+R_{\mathrm{exc}}^l}
\Gamma_{\mathrm{flip}}^l.
\end{equation}

Using
\(R_{\mathrm{exc}}^l
=\Gamma_{\mathrm{rad}}P/P_{\mathrm{sat}}^l\),
Eq.~(\ref{eq:Gammaeff}) becomes
\begin{equation}
\label{eq:Gammaeff_power}
\Gamma_{\mathrm{eff}}^l(P)
=
\frac{P}
{P+P_{\mathrm{sat}}^l}
\Gamma_{\mathrm{flip}}^l.
\end{equation}

Thus, near saturation
(\(P\simeq P_{\mathrm{sat}}^l\)),
\(\Gamma_{\mathrm{eff}}^l\simeq
\Gamma_{\mathrm{flip}}^l/2\), whereas in the high-power limit
(\(P\gg P_{\mathrm{sat}}^l\)),
\(\Gamma_{\mathrm{eff}}^l\rightarrow
\Gamma_{\mathrm{flip}}^l\).
In the experiment, the resonant excitation power is varied from
\(400~\mathrm{nW}\) to \(30~\mu\mathrm{W}\), while
\(P_{\mathrm{sat}}^{E_y}=446.5\pm36.5~\mathrm{nW}\) and
\(P_{\mathrm{sat}}^{E_x}=459.4\pm27.9~\mathrm{nW}\).
The high-power plateau of the measured single-exponential decay rate
therefore directly yields the \(E_x\) and \(E_y\) spin-flip rates.

\subsection*{Extraction of the \(A_2\) and \(E_{1,2}\) spin-flip rates in Fig.~3(b)}

For the \(A_2\) and \(E_{1,2}\) transitions, the fluorescence transients
measured at a resonant excitation power of \(30~\mathrm{nW}\) exhibit
two distinct dynamical components and are described by
\begin{equation}
\label{eq:biexp}
I(t)
=
A_{\mathrm{fast}}
e^{-k_{\mathrm{fast}}t}
+
A_{\mathrm{slow}}
e^{-k_{\mathrm{slow}}t}.
\end{equation}

In contrast to the weak-ISC \(E_x/E_y\) case, these components contain
the combined effects of resonant optical cycling, ISC-mediated shelving,
and spin-changing population transfer. We therefore use the experimentally
resolved slow component to quantify the long-timescale spin-changing
dynamics, rather than assigning the two fitted decay constants individually
to eigenmodes of the full microscopic rate matrix.

The fluorescence transients are measured under continuous resonant
excitation. Here, \(\Gamma_{\mathrm{obs}}^i\) denotes the experimentally
extracted effective rate associated with the resolved long-timescale
spin-changing dynamics for transition \(i\). During resonant optical
cycling, the addressed population is distributed between the ground and
excited states. Neglecting the much slower spin-changing dynamics when
determining this optical-cycling population distribution, the quasi-steady
state condition gives

\begin{equation}
R_{\mathrm{exc}}^i N_g
\simeq
\Gamma_{\mathrm{rad}} N_e ,
\end{equation}

and hence the fraction of the addressed population available in the
ground state for resonant excitation is
\begin{equation}
\frac{N_g}{N_g+N_e}
=
\frac{\Gamma_{\mathrm{rad}}}
{\Gamma_{\mathrm{rad}}+R_{\mathrm{exc}}^i}.
\end{equation}

Accordingly, correcting the experimentally observed rate for this
optical-cycling population factor gives the branch-resolved spin-flip rate
\begin{equation}
\label{eq:optical_cycling}
\Gamma_{\mathrm{flip}}^i
=
\Gamma_{\mathrm{obs}}^i
\frac{\Gamma_{\mathrm{rad}}+R_{\mathrm{exc}}^i}
{\Gamma_{\mathrm{rad}}}
=
\Gamma_{\mathrm{obs}}^i
\left(
1+
\frac{R_{\mathrm{exc}}^i}
{\Gamma_{\mathrm{rad}}}
\right).
\end{equation}

For the \(30~\mathrm{nW}\) measurements considered here,
\(R_{\mathrm{exc}}^i/\Gamma_{\mathrm{rad}}\simeq0.08\), so that this
optical-cycling correction changes the experimentally extracted rate by
only approximately \(8\%\).

For the \(A_2\) transition, the biexponential fit gives
\(k_{\mathrm{fast}}=6.266~\mathrm{MHz}\) and
\(k_{\mathrm{slow}}=0.499~\mathrm{MHz}\), corresponding to
\(\tau_{\mathrm{fast}}=159.60~\mathrm{ns}\) and
\(\tau_{\mathrm{slow}}=2.003~\mu\mathrm{s}\), respectively.
The slow component characterizes the experimentally resolved
long-timescale spin-changing dynamics, and we therefore take
\begin{equation}
\Gamma_{\mathrm{obs}}^{A_2}
=
k_{\mathrm{slow}}
=
0.499~\mathrm{MHz}.
\end{equation}

The measured saturation power
\(P_{\mathrm{sat}}^{A_2}=396.8\pm48.0~\mathrm{nW}\) gives
\begin{equation}
R_{\mathrm{exc}}^{A_2}
=
\Gamma_{\mathrm{rad}}
\frac{P}{P_{\mathrm{sat}}^{A_2}}
=
62.5\times\frac{30}{396.8}
=
4.73~\mathrm{MHz}
\end{equation}

at \(P=30~\mathrm{nW}\). Using Eq.~(\ref{eq:optical_cycling}), the
corresponding spin-flip rate is
\begin{equation}
\label{eq:A2flip}
\begin{aligned}
\Gamma_{\mathrm{flip}}^{A_2}
&=
k_{\mathrm{slow}}
\left(
1+
\frac{R_{\mathrm{exc}}^{A_2}}
{\Gamma_{\mathrm{rad}}}
\right)\simeq
0.54~\mathrm{MHz}.
\end{aligned}
\end{equation}

For the \(E_{1,2}\) transition, the biexponential fit gives
\(k_{\mathrm{fast}}=5.056~\mathrm{MHz}\) and
\(k_{\mathrm{slow}}=0.522~\mathrm{MHz}\), corresponding to
\(\tau_{\mathrm{fast}}=197.77~\mathrm{ns}\) and
\(\tau_{\mathrm{slow}}=1.914~\mu\mathrm{s}\), respectively.
Both components contribute appreciably to the measured transient.
We therefore use the normalized amplitude of the slow component,
\begin{equation}
\label{eq:fslow}
f_{\mathrm{slow}}
=
\frac{A_{\mathrm{slow}}}
{A_{\mathrm{fast}}+A_{\mathrm{slow}}}
=
0.271.
\end{equation}

Within this experimental decomposition,
\(f_{\mathrm{slow}}\) quantifies the fraction of the resonantly driven
population dynamics associated with the resolved slow spin-changing
channel. Since resonant excitation events occur at a rate
\(R_{\mathrm{exc}}^{E_{1,2}}\), the corresponding experimentally observed
spin-changing rate is
\begin{equation}
\label{eq:E12obs}
\Gamma_{\mathrm{obs}}^{E_{1,2}}
=
R_{\mathrm{exc}}^{E_{1,2}}
f_{\mathrm{slow}}.
\end{equation}

The measured saturation power
\(P_{\mathrm{sat}}^{E_{1,2}}=368.3\pm53.9~\mathrm{nW}\) gives
\begin{equation}
R_{\mathrm{exc}}^{E_{1,2}}
=
\Gamma_{\mathrm{rad}}
\frac{P}{P_{\mathrm{sat}}^{E_{1,2}}}
=
62.5\times\frac{30}{368.3}
=
5.09~\mathrm{MHz}
\end{equation}

at \(P=30~\mathrm{nW}\).

Applying the same finite-excitation normalization as for \(A_2\) gives
\begin{equation}
\label{eq:E12flip}
\begin{aligned}
\Gamma_{\mathrm{flip}}^{E_{1,2}}=\Gamma_{\mathrm{obs}}^{E_{1,2}}\left(
1+\frac{R_{\mathrm{exc}}^{E_{1,2}}}
{\Gamma_{\mathrm{rad}}}
\right)
&=
R_{\mathrm{exc}}^{E_{1,2}}
f_{\mathrm{slow}}
\left(
1+
\frac{R_{\mathrm{exc}}^{E_{1,2}}}
{\Gamma_{\mathrm{rad}}}
\right)\simeq
1.49~\mathrm{MHz}.
\end{aligned}
\end{equation}

We therefore obtain
\(\Gamma_{\mathrm{flip}}^{E_{1,2}}=1.49~\mathrm{MHz}\) and
\(\Gamma_{\mathrm{flip}}^{A_2}=0.54~\mathrm{MHz}\) at a resonant
excitation power of \(30~\mathrm{nW}\). The approximately threefold
smaller spin-flip rate of \(A_2\) indicates substantially weaker
spin-changing dynamics during resonant optical cycling, consistent with
its favorable role as the polarization-selective optical interface
discussed in the main text.

\clearpage
\section*{Supplementary Note 9. Prediction of spin mixing and cycling photons from the excited-state Hamiltonian}
\subsection*{Strain-dependent spin purity and excited-state spin mixing}

The spin-conserving character of a resonant optical transition is determined
not only by its orbital symmetry but also by the spin composition of the
corresponding excited-state eigenstate. In the absence of excited-state spin
mixing, resonant excitation from the \(m_s=0\) ground state to the nominal
\(E_x\) or \(E_y\) excited state preserves the spin projection. By contrast,
admixture of \(m_s=\pm1\) components into these excited states opens an
intrinsic spin-changing pathway during repeated optical excitation and
spontaneous emission. The degree of excited-state spin mixing therefore
provides a direct microscopic measure of the intrinsic spin-conserving
character of the optical transition.

To examine how the excited-state spin-mixing properties of PL6 differ from
those of structurally related divacancies, we compare PL6 with the two axial
divacancy configurations PL1 (\(hh\)) and PL2 (\(kk\)).
This comparison serves two distinct purposes. PL1 is particularly important
because it has the same \(hh\) lattice configuration as PL6: PL1 is the
\(hh\)-DV, whereas PL6 corresponds to the \(hh\)-OV
complex in which one of the vacancy sites is replaced by oxygen. The PL1--PL6 comparison therefore provides a structurally matched reference
for examining the modification of the excited-state structure associated with
oxygen substitution. PL2, corresponding to the \(kk\)-DV, is included
as an additional axial-divacancy benchmark and provides a broader comparison
with the configuration dependence of the divacancy excited-state manifold.

For all three defects, we use the same six-level effective Hamiltonian for the
orbital-doublet excited-state manifold. The Hamiltonian contains the axial
spin-orbit interaction, the excited-state spin-spin interaction, and the
transverse-strain term \(\delta_\perp\), together with the off-diagonal
spin-spin terms responsible for coupling states of different spin character.
The Hamiltonian parameters for PL6 are obtained from the strain-dependent
spectroscopy described in the main text, while those for PL1 and PL2 are
taken from the corresponding divacancy excited-state Hamiltonians. Using the
same Hamiltonian convention for the three defects allows the resulting
eigenstate compositions to be compared directly.

At each value of \(\delta_\perp\), we diagonalize the
excited-state Hamiltonian and identify the eigenstates
with the largest overlap with the nominal \(E_x\) and
\(E_y\) orbital basis states, respectively. We quantify
their spin-conserving character by the \(m_s=0\) spin purity
\begin{equation}
P_0^{(x,y)}
=
\left|
\left\langle e_{x,y}^{(0)}
\middle|
\psi_{x,y}
\right\rangle
\right|^2,
\label{eq:spin_purity}
\end{equation}
where \(|e_x^{(0)}\rangle\) and \(|e_y^{(0)}\rangle\)
denote the nominal \(m_s=0\) basis states associated
with the \(E_x\) and \(E_y\) orbital branches, respectively,
and \(|\psi_{x,y}\rangle\) are the corresponding
excited-state eigenstates.

Here, \(P_0^{(x,y)}=1\) corresponds to a pure \(m_s=0\)
excited state, whereas \(1-P_0^{(x,y)}\) quantifies the
admixture of other spin components. The latter quantity
therefore provides the Hamiltonian contribution to the
spin-mixing probability. Importantly, this analysis concerns
only the composition of the excited-state eigenstates and does
not yet include dynamical spin-changing processes such as
intersystem crossing (ISC), which are considered separately below.

\begin{figure}[htbp]
    \centering
    \includegraphics[width=\linewidth]
    {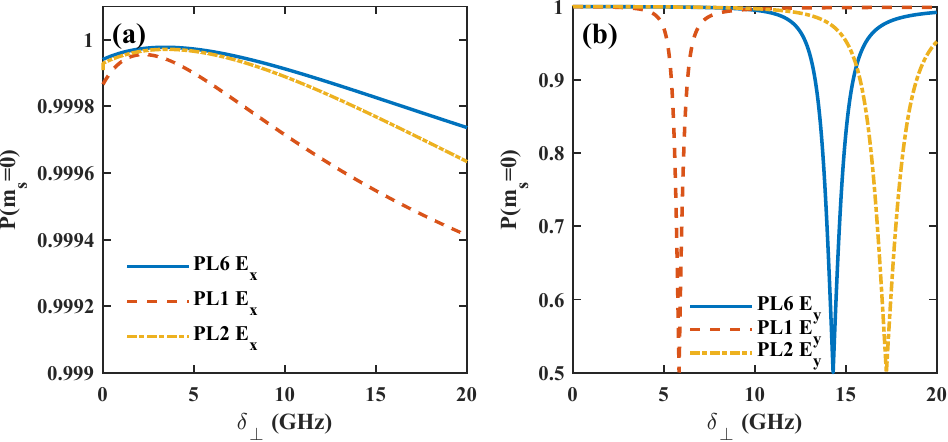}
    \caption{
    \textbf{Strain-dependent excited-state spin purity of PL1, PL2, and PL6.}
    (a) Calculated \(m_s=0\) spin purity \(P(m_s=0)\) of the \(E_x\)
    branch as a function of transverse strain \(\delta_\perp\).
    PL6 maintains the highest \(E_x\) spin purity throughout the investigated
    \(0\)--\(20~\mathrm{GHz}\) range, followed by PL2 and PL1.
    (b) Corresponding \(E_y\) spin purity. Strong spin-mixing features,
    characterized by a pronounced reduction of the \(m_s=0\) character,
    occur near \(\delta_\perp=5.83\), \(14.30\), and
    \(17.23~\mathrm{GHz}\) for PL1, PL6, and PL2, respectively.
    Relative to the structurally matched PL1 divacancy, the PL6 center therefore
exhibits a pronounced shift of the strain region associated with
\(E_y\) spin mixing, reflecting the modification of the excited-state
Hamiltonian associated with oxygen substitution.
    }
    \label{fig:S11}
\end{figure}

Figure~S\ref{fig:S11}(a) compares the calculated \(E_x\) spin purity over the transverse
strain range \(0\leq\delta_\perp\leq20~\mathrm{GHz}\). All three defects
retain a predominantly \(m_s=0\) \(E_x\) state over this range, with
\(P_0^{(x)}\) remaining close to unity. Nevertheless, a clear and systematic
difference is visible: PL6 maintains the highest \(E_x\) spin purity
throughout the entire investigated strain range, PL2 is intermediate, and
PL1 exhibits the strongest spin admixture. The difference is small on the
absolute scale of \(P_0^{(x)}\), but it is physically significant for
repeated resonant optical cycling because the relevant spin-changing
probability is the much smaller quantity \(1-P_0^{(x)}\). Consequently,
small differences in a spin purity close to unity can correspond to large
differences in the number of spin-conserving optical cycles, as quantified
in Figure~S\ref{fig:S12}.

The strain dependence of the \(E_y\) state is qualitatively different
[Fig.~S\ref{fig:S11}(b)]. In contrast to the weak and smooth variation of the \(E_x\)
spin purity, each defect exhibits a pronounced region in which the
\(m_s=0\) character of \(E_y\) is strongly reduced. These features arise
when transverse strain shifts the nominal \(E_y\) branch into energetic
proximity with excited states carrying substantial \(m_s=\pm1\) character.
The off-diagonal spin-spin interaction then produces strong hybridization,
resulting in a rapid loss of \(m_s=0\) purity. The strain position of this
mixing feature is therefore determined by the complete excited-state
Hamiltonian rather than by the transverse strain alone.

The calculated strong-mixing regions occur at markedly different transverse
strains for the three defects. For PL1, the \(E_y\) spin purity reaches its
minimum near \(\delta_\perp\simeq5.83~\mathrm{GHz}\). In PL6, the
corresponding feature is shifted to approximately
\(14.30~\mathrm{GHz}\), while for PL2 it occurs at still higher strain,
near \(17.23~\mathrm{GHz}\). Thus, relative to the structurally matched PL1 reference, the prominent
\(E_y\) spin-mixing feature of PL6 is shifted by approximately
\(8.5~\mathrm{GHz}\). This large displacement reflects a substantial
modification of the excited-state spin-orbital Hamiltonian associated with
oxygen substitution.

The PL2 comparison further shows that this modification should not be
interpreted as a uniform improvement of all excited-state spin purities.
Although PL6 has the highest \(E_x\) spin purity over the entire calculated
range, the prominent \(E_y\) mixing feature of PL2 occurs at a still higher
strain than that of PL6. The modification observed in PL6 is therefore state dependent: it is
associated with both a reduced intrinsic \(E_x\) spin admixture relative to
PL1 and a pronounced displacement of the strain-dependent \(E_y\) mixing
region, without necessarily maximizing the spin purity of every excited-state
branch relative to all divacancy configurations.

Taken together, these calculations establish a clear connection between local atomic structure and excited-state spin mixing. The most relevant comparison is between PL1 and PL6 because the two centers share the same \(hh\) lattice configuration but possess distinct local defect structures, corresponding to hh-DV and hh-OV, respectively. Within this structurally matched pair, PL6 exhibits
both a systematically higher \(E_x\) spin purity and a pronounced shift of the
\(E_y\) spin-mixing feature from approximately \(5.8\) to
\(14.3~\mathrm{GHz}\), demonstrating the substantial modification of the
excited-state Hamiltonian associated with oxygen substitution. PL2 provides a
complementary benchmark showing the range of behavior accessible through a
change of lattice configuration. These differences in the excited-state
eigenstate composition provide the microscopic basis for the distinct
Hamiltonian-limited optical cycling properties discussed in the following
section.

\subsection*{Hamiltonian-limited optical cycling}

The spin purity provides a microscopic measure of excited-state spin mixing.
Its consequence for resonant optical cycling can be expressed more directly
in terms of the mean number of photons that can be emitted before a
spin-changing event induced by excited-state spin mixing occurs. In this
section, we consider only spin mixing contained in the excited-state
Hamiltonian and neglect additional dynamical loss channels such as
intersystem crossing (ISC).

For an excited state with \(m_s=0\) character \(P_0\), we associate
\(P_0\) with the spin-conserving probability of an optical cycle and
\(1-P_0\) with the spin-mixing probability. Assuming independent optical
cycles, the corresponding mean number of spin-conserving photons before a
spin-changing event is
\begin{equation}
\left\langle N_{\mathrm{ph}}\right\rangle_{\mathrm{mix}}
=
\frac{P_0}{1-P_0}.
\label{eq:Nph_mix}
\end{equation}

Here, \(\langle N_{\mathrm{ph}}\rangle_{\mathrm{mix}}\) represents the
Hamiltonian-limited, or spin-mixing-limited, optical cycling number. It should
therefore be regarded as an intrinsic upper bound rather than the experimentally
accessible cycling number, because ISC and other dynamical depolarization or
loss mechanisms are not included.

\begin{figure}[htbp]
    \centering
    \includegraphics[width=\linewidth]
    {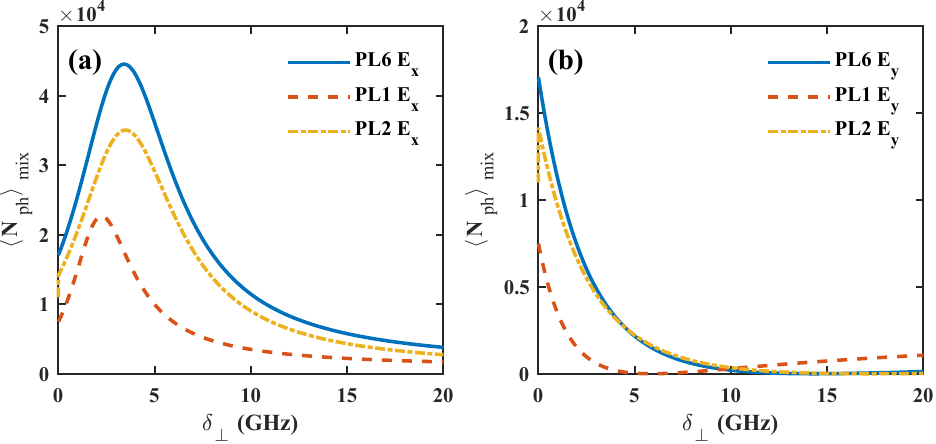}
    \caption{
    \textbf{Hamiltonian-limited optical cycling of PL1, PL2, and PL6.}
    (a) Mean spin-mixing-limited cycling photon number
    \(\langle N_{\mathrm{ph}}\rangle_{\mathrm{mix}}
    =P_0/(1-P_0)\) for the \(E_x\) transition.
    PL6 exhibits the largest intrinsic cycling number throughout the
    investigated strain range.
    The maximum calculated values are approximately
    \(4.45\times10^4\), \(3.51\times10^4\), and
    \(2.26\times10^4\) for PL6, PL2, and PL1, respectively.
    (b) Corresponding result for the \(E_y\) transition. The strong
    reductions of \(\langle N_{\mathrm{ph}}\rangle_{\mathrm{mix}}\)
    track the strain-dependent spin-mixing features shown in
    Fig.~S\ref{fig:S11}(b).
    These calculations include only excited-state Hamiltonian-induced spin
    mixing and therefore represent intrinsic upper bounds; ISC and other
    dynamical optical-cycle loss mechanisms are not included.
    }
    \label{fig:S12}
\end{figure}

Figure~S\ref{fig:S12}(a) shows
\(\langle N_{\mathrm{ph}}\rangle_{\mathrm{mix}}\) for the \(E_x\)
transitions of PL1, PL2, and PL6. Because
\(\langle N_{\mathrm{ph}}\rangle_{\mathrm{mix}}\) increases monotonically
with the spin purity, the ordering follows directly from Fig.~S\ref{fig:S11}(a):
PL6 exhibits the largest Hamiltonian-limited cycling number throughout the
investigated strain range, PL2 is intermediate, and PL1 gives the smallest
value. The maximum calculated values are approximately
\(4.45\times10^4\), \(3.51\times10^4\), and \(2.26\times10^4\) for
PL6, PL2, and PL1, respectively.

Notably, the maximum Hamiltonian-limited cycling photon number of
PL6 is approximately \(27\%\) larger than that of PL2, demonstrating
that the enhanced \(E_x\) spin purity of PL6 is also reflected in
its intrinsic optical cycling capability. This comparison provides
a direct quantitative benchmark of the excited-state spin-mixing
properties of the oxygen-vacancy center against the kk divacancy
under the same Hamiltonian-based framework.

Thus, even the relatively small differences in \(E_x\) spin purity
translate into substantial differences in the intrinsic number of
spin-conserving optical cycles.

The \(E_y\) branch exhibits a much stronger strain dependence
[Fig.~S\ref{fig:S12}(b)]. At low transverse strain, PL6 provides the largest
\(\langle N_{\mathrm{ph}}\rangle_{\mathrm{mix}}\), followed by PL2 and
PL1. As the strain approaches the respective spin-mixing regions, however,
the cycling number decreases rapidly. This reduction occurs first for PL1
near \(5.8~\mathrm{GHz}\), then for PL6 near \(14.3~\mathrm{GHz}\), and
finally for PL2 near \(17.2~\mathrm{GHz}\). Therefore, unlike the \(E_x\)
branch, no single defect maintains the largest \(E_y\) cycling number over
the entire \(0\)--\(20~\mathrm{GHz}\) interval. Instead, the principal
difference is the strain range over which a highly spin-conserving \(E_y\)
transition is preserved.

For a direct comparison under the experimentally relevant
conditions, we further evaluate the Hamiltonian-limited
cycling photon numbers at
\(\delta_\perp=0.688~\mathrm{GHz}\).
For the \(E_x\) transition, the calculated values are
approximately \(2.26\times10^4\), \(1.84\times10^4\),
and \(1.23\times10^4\) for PL6, PL2, and PL1,
respectively. For the \(E_y\) transition, the corresponding
values are approximately \(1.28\times10^4\),
\(1.09\times10^4\), and \(4.43\times10^3\).
Thus, at the same transverse strain, PL6 exhibits
approximately 1.23- and 1.18-fold higher \(E_x\) and
\(E_y\) cycling photon numbers than PL2, and
1.84- and 2.89-fold higher values than PL1,
respectively.

Taken together, the comparison demonstrates that PL6 combines the
largest intrinsic \(E_x\) cycling photon number among the three
investigated centers with a substantial displacement of the
\(E_y\) spin-mixing region relative to the structurally matched
PL1 divacancy. The PL2 comparison further establishes that the
enhanced \(E_x\) cycling capability of PL6 is not restricted to
the comparison with hh-DV, but is also evident relative to the
kk-DV configuration. These results highlight the favorable
excited-state spin properties associated with oxygen substitution
while retaining the configuration-dependent character of the
\(E_y\) response.

\subsection*{Experimentally accessible optical cycling}

The Hamiltonian-limited cycling numbers discussed above quantify the
intrinsic limit imposed solely by coherent spin mixing within the excited-state
manifold. They do not include additional dynamical spin-depolarization
processes that may occur during resonant optical excitation. We therefore
compare these intrinsic limits with the experimentally accessible optical
cycling inferred directly from the measured spin-flip dynamics and resonant
saturation fluorescence.

For the spin-flip measurements, the time-resolved fluorescence is recorded
with a 50-ns time bin. At each resonant excitation power, the fluorescence
count in the initial bin of the exponential spin-flip decay is normalized by
the number of experimental repetitions and the bin duration, yielding the
initial detected photon rate \(I_0(P)\). Its power dependence is described by
the saturation model

\[
{
I_0(P)=I_S\frac{P}{P+P_S},
}
\]

where \(I_S\) is the detected photon rate in the saturation limit. For the
\(E_x\) and \(E_y\) transitions, we obtain
\(I_S=425.3~\mathrm{kcps}\) and \(304.3~\mathrm{kcps}\), respectively,
as summarized in Supplementary Table~1.

The resonant fluorescence decays approximately exponentially as the spin
leaves the optically bright state,

\[
{
I(t)=I_0 e^{-\Gamma_{\rm flip}t},
}
\]
where \(\Gamma_{\rm flip}\) is the measured spin-depolarization rate.
Consequently, the mean number of detected photons within a spin-preserving
readout interval is given by the integrated fluorescence,

\[
{
\bar n_{\rm det}
=
\int_0^\infty I(t)\,dt
=
\frac{I_0}{\Gamma_{\rm flip}}.
}
\]

In the saturation limit, using
\(\Gamma_{\rm flip}=105~\mathrm{kHz}\) for \(E_x\) and
\(130~\mathrm{kHz}\) for \(E_y\), we obtain

\[
{
\bar n_{\rm det}^{\rm sat}
=
\frac{I_S}{\Gamma_{\rm flip}}
=
4.05 \quad (E_x),
\qquad
2.34 \quad (E_y).
}
\]

We further estimate the corresponding number of emitted photons using the
radiative decay rate determined from the lifetime measurements.
Using \(\Gamma_{\rm rad}=62.5~\mathrm{MHz}\), and assuming an excited-state
population of \(1/2\) in the saturation limit, the radiative photon-emission
rate is therefore

\[
{
R_{\rm em}^{\rm sat}
=
\frac{\Gamma_{\rm rad}}{2}
=
31.25~\mathrm{MHz}.
}
\]

Combining this rate with the measured spin-depolarization rates gives the
estimated mean number of emitted photons before spin depolarization,

\[
{
\bar n_{\rm em}^{\rm sat}
=
\frac{R_{\rm em}^{\rm sat}}{\Gamma_{\rm flip}}
=
298 \quad (E_x),
\qquad
240 \quad (E_y).
}
\]

These experimentally accessible cycling numbers are substantially smaller
than the intrinsic Hamiltonian-limited values. At the experimentally relevant
transverse strain of \(\delta_\perp=0.688~\mathrm{GHz}\), the excited-state
Hamiltonian gives
\(\langle N_{\rm ph}\rangle_{\rm mix}\approx2.25\times10^4\) for
\(E_x\) and \(1.28\times10^4\) for \(E_y\), whereas the measured
spin-depolarization dynamics correspond to only approximately \(3\times10^2\)
emitted photons before spin depolarization. The Hamiltonian calculation
nevertheless reproduces the observed branch dependence: \(E_x\) is predicted
to have a larger intrinsic cyclicity than \(E_y\), consistent with its smaller
measured spin-depolarization rate and larger detected photon number.

The difference in absolute magnitude shows that coherent excited-state spin
mixing is not the dominant limitation to optical cycling under the present
experimental conditions. Additional dynamical spin-depolarization channels,
which may include ISC and other processes not captured by the coherent
excited-state Hamiltonian, reduce the experimentally accessible cycling well
below the intrinsic spin-mixing limit. The Hamiltonian-derived
\(\langle N_{\mathrm{ph}}\rangle_{\mathrm{mix}}\) should therefore be
interpreted as an intrinsic upper bound, whereas
\(\bar{n}_{\mathrm{em}}^{\mathrm{sat}}\) and
\(\bar{n}_{\mathrm{det}}^{\mathrm{sat}}\) characterize the estimated emitted
optical cycling and experimentally detected fluorescence, respectively.

\clearpage
\section*{Supplementary Note 10. Additional spin-coherence measurements at 5.7 mT}

To complement the spin-coherence measurements performed at
\(B\approx17.1\ \mathrm{mT}\) in the main text (Fig.~4), we performed
Hahn-echo and dynamical-decoupling measurements on the same low-strain PL6
center (\(\delta_\perp=0.688\ \mathrm{GHz}\)) at a lower axial magnetic
field of \(B\approx5.7\ \mathrm{mT}\). The experimental conditions,
including the temperature of \(6.35\ \mathrm{K}\), resonant spin
initialization and readout, and pulse sequences, were otherwise unchanged.

\begin{figure}[htbp]
\centering
\includegraphics[scale=1]{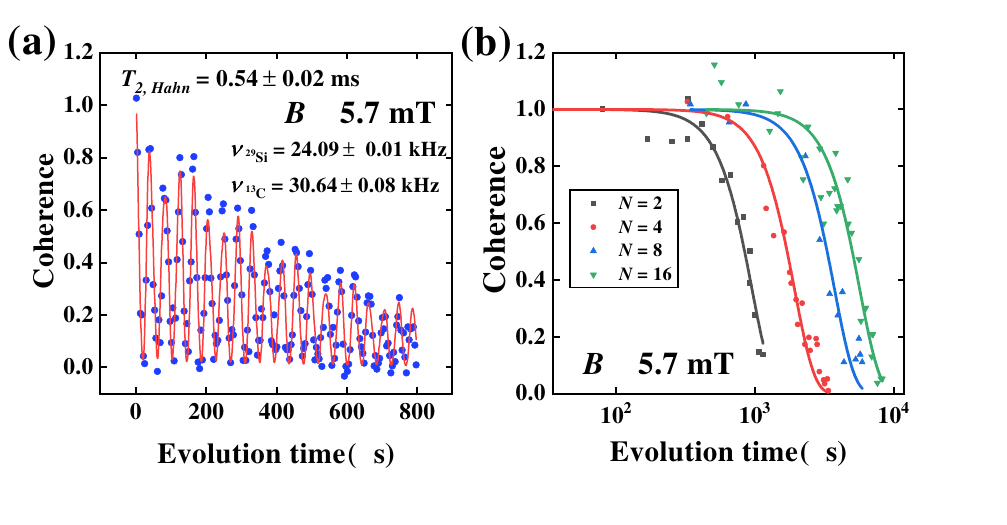}
\caption{
\textbf{Spin coherence and dynamical decoupling at 5.7 mT.}
(a) Hahn-echo decay measured at \(B\approx5.7\ \mathrm{mT}\), showing
ESEEM from the surrounding nuclear-spin bath. The solid curve is a fit
using Eq.~(\ref{eq:Hahn_fit}).
(b) Coherence decay curves measured using XY8 dynamical-decoupling
sequences with \(N=2,4,8,\) and \(16\) \(\pi\) pulses. Solid curves are
fits using Eq.~(\ref{eq:XY8_fit}).
}
\label{fig:spin_5p7}
\end{figure}

Figure~S\ref{fig:spin_5p7}(a) shows the Hahn-echo decay measured at
\(B\approx5.7\ \mathrm{mT}\). The Hahn-echo data are fitted using an
exponential decay modulated by electron-spin-echo envelope modulation
(ESEEM) arising from weakly coupled \(^{29}\mathrm{Si}\) and
\(^{13}\mathrm{C}\) nuclear spins \cite{koehl2011room},
\begin{equation}
\label{eq:Hahn_fit}
I_{\mathrm{Hahn}}(\tau)
=
A\exp\left(-\frac{\tau}{T_2}\right)
\left[
1-K_1\sin^2\left(\pi\nu_{^{29}\mathrm{Si}}\tau\right)
\right]
\left[
1-K_2\sin^2\left(\pi\nu_{^{13}\mathrm{C}}\tau\right)
\right],
\end{equation}
where \(A\) is the signal amplitude, \(T_2\) is the Hahn-echo coherence
time, \(K_1\) and \(K_2\) are the modulation depths, and
\(\nu_{^{29}\mathrm{Si}}\) and \(\nu_{^{13}\mathrm{C}}\) are the
modulation frequencies associated with \(^{29}\mathrm{Si}\) and
\(^{13}\mathrm{C}\), respectively. The same fitting function is used for
the Hahn-echo data measured at \(B\approx17.1\ \mathrm{mT}\) shown in
main-text Fig.~4(b).

The fit at \(B\approx5.7\ \mathrm{mT}\) yields
\(T_{2,\mathrm{Hahn}}=0.54\pm0.02\ \mathrm{ms}\), with modulation
frequencies
\(\nu_{^{29}\mathrm{Si}}=24.09\pm0.01\ \mathrm{kHz}\) and
\(\nu_{^{13}\mathrm{C}}=30.64\pm0.08\ \mathrm{kHz}\).
Their proportional reduction relative to the corresponding frequencies
measured at \(17.1\ \mathrm{mT}\) is consistent with the expected
magnetic-field scaling of the nuclear Larmor frequencies, further
supporting their assignment to \(^{29}\mathrm{Si}\) and
\(^{13}\mathrm{C}\) nuclear spins.

We further performed XY8 dynamical-decoupling measurements
\cite{de2008universal} at \(B\approx5.7\ \mathrm{mT}\) with
\(N=2,4,8,\) and \(16\) \(\pi\) pulses. The corresponding coherence
decay curves are shown in Fig.~S\ref{fig:spin_5p7}(b). For each pulse number, the decay
is fitted using a stretched-exponential function,
\begin{equation}
\label{eq:XY8_fit}
I_{\mathrm{XY8}}(t)
=
A\exp\left[
-\left(\frac{t}{T_2}\right)^n
\right],
\end{equation}
where \(A\) is the signal amplitude, \(T_2\) is the coherence time under
dynamical decoupling, and \(n\) is the stretching exponent. The same
stretched-exponential function is used to fit the XY8 coherence decays
measured at \(B\approx17.1\ \mathrm{mT}\) shown in main-text
Fig.~4(c).

The extracted coherence times \(T_2(N)\), together with those measured at
\(17.1\ \mathrm{mT}\), are included in the power-law scaling analysis in
main-text Fig.~4(d), where they are described by
\(T_2(N)=\alpha N^\beta\). The comparable Hahn-echo coherence times and
similar dynamical-decoupling scaling observed at the two magnetic fields
indicate that the coherence behavior is not governed solely by the
magnetic-field-dependent nuclear-spin bath. The measurements further show
that the dynamical-decoupling enhancement persists at both magnetic fields
investigated.

.